\documentclass[11pt]{article}
\usepackage{amssymb}
\usepackage{amsmath}
\usepackage{dsfont}
\usepackage{slashed}
\usepackage{mathrsfs}
\usepackage[utf8]{inputenc}
\usepackage[british]{babel}
\usepackage{lmodern}

\usepackage{appendix}
\usepackage{xcolor}
\definecolor{darkblue}{rgb}{0.1,0.1,.7}
\definecolor{darkgreen}{rgb}{0,.5,0}
\usepackage[pdftex,breaklinks,colorlinks,linkcolor=darkblue,citecolor=darkgreen]{hyperref}

\usepackage{longtable}
\usepackage{everypage}

\newcommand{\Lpagenumber}{\ifdim\textwidth=\linewidth\else\bgroup
  \dimendef\margin=0 
  \ifodd\value{page}\margin=\oddsidemargin
  \else\margin=\evensidemargin
  \fi
  \raisebox{\dimexpr -\topmargin-\headheight-\headsep-0.5\linewidth}[0pt][0pt]{%
    \rlap{\hspace{\dimexpr \margin+\textheight+\footskip}%
    \llap{\rotatebox{90}{\thepage}}}}%
\egroup\fi}
\AddEverypageHook{\Lpagenumber}%

\newenvironment{acknowledgments}{\vspace{12pt}\begin{center}
\textbf{Acknowledgments}\end{center}\vspace{-12pt}}{}
\newcommand{\ack}[1]{\begin{samepage}\begin{acknowledgments} {#1} \end{acknowledgments}\end{samepage}}

\usepackage{adjustbox}
\usepackage{makecell}
\usepackage{rotating}
\usepackage{pdflscape}
\usepackage{adjustbox}
\usepackage{tikz}
\usepackage{float}
\DeclareMathOperator{\tr}{tr}
\newcommand{\be}{\begin{equation}}\newcommand{\ee}{\end{equation}}
\newcommand{\nn}{\nonumber}
\newcommand{\vep}{\varepsilon}
\newcommand{\vphi}{\varphi}

\renewcommand{\S}{{\cal S}}
\newcommand{\pr}{\partial}

\newcommand{\ts}{\textstyle}
\newcommand{\hdelta}{{\hat \delta}}
\newcommand{\hd}{{\hat d}}
\newcommand{\ha}{{\hat a}}
\newcommand{\hb}{{\hat b}}
\newcommand{\hc}{{\hat c}}
\newcommand{\bG}{{\mathbb G}}

\renewcommand{\geq}{\geqslant}
\renewcommand{\ge}{\geqslant}

\renewcommand{\le}{\leqslant}

\begin{document}

\numberwithin{equation}{section}

\begin{titlepage}

\begin{flushright}
\small
LA-UR-20-27569\\
October 2020 \\
\normalsize
\end{flushright}

\vspace{1cm}
\begin{center}

{ \LARGE Heavy Handed Quest for Fixed Points in Multiple Coupling\\
\vspace{6pt}
Scalar Theories in the $\vep$ Expansion}
\end{center}

\vspace{1cm}
\begin{center}

{Hugh Osborn$^{a}$ and Andreas Stergiou$^{b}$}
\vskip 1cm

{$^a$Department of Applied Mathematics and Theoretical
Physics, Wilberforce Road,\\ Cambridge CB3 0WA, England\\
\vspace{3pt}
$^b$ Theoretical Division, MS B285, Los Alamos National Laboratory, \\ Los Alamos, NM 87545, USA
}
\end{center}

\begin{abstract}

{The tensorial equations for non trivial fully interacting fixed points at
lowest order in the $\vep$ expansion in $4-\vep$ and $3-\vep$ dimensions
are analysed for $N$-component fields and corresponding multi-index
couplings  $\lambda$ which are symmetric tensors with four or six indices.
Both analytic and numerical methods are used. For $N=5,6,7$  in the
four-index case large numbers of irrational fixed points  are found
numerically where  $||\lambda ||^2$ is close to the bound found by Rychkov
and Stergiou~\cite{RychkovS}. No solutions, other than those already known,
are found which saturate the bound.  These examples in general do not have
unique quadratic invariants in the fields. For $N \ge 6$ the stability
matrix in the full space of couplings always has negative eigenvalues.
In the six index case the  numerical search generates a very large number of solutions
for $N=5$.
}

\end{abstract}

\end{titlepage}
\pagenumbering{roman}
\newpage

\pagenumbering{arabic}

\setcounter{footnote}{0}

 \section{Introduction}

 There is a huge literature devoted to analysing fixed points using the $\vep$ expansion for a very
 large number of physical systems and determining their critical exponents.
 A review covering many of the cases that have appeared in the literature
 is found in \cite{RGrev} and
a wide range of known fixed points were also discussed by us with a
different perspective in \cite{Seeking}.  In many respects the
$\vep$ expansion is a universal solvent for understanding critical
phenomena and builds on and extends the historic analysis based on Landau
mean field theory. In practice it reduces to determining $\beta$-functions and
anomalous dimensions in a loop expansion based on Feynman graphs. Results
have  been recently extended to seven loops in \cite{Schnetz}, as applied
in \cite{Ryttov}, and with $O(N)$ symmetry to six \cite{SixLoop} which have
been extended to when the symmetry is reduced to $O(m) \times O(n)$
\cite{SixLoop2} and also cubic symmetry \cite{SixLoop3}.  Using
sophisticated resummation techniques to extend the $\vep$ expansion to
$\vep=1$ there is remarkable agreement with results obtained by the
bootstrap in three dimensions \cite{Bootstrap}, although some tension also
exists~\cite{Stergiou:2019dcv, Henriksson:2020fqi}.

The discovery of possible fixed points  in any $\vep$ expansion reduces to finding the zeros of the one loop $\beta$-functions in $4-\vep$
dimensions. Higher loops provide perturbative corrections but do not generally eliminate the fixed point except at possible
bifurcation points.
Here the notionally small parameter $\vep$ scales out and it is necessary to solve a tensorial quadratic
equation for the symmetric couplings $\lambda_{ijkl}$ where the indices range from $1$ to $N$.
The dimension of the space of symmetric 4-index couplings, $\tfrac{1}{24} N(N+1)(N+2)(N+3)$, increases rapidly
with $N$ and the determination of possible fixed points consequently becomes non trivial for quite low $N$ without
additional assumptions such as imposing a symmetry to reduce the number of independent couplings.
Of course for $N=1$ the fixed point equations  become totally trivial, giving rise to just the Ising fixed point.
For $N=2,3$ historic discussions are contained in \cite{Gufan,ZiaW2}
and a detailed analysis for $N=2$ is contained in \cite{Seeking}.  More recently a careful analysis of the $N=3$ case is
 contained in \cite{Safari} and  this has also been extended to $N=4$ in \cite{Codello4}.

There exist  several examples of fixed points which appear
for any $N$, most simply when there is $O(N)$ symmetry and just a single coupling, but there were until very recently no  complete general
results even when $N=3$.  Various theorems for fixed points and their stability properties have been obtained in
\cite{Michel,Michel3,MichelT, VicariZ}.
More recently a fundamental bound was proved by Rychkov and Stergiou  \cite{RychkovS} which takes the form,
after scaling out $\vep$ and the usual factors of $16\pi^2$ which arise in a loop expansion,
\be
S_N = || \lambda ||^2  = \lambda_{ijkl}\, \lambda_{ijkl} \le \tfrac18 N \, , \quad N \ge 4 \, .
\label{Sbound}
\ee
For $N=2,3$ there are stronger bounds. When the bound is saturated there is a bifurcation point and the stability matrix
develops a zero eigenvalue. Further bounds have been obtained by Hogervorst
and Toldo~\cite{Hogervorst, HogervorstToldo}
including tighter results for $S_N$ when $N=2,3$.
In terms of an $a$-function where  the lowest order $\beta$-functions are expressed as a gradient \cite{WallaceG,Seeking},
$\beta_{ijkl} = \pr A /\pr_{\lambda_{ijkl} } = -\vep \, \lambda_{ijkl} +{\rm O} (\lambda^2) $, then at a fixed point
\be
A_* = - \tfrac16\, S_N \, \vep^3 \, .
\ee
Away from a fixed point with a renormalization group (RG) flow given by
${\dot \lambda}_{ijkl} = - \beta_{ijkl}$ then ${\dot A}= - || \beta ||^2
<0$. For given $N$ an RG flow from $\text{CFT}_1$ to $\text{CFT}_2$ cannot
therefore exist unless $S_{N,\text{CFT}_2}>S_{N,\text{CFT}_1}$.

A related issue is the question of a lower bound for $S_N$. Of course for
the Gaussian  theory with $N$ free fields $S_N=0$.  For two decoupled
theories $S_N=S_{N_1}+S_{N_2}$ where $N=N_1 + N_2$ and $S_{N_1}, \,
S_{N_2}$ correspond to fixed points with $N_1, \, N_2$ fields. For $N$
decoupled Ising fixed points in our conventions $S_N= \frac19 N$ and
assuming any  perturbation of  the $N$ decoupled  Ising fixed point theory
so as to generate a fully interacting theory  decreases $A$ and hence
increases $S_N$, then at any resulting fixed point \be S_{N,{\rm fully\,
interacting}} > \tfrac19 N\, .  \ee However this is violated by $S_N$ for
the fully interacting $O(N)$ symmetric fixed point if $N\ge 10$ (as $N\to
\infty$ in this case $S_N\to 3$), which implies that for $N\ge 10$
there is no RG flow from the decoupled Ising to the $O(N)$ theory. Without
imposing any condition that the fixed point not contain decoupled free
theories \cite{ HogervorstToldo} obtained $S_N> \frac19$.

In previous literature the starting point has  usually been the determination of all quartic polynomials  in the scalar fields $\phi_i$
invariant under some
subgroup $H$ of $O(N)$ for particular $N$. The choice of $H$ depends on the particular physical system for which
critical exponents are to be found. For $N=4$ \cite{Michel2,Brezin2} and $N=6$ \cite{Hatch,Hatch1,Hatch2}
detailed investigations for all possible subgroups of $O(N)$,  the corresponding spaces of quartic polynomials and associated
fixed points was undertaken. Further analysis  for low $N$ in $4-\vep$ and $3-\vep$ dimensions based on the symmetry groups
for regular solids was described in \cite{Platonic}.
In these discussions the condition that there is a unique quadratic polynomial, which
may be taken as $\phi^2=\phi_i\phi_i$, is imposed. The fixed points found in this fashion are rational and generally have
rational critical exponents. An emphasis in these papers is whether the fixed points in a particular symmetry class  are
stable or not. A fixed point is stable if there are no marginally relevant quartic operators, or equivalently the eigenvalues
of the stability matrix, formed from the derivative of the $\beta$-function at the point where it vanishes, are all positive.
Although not entirely evident in  \cite{Michel2,Brezin2}  and \cite{Hatch,Hatch1,Hatch2} different symmetry groups
may lead to the same fixed point.

Fixed points may apparently have different couplings corresponding to restrictions to different bases of quartic polynomials
but if they are related by an $O(N)$ rotation or reflection they are equivalent. The fixed points  which are determined
by the zeros of the multi coupling $\beta$-functions may also correspond
to decoupled theories, reducible to fixed point theories, including free theories, with lower $N$. Although the fixed points
may be the same, the number of couplings and hence the dimension of the stability matrix may differ. This ensures that
the question as to whether a fixed point is stable depends on the symmetry group which is initially imposed. A particular
fixed point has identical values of $O(N)$ invariants such as $S_N$ in \eqref{Sbound} although equality of $S_N$ by itself
does not suffice to guarantee the same fixed point. This will be shown by particular examples later.

In this paper we follow an orthogonal approach by looking for solutions of
the basic equations numerically for low $N$ where $S_N$ is close to the bound
in \eqref{Sbound}. We focussed on such $S_N$ since we initially hoped to find fixed points
where the bound was saturated. Although no such cases were found there are generically many fixed
points wih increasing $N$ where $S_N$ is very close to $\tfrac18 N$.
Our numerical search used the optimization algorithm
\href{https://coin-or.github.io/Ipopt/}{\texttt{Ipopt}}~\cite{ipopt}
through \href{https://github.com/esa/pygmo2}{\texttt{pygmo}}, which
provides Python bindings of the C++ library
\href{https://github.com/esa/pagmo2/}{\texttt{pagmo}}~\cite{pagmo}.
\href{https://coin-or.github.io/Ipopt/}{\texttt{Ipopt}} can perform
constrained non linear optimization. The quantity $S_N$ was given as an
objective to \href{https://coin-or.github.io/Ipopt/}{\texttt{Ipopt}}, while
the $\beta$-function equations were given as constraints.
\href{https://coin-or.github.io/Ipopt/}{\texttt{Ipopt}} was then called
many times on a cluster, with a random initial point provided by
\href{https://github.com/esa/pagmo2/}{\texttt{pygmo}}. In many runs the
algorithm failed to find a solution, but feasible solutions were frequently
found too.  Since our problem is non convex, different solutions were
generally obtained in different runs. In our runs
\href{https://coin-or.github.io/Ipopt/}{\texttt{Ipopt}} entered the
so-called ``restoration phase''~\cite{ipopt}, in which the objective
function $S_N$ was ignored and only the $\beta$-function equation violations
were minimized.  Thus, for a feasible solution, $S_N$ was not attempted to be
maximised by \href{https://coin-or.github.io/Ipopt/}{\texttt{Ipopt}}. Only
the $\beta$-function equations were numerically satisfied, with a tolerance
we chose to be $10^{-10}$. We subsequently improved the solutions obtained
using \href{https://www.sympy.org/en/index.html}{\texttt{SymPy}}'s
\href{https://docs.sympy.org/latest/modules/solvers/solvers.html}{\texttt{nsolve}}
or \emph{Mathematica}'s
\href{https://reference.wolfram.com/language/ref/FindRoot.html}{\texttt{FindRoot}},
in some cases to a tolerance of $10^{-200}$. We note here that it is not
guaranteed that our method will find all possible solutions.

This method reproduces the known fixed points where $S_N$ is rational but
also produces many irrational fixed points. Of course it is necessary to
isolate those which are decoupled theories.  For decoupled interacting
fixed points the  stability matrix has the eigenvalue 1 with degeneracy
greater than one and for a free theory there are eigenvalues $-1$.
Although our search is restricted to fixed points with $S_N$ close to the
bound, we believe that it is possible to find all non trivial fixed points
for $N=3,4$ and perhaps $N=5,6$. In the space of all quartic couplings the
stability matrix has negative eigenvalues  always when $N\ge 6$.

The irrational fixed points generally correspond to theories with two or
more quadratic invariants.  Such cases are not commonly considered but have
recently been found to have relevance in discussions of conformal field
theories (CFTs) at non zero
temperature \cite{Komar}. In many cases we are able to match the fixed
points found numerically with results for fixed points in so called
biconical theories, or various generalisations thereof. In the simplest
case two theories which are separately $O(n), \, O(m)$ invariant are linked
by a product of two singlet quadratic operators so the symmetry
$O(n)\times O(m)$ is preserved. The fixed points arising from RG flow
starting from the decoupled theory perturbed by the product of quadratic
operators include one with the  maximal  $O(n+m)$ symmetry, but for $n\ne
m$, and with suitable restrictions on $n,\, m$,   they also lead to
irrational fixed points with two quadratic invariants. The generalisations
discussed here allow for several quadratic invariants.  Such irrational
fixed points were recognised long ago \cite{Biconical}, for other
literature see \cite{VicariB,Folk1,Folk2,Folk3}.

In this paper in the next section we discuss the general features of the
lowest order equations for the couplings $\lambda_{ijkl}$ and their
decomposition under $O(N)$ and introduce $O(N)$ invariants $a_0,a_2,a_4$.
Along with $S_N$ in \eqref{Sbound} these serve to characterise different
fixed points in an invariant fashion.  Some bounds are obtained together
with properties of stability matrix eigenvalues.  In section \ref{sec:anyN}
the fixed point equations are solved analytically for general $N$ assuming
various symmetries, in particular the case when the symmetry is $\S_N$, the
permutation group of $N$ objects. This includes previously known examples
with cubic and tetrahedral symmetry. In section \ref{sec:doubletrace} fixed
points which arise when two decoupled theories are perturbed by what can be
regarded as double trace operators are described. These include so called
biconical fixed points. They typically generate irrational fixed points
although in some cases rational ones as well. Detailed results for all $N$
up to $N=7$ are presented in section \ref{sec:lowN}. Where possible,
numerical results are related to the analytic discussion earlier.  Irrational fixed points for $N=4$
were found previously in \cite{Codello4}. This case is special in that fixed points which are
degenerate at lowest order split at higher orders in $\vep$.
A similar discussion relating to the fixed point solutions for
$\lambda_{ijklmn}$ is undertaken in section \ref{sec:sixindex}.

\section{General \texorpdfstring{$N$}{N} Fixed Points}

The basic algebraic equation determining fixed points at lowest order in the $\vep$ expansion in $4-\vep$ dimensions is
\be
\lambda_{ijkl} = \lambda_{ijmn} \, \lambda_{klmn} + \lambda_{ikmn} \, \lambda_{jlmn} + \lambda_{ilmn} \, \lambda_{jkmn}
\equiv S_{3,ijkl} \, \lambda_{ijmn} \, \lambda_{klmn}  \, \, ,
\label{RG1}
\ee
for $\lambda_{ijkl} $ the symmetric tensor  determining the renormalisable couplings and we assume  $i,j,k,l= 1, \dots, N$.
For any tensor $X_{ijkl}$, the action of the symmetriser $S_n$ is defined such  $S_{n,ijkl}\,  X_{ijkl}$
denotes the sum over  the $n$ terms, with unit weight, which are the minimal necessary to ensure the sum is a fully symmetric four index tensor,
if $X_{ijkl}$ is invariant under a subgroup $G_X \subseteq\S_4$ then $n= 24/|G_X|$.
With the potential
\be
V(\phi) = \tfrac{1}{24} \, \lambda_{ijkl} \, \phi_i \phi_j \phi_k \phi_l \, ,
\ee
\eqref{RG1} can be expressed more succinctly as
\be
V(\phi) = \tfrac12 \, V_{ij}(\phi) V_{ij}(\phi)\, , \qquad V_{ij}(\phi) = \pr_i \pr_j V(\phi) \, .
\label{RG2}
\ee

The fixed point equation \eqref{RG1}  is manifestly covariant under $O(N)$ and for any solution $\lambda_{ijkl}$ equivalent solutions are
acting on $\lambda_{ijkl}$ with $O(N)$ rotations.
Up to quadratic order there are $O(N)$  invariants in addition to $S_N$ in \eqref{Sbound} given by
\be
a_0 = \lambda_{iijj} \, , \qquad a_1 = \lambda_{ijkk}\, \lambda_{ijll} \,.
\ee
The symmetric coupling $\lambda_{ijkl}$ transforms as a four index tensor  can be decomposed into the three possible $O(N)$
irreducible representations by writing
\begin{align}
\lambda_{ijkl} = {}& d_0 \, S_{3,ijkl} \, \delta_{ij}\, \delta_{kl}  + S_{6,ijkl} \, \delta_{ij} \, d_{2,kl}
 + d_{4,ijkl} \, ,
\label{ldec}
\end{align}
where $d_{2,ij}$ and $d_{4,ijkl}$ are symmetric and traceless rank two and rank four tensors; $d_{2,ij}$ is determined by
\be
(N+4)\, d_{2,ij} = \lambda_{ijkk} - (N+2) \, d_0 \, \delta_{ij} \, ,
\ee
and then $d_{4,ijkl} $ is given in terms of $\lambda_{ijkl} $ by subtraction of the $d_0, \, d_2$ pieces.
Following Hogervorst and Toldo \cite{Hogervorst, HogervorstToldo}
\begin{align}
a_0 ={}& N (N+2) d_0 \, , \qquad a_2 = (N+4)^2  || d_2 ||^2  = a_1 - \tfrac{1}{N} \, a_0{\!}^2 \, , \nn \\
a_4 = {}& || d_4 ||^2  = S_N - \tfrac{6}{N+4} \, a_2 -  \tfrac{3}{N(N+2)}\,  a_0{\!}^2 \, .
\label{adef}
\end{align}
If $\alpha_r$ are the eigenvalues of $\lambda_{ijkk}$  then $a_2 = \sum_{r<s} (\alpha_r -\alpha_s)^2/N $ and crucially
it follows that
\be
a_2 = 0 \quad \Leftrightarrow  \quad \lambda_{ijkk} = \alpha \, \delta_{ij} \, , \quad \alpha= \tfrac{1}{N} \, a_0\, .
\label{a2alpha}
\ee

For decoupled theories $\lambda_{1\cup2,ijkl} = \lambda_{1,ijkl} + \lambda_{2,ijkl} $ where $\lambda_1 \cdot \lambda_2 = 0$.
In this case $a_0,\, a_1, \, S_N$ are additive though $a_2, \, a_4$ are not,
\be
a_{2,1\cup 2} = a_{2,1}+a_{2,2} + \frac{1}{N_1N_2 \, N} \big ( N_2 \, a_{0,1} - N_1 \, a_{0,2} \big )^2 \, , \quad N=N_1 + N_2 \, .
\label{a2}
\ee

Other $O(N)$ invariants  are given by the $N$  eigenvalues  $\{\gamma\}$  of $[\Gamma_{ij}]$ where
\be
\Gamma_{ij} = \lambda_{iklm} \, \lambda_{jklm} \, ,   \qquad \Gamma_{ij} v_j = \gamma\, v_i \, , \quad {\ts \sum} \gamma = S_N \, .
\label{defGam}
\ee
These determine the  lowest order anomalous dimensions for the $N$-component fields $\phi_i$ at the fixed point which are
$\tfrac{1}{12} \gamma \, \vep$.
If  \eqref{a2alpha} is satisfied then from \eqref{RG1}
\be
2 \,\Gamma_{ij}  = (\alpha - \alpha^2) \,\delta_{ij} \, .
\label{Galpha}
\ee
Conversely for $a_2$ non zero $\Gamma_{ij}$ should  not  be proportional to $\delta_{ij}$ and
there are necessarily different eigenvalues $\gamma$. For all eigenvalues $\gamma$ equal then
since $[ \Gamma_{ij} ]$ is a positive matrix it is necessary that $0 \le \gamma \le \tfrac18$; saturating the upper bound is
equivalent to \eqref{Sbound}.

Further invariants are  also given by the $\frac{1}{24}N(N+1)(N+2)(N+3)$ eigenvalues $\{ \kappa \}$ of the stability matrix
at the fixed point which are determined  by
\begin{align}
(\kappa+1) \, v_{ijkl} =  S_{6,ijkl} \, \lambda_{ijmn} \, v_{klmn}
={}& \lambda_{ijmn} \, v_{klmn} + \lambda_{ikmn} \, v_{jlmn} + \lambda_{ilmn} \, v_{jkmn}\nn \\
&{} + \lambda_{klmn} \, v_{ijmn} + \lambda_{jlmn} \, v_{ikmn} + \lambda_{jkmn} \, v_{ilmn} \, ,
\label{keig}
\end{align}
for $v_{ijkl}$ symmetric. In this case
\be
{\ts \sum} \, \kappa = \tfrac12 ( N+2)(N+3) a_0 - \tfrac{1}{24} N (N+1)(N+2)(N+3) \, .
\label{ksum}
\ee
Furthermore
\be
{\ts \sum} \, (\kappa + 1 )^2   = \tfrac12 ( N+4)(N+5) S_N  + 4 (N+4)  a_1 +  a_0{\!}^2 \, .
\label{ksum2}
\ee
A solution is always obtained by taking $v_{ijkl} \to \lambda_{ijkl}$ giving,  by virtue of \eqref{RG1}, $\kappa=1$.
This is in general non degenerate except for decoupled theories. Directly from \eqref{keig} and \eqref{RG1}
\be
( 1 -\kappa ) \lambda_{ijkl} v_{ijkl}= 0 \, .
\ee
Any non zero
\be
v_{ijkl} = \omega_{ir} \, \lambda_{rjkl} +  \omega_{jr} \, \lambda_{irkl} +  \omega_{kr} \, \lambda_{ijrl} +  \omega_{kr} \, \lambda_{ijkr} \, , \quad
\omega_{ij} = - \omega_{ji} \, ,
\label{vomega}
\ee
gives solutions of \eqref{keig} with $\kappa=0$. In general $\omega_{ij}$ corresponds to elements of the Lie algebra of
$O(N)$. The number of zero modes \eqref{vomega} is given by $ \tfrac12 N (N-1) -  \dim \mathfrak {H}$ where $\mathfrak {H}$
is the Lie algebra of the unbroken subgroup $H\subset O(N)$ which leaves $\lambda_{ijkl}$ invariant.

The eigenvalue equations may be extended to $\phi^2$ and $\phi^3$ operators where the anomalous dimensions are determined by
\be\
\mu \, v_{ij} = \lambda_{ijkl} \, v_{kl} \, , \qquad  \nu \, v_{ijk} =
\lambda_{ijlm} \, v_{klm} +  \lambda_{jklm} \, v_{ilm} +  \lambda_{kilm} \, v_{jlm}  \, ,
\label{eigmn}
\ee
where
\be
{\ts \sum} \, \mu = a_0 \, , \qquad {\ts \sum} \, \nu =  (N+2)  a_0  \, .
\ee
A solution of the eigenvalue equation for $\nu$ is obtained with $\nu =1$ by taking $v_{ijk} \to \lambda_{ijkl} v_l$, for any vector $v_l$,
so this is $N$-fold degenerate.

For decoupled theories 1 and 2
\begin{align}
\{ \kappa \}_{1\cup 2} = {}& \{ \kappa \}_1 \cup \{ \kappa \}_2 \cup \big (  \{ \nu \}_1 {\mathds 1}_{N_2} - 1 \big  )
\cup \big (  {\mathds 1}_{N_1}  \{ \nu \}_2 - 1 \big  ) \nn \\
&{} \cup \big (  \{ \mu \}_1 {\mathds 1}_{\frac12 N_2(N_2+1) } +  {\mathds 1}_{\frac12 N_1(N_1+1) }  \{ \mu \}_2 - 1 \big  ) \, ,
\end{align}
where we have used the result that the anomalous dimension of $\phi$ is zero at lowest order. This satisfies, by using \eqref{ksum} and
\eqref{ksum2},
\begin{align}
{\ts \sum_{1\cup 2} } \, \kappa =  {}& \big ( \tfrac12 (N_1+2)(N_1+3) + (N_1 + 1)N_2  + \tfrac12 N_2(N_2+1) \big ) a_{0,1}
+  1 \leftrightarrow 2\nn \\
&{}- \big ( \tfrac{1}{24} N_1(N_1+1)(N_1+2) (N_1+3)+ \tfrac16 N_1(N_1+1) (N_1 + 2) N_2 +  1 \leftrightarrow 2 \big ) \nn \\
&{}- \tfrac14 N_1(N_1+1) N_2 (N_2+1) \, ,
\end{align}
which satisfies \eqref{ksum} with $N=N_1+N_2$ since $a_0= a_{0,1}+
a_{0,2}$.  For the $n$-fold free theory $\kappa = -1(\frac{1}{24}
n(n+1)(n+2)(n+3)), \ \nu = 0 (\frac16 n(n+1)(n+2)), \ \mu = 0 ( \frac12
n(n+1))$.\footnote{Here and below we use the notation
eigenvalue(degeneracy).} As a consequence of the results for eigenvalues
with $\nu =1$ there will be  $2 N_1 N_2 $ zero eigenvalues for $\{ \kappa
\}_{1\cup 2} $,  in addition to those arising from $ \{ \kappa \}_1$ or $
\{ \kappa \}_2 $,  for two decoupled  interacting theories. For a decoupled fixed point formed by $n$-fold
free fields and a $N$-component interacting fixed point $FP_N$ there are $N n$ extra zero modes besides
those present in $FP_N$.
For $n$ decoupled interacting theories $\{ \kappa \}_{1\cup 2 \cup\dots \cup n} $
contains $2 \sum_{1 \le i < j \le n}  N_i N_j $ extra zero modes.

\subsection{Decomposition of Fixed Point Equation}

Applying \eqref{ldec} in \eqref{RG1} gives rise to three separate equations
 \begin{subequations}
\begin{align} \hskip - 1cm
 d_0 = {}& (N+8)\,  d_0{}^2 + \tfrac{(N+4)(N+16)}{N(N+2)} \, || d_2 ||^2 +  \tfrac{2}{N(N+2)} \, || d_4 ||^2 \, , \label{RG3a}\\
\noalign{\vskip 2pt}
d_{2,ij} ={}& (N+16) \, d_0 \, d_{2,ij} + \tfrac{2 }{N+4}(5N+32) \, \big ( d_{2,ik} \, d_{2,jk} - \tfrac1N \, \delta_{ij} \, || d_2 ||^2  \big ) \nn \\
&{}+ \tfrac{1}{N+4} (N+16) \, d_{4,ijkl}\,  d_{2,kl} +  \tfrac{2}{N+4} \,  \big ( d_{4,iklm}\,  d_{4,jklm} - \tfrac1N \, \delta_{ij} \, || d_4 ||^2  \big ) \, ,  \label{RG3b}  \\
\noalign{\vskip 2pt}
d_{4,ijkl} = {}&  12 \, d_0 \, d_{4,ijkl} \nn \\
&{}+ (N+ 16) \, S_{3,ijkl} \big ( d_{2,ij} \, d_{2,kl} - \tfrac{2}{N+4} ( \delta_{ij} \, d_{2,km} \, d_{2,lm} + \delta_{kl} \, d_{2,im} \, d_{2,jm} ) \nn \\
\noalign{\vskip -2pt}
& \hskip 6.8 cm {} +  \tfrac{2}{(N+2)(N+4)} \, \delta_{ij} \, \delta_{kl} \, || d_2 ||^2 \big ) \nn \\
&{}+     S_{3,ijkl} \big ( d_{4,ijmn}\,  d_{4,klmn} - \tfrac{2}{N+4} ( \delta_{ij} \, d_{4,kmnp}\,  d_{4,lmnp} + \delta_{kl} \, d_{4,imnp} \, d_{4,jmnp} ) \nn \\
\noalign{\vskip -2pt}
& \hskip 7.5 cm {} +  \tfrac{2}{(N+2)(N+4)} \, \delta_{ij} \, \delta_{kl} \, || d_4 ||^2 \big ) \nn \\
&{}+ 6 \big ( S_{4,ijkl} \, d_{4,ijkm}\, d_{2,lm} - \tfrac{2}{N+4} \, S_{6,ijkl} \, d_{4,ijmn} \, d_{2,mn} \, \delta_{kl} \big )  \, .
\label{RG3c}
\end{align}
 \end{subequations}
From \eqref{RG1} or  \eqref{RG3a}
\be
a_0 = a_1 + 2 \, S_N \, , \qquad a_0 \big ( 1 - \tfrac{N+8}{N(N+2)\,} a_0 \big )  = 2 \, a_4  + \tfrac{N+16}{N+4}\, a_2 \, .
\label{Sfp}
\ee
Hence using  \eqref{adef}, \eqref{Sfp} we obtain the bounds
\be
 a_0 < \tfrac{N(N+2)}{N+8} \, , \quad S_N + \tfrac12 \, a_2 =  \tfrac18 N - \tfrac{1}{2N} \big ( a_0 - \tfrac12 N \big )^2   \le
 \begin{cases}\tfrac18 N \, ,  \ \ &N\ge 4 \\  \tfrac{3N(N+2)}{(N+8)^2} \,  , & N<4 \end{cases} \, .
 \label{bounds}
 \ee
 This is just the Rychkov--Stergiou bound  \cite{RychkovS} as extended by
 Hogervorst--Toldo \cite{Hogervorst, HogervorstToldo}; the Rychkov--Stergiou
 case arises for $N\ge 4$ when $a_2=0$.
 This bound is saturated for $a_0 = \tfrac12 N$, which is necessary for the sum of two decoupled theories saturating the
 bound to also satisfy the bound since \eqref{a2} requires  $N_2 \, a_{0,1} - N_1 \, a_{0,2} =0$. The bound on  $a_0$ is saturated
 for the $O(N)$ symmetric theory. With this bound on $a_0$ then
 \eqref{ksum} requires $\sum \kappa < 0 $ when $N\ge 6$. If \eqref{Sbound} is saturated then necessarily $a_2=0$ and from
 \eqref{a2alpha} and \eqref{Galpha}  then $\Gamma_{ij} = \tfrac18\, \delta_{ij}$.

 The fixed point equations can be reduced for any basis of 4 and 2 index symmetric traceless tensors
 $\{ d_{r,ijkl}, \, \hd_{a,ij}\} $, $r=1,\dots ,p, \ a= 1,\dots ,q$,  satisfying
 \begin{subequations}
 \begin{align}
&  \tfrac12 S_{6,ijkl} \,d_{r,ijmn}\,  d_{s,klmn} =  a_{rs}\, S_{3,ijkl} \, \delta_{ij}\delta_{kl} +
   {\ts \sum_t}\, b_{rs}{}^t d_{t,ijkl} + {\ts \sum_b}\,  c_{rs}{}^a\,  S_{6,ijkl} \, \delta_{ij}  \hd_{a,kl} \, ,
   \label{dda} \\
&   S_{4,ijkl} \, d_{r,ijkm}\, \hd_{a,lm} =   {\ts \sum_b}\, e_{ra}{}^b\,  S_{6,ijkl} \, \delta_{ij}  \hd_{b,kl} +  {\ts \sum_s}\, f_{ra}{}^s \,  d_{s,ijkl} \, ,\label{ddb}  \\
&  \tfrac12 S_{6,ijkl}\,   \hd_{a,ij} \, \hd_{b,kl} = \tfrac{2}{N+4}\,
  {\hat  a}_{ab}\, S_{3,ijkl} \, \delta_{ij}\delta_{kl} + {\ts \sum_r}\,  {\hat b}_{ab}{}^r d_{r,ijkl} +
   {\ts \sum_c}\, {\hat c}_{ab}{}^c\,  S_{6,ijkl} \, \delta_{ij}  \hd_{c,kl} \, , \label{ddc}
     \end{align}
\end{subequations}
where it is necessary that $a_{rs}, \,  b_{rs}{}^t ,  \, c_{rs}{}^a$ are symmetric under $r \leftrightarrow s$  and similarly
$ {\hat  a}_{ab}, \, {\hat b}_{ab}{}^r , \,  {\hat c}_{ab}{}^c$ for $a\leftrightarrow b$.
 Consistency between \eqref{dda}, \eqref{ddb} and \eqref{ddc} requires
\begin{align}
& {\ts \sum_u}\, b_{rs}{}^u a_{ut} = b_{rst} = b_{(rst)} \, , \quad   {\ts \sum_d}\,  {\hat c}_{ab}{}^d {\hat a}_{dc} = {\hat c}_{abc}=  {\hat c}_{(abc)}  \, , \nn \\
&{\hat b}_{ab\hskip 0.5pt r} = {\ts \sum_s} \, {\hat b}_{ab}{}^s{} a_{sr} = 3 \, e_{ra}{}^c {\hat a}_{cb} =  3 \, e_{r\hskip 0.5pt ab}
\, , \quad 4 \, c_{rs\hskip 0.5pt a }= 4 \, {\ts \sum_b}\, c_{rs}{}^b \,{\hat a}_{ba} = f_{ra}{}^t a_{ts} =  f_{ra\hskip 0.5pt  s} \, .
\label{symbbc}
\end{align}
From \eqref{ddc}
\be
(\hd_{a} \vee_2 \hd_{b})=  \tfrac12 S_{2,ij}\,   \hd_{a,ik} \, \hd_{b,jk} = \tfrac{N+2}{N+4}\,  {\hat  a}_{ab}\, \delta_{ij} + \tfrac12(N+4) \,
  {\ts \sum_c}\, {\hat c}_{ab}{}^c\,  \hd_{c,ij}  \, ,
  \label{d2d2}
 \ee
 The non associative algebra of symmetric matrices defined the symmetric product $\vee_2$  is a Jordan algebra since, for $U,V$
 symmetric, $U\vee_2 ( V\vee_2(U\vee_2 U)) = ( U\vee_2  V) \vee_2(U\vee_2 U)$. If
 \eqref{d2d2} is reducible to the form
\be
{\hat d}_{a,ik} \, {\hat d}_{b,jk} = \delta_{ab} \big ( \delta_{ij} + {\hat c}_a \, {\hat d}_{a,ij} \big )  \, .
\ee
then all solutions are expressible in terms of projection operators with
\be
{\hat c}_a  = \frac{n_a-m_a}{\sqrt{ n_a \, m_a}} \, , \qquad n_a + m_a = N\, .
\ee

If \eqref{ddc}  is regarded as defining ${\ts \sum_r}\,  {\hat b}_{ab}{}^r d_{r,ijkl}$ then inserting in \eqref{ddb} and using \eqref{d2d2} gives
\begin{align}
{\ts \sum_s} \, {\hat b}_{ab}{}^s \, f_{sc}{}^r = {}&  {\ts \sum_d} \big (2 (N+4)  \, {\hat c}_{c(a}{}^d \, {\hat b}_{b)d}{}^r   -
4\,  {\hat c}_{ab}{}^d \, {\hat b}_{dc}{}^r \big ) \, , \nn \\
\noalign{\vskip 4pt}
 {\ts \sum_r} \, {\hat b}_{ab}{}^r \,{e}_{r \hskip 0.1pt c}{}^d  = {}&  \tfrac{1}{N+4}  \big (2 (N+2) \,  \delta_{(a}{}^d \, \ha_{b)c}\,
 - 4\,  \delta_c{}^d \, {\hat a}_{ab}  \big )  \nn \\
&  {} +  {\ts \sum_e} \big ( 2 (N+4)  \, {\hat c}_{c(a}{}^e \, {\hat c}_{b)e}{}^d  - (N+8) \,  {\hat c}_{ab}{}^e\, {\hat c}_{ec}{}^{d} \big ) \, .
\end{align}
Using \eqref{symbbc}
\begin{align}
{\ts \sum_{r,s}} \, \hb_{ab}{}^r \,\hb_{cd} {}^s \, a_{rs}{}= {}& 3\, \tfrac{N+2}{N+4} ( \ha_{ac}\, \ha_{bd} + \ha_{ad}\, \ha_{bc} ) -  \tfrac{12}{N+4} \,
\ha_{ab}\, \ha_{cd} \nn \\
&{}+ 3(N+4) \big ( \hc_{ac}{}^e \, \hc_{ebd} +  \hc_{bc}{}^e \, \hc_{ead} \big ) - 3(N+8) \,  \hc_{ab}{}^e \, \hc_{ecd} \, .
\end{align}

Writing
\be
d_0=\lambda \, , \qquad d_{2,ij} = {\ts \sum_a} \, h^a \, \hd_{a,ij} \, , \qquad  d_{4,ijkl} = {\ts \sum_r} \, g^r \, d_{r,ijkl} \, ,
\ee
so that
\be
|| d_2 ||^2 = \tfrac{1}{N+4} N ( N+2)\,  {\ts \sum_{ab}}\, {\hat a}_{ab}h^a h^b  \, , \qquad
|| d_4||^2 = \tfrac12 N ( N+2) \, {\ts \sum_{rs}} \, a_{rs} g^r g^s   \, ,
\ee
the lowest order fixed point equations for the $p+q+1$ couplings $\lambda, \, g^r, \, h^a$  from \eqref{RG3a}, \eqref{RG3b}, \eqref{RG3c} are then
\begin{align}
\lambda ={} & (N+8) \lambda^2 +  {\ts \sum_{r,s}} \, a_{rs}\,  g^r g^s   + (N+16) \, {\ts \sum_{a,b}}\, {\hat a}_{ab}\, h^a h^b   \, , \nn \\
g^r = {}& 12 \, \lambda\, g^r + {\ts \sum_{s,t}} \, b_{st}{}^r g^s g^t + (N+16) \,  {\ts \sum_{a,b}} \, {\hat b}_{ab}{}^r h^a h^b
+ 6 \, {\ts \sum_{s,a}} \, f_{sa}{}^r \, g^s h^a \, , \nn \\
h^a ={}& (N+16) \big ( \lambda\, h^a + \tfrac12 \, {\ts \sum_{r,b}} \, e_{r b}{}^a \, h^b g^r \big )
 + (5N+32) \, {\ts \sum_{b,c}} \, {\hat c}_{bc}{}^a h^b h^c + {\ts \sum_{r,s}} \, c_{rs}{}^a g^r g^s \, ,
 \label{RG4}
\end{align}
where  $f_{sa}{}^r, \,   e_{r b}{}^a$ can be eliminated using \eqref{symbbc}.

 \subsection{Further Bounds}

 A general analysis of \eqref{RG3a}, \eqref{RG3b}, \eqref{RG3c} is not straightforward. In general $d_0
 \le \frac{1}{N+8}$, which is equivalent to the $a_0$ bounds
 in \eqref{bounds}.
 If \eqref{Sbound} is to be saturated, then $d_{2,ij}=0$ and  the equations simplify to
 \begin{align}
&  d_0 =  (N+8)\,  d_0{}^2 + \tfrac { 2}{N(N+2)} \, || d_4 ||^2 \, ,  \quad
 d_{4,iklm}\,  d_{4,jklm} = \tfrac1N \, \delta_{ij} \, || d_4 ||^2  \, , \nn \\
\noalign{\vskip 2pt}
& ( 1- 12 \, d_0 )\,  d_{4,ijkl} =  S_{3,ijkl} \big ( d_{4,ijmn}\,  d_{4,klmn} -   \tfrac{2}{N(N+2)} \, \delta_{ij} \, \delta_{kl} \, || d_4 ||^2 \big )  \, .
\end{align}
The last  two equations are equivalent to writing
\be
 d_{4,ijmn} \, d_{4, klmn} = \tfrac{1}{N-1} \, a
\big ( \tfrac12 N (\delta_{ik} \delta_{jl}   +  \delta_{il} \delta_{jk})   - \delta_{ij} \delta_{kl}  \big )
+\tfrac13\,  b \, d_{4,ijkl} + w_{ijkl}\, ,
\label{ddw}
\ee
where
\begin{align}
 w_{ijkl} = {}& \tfrac13\big  (2 \, d_{4,ijmn} \, d_{4,klmn} - d_{4,ikmn} \, d_{4,jlmn} - d_{4,ilmn} \, d_{4,jkmn}\big  )\nn \\
&{} + \tfrac{1}{3N(N-1)}\, || d_4 ||^2 \big  ( 2 \, \delta_{ij} \delta_{kl} - \delta_{ik} \delta_{jl} - \delta_{il} \delta_{jk} \big ) \, , \nn \\
& a = \tfrac{2} { N(N+2)} \,   || d_4 ||^2  \, , \qquad b =  1- 12 \, d_0 \, .
\label{wdef}
\end{align}
$w_{ijkl}$ is a mixed symmetry tensor satisfying $w_{ijkl} = w_{(ij)kl} = w_{klij}$, $w_{i(jkl)}=0$  and $w_{ijkk}=0$. With this definition
for $w_{ijkl}$
\be
|| w||^2 = w_{ijkl}w_{ijkl} = \tfrac23 \, d_{4,ijmn} d_{4,klmn}\big (  d_{4,ijpq} d_{4,klpq} -  d_{4,ikpq} d_{4,jlpq}\big )
- \tfrac{2}{3N(N-1)} \, || d_4 ||^4 \,  .
\ee
Substituting \eqref{ddw} ensures this is just an identity.

Higher rank tensors can be formed from $d_{4,ijkl}$. There is a symmetric 6-index tensor
\be
S_{6,ijklmn} = d_{(ijk|p} \,d_{lmn)p} - \frac{3\, b}{N+8} \, \delta_{(ij} \, d_{klmn)}
- \frac{3\, a}{N+4} \, \delta_{(ij} \, \delta_{kl}\, \delta_{mn)} \, ,
\ee
and a corresponding  mixed symmetry symmetry tensor
\begin{align}
M_{6,ijklmn} = {}& d_{ij(k|p} \,d_{lmn)p} - d_{i(kl|p} \,d_{mn)jp} \nn \\
\noalign{\vskip -3pt}
&{}+ \frac{b}{3(N-2)} \big (  \delta_{ij} \, d_{klmn} -  \delta_{i(k} \, d_{lmn)j} -  \delta_{j(k} \, d_{lmn)i}
+  \delta_{(kl} \, d_{mn)ij} \big ) \nn \\
\noalign{\vskip -3pt}
&{} + \frac{a}{N-1} \big ( \delta_{ij} \, \delta_{(kl}\, \delta_{mn)} - \delta_{i(k} \, \delta_{lm}\, \delta_{n)j} \big )
- \frac{5}{N+4}  \, w_{ij(kl}\, \delta_{mn)}  \, .
\end{align}
Then
\be
|| S_6 ||^2 = \frac{1}{40} \, N(N+2)^2  \bigg ( \frac{(N-2)(N+14)}{(N-1)(N+4)} \, a^2  +
\frac{2}{N+8} \, a \, b^2 \bigg )  - \frac{9}{20}   \, || w ||^2   \,,
\label{S6bound}
\ee
and
\be
|| M_6 ||^2= \frac{5}{48} \, N(N+2) ^2  \bigg (   \frac{N-2}{N-1} \, a^2 -
\frac{2}{9(N-2)} \, a \,b^2 \bigg ) + \frac{5(N-16)}{24(N+4)}  \, || w||^2     \, .
\label{Mbound}
\ee
In general \eqref{Mbound} gives an upper bound on $b$ and hence  bounds on $d_0$, clearly
\be
b^2 \le \tfrac{9(N-2)^2}{2(N-1)}\, a \, , \quad N\le 16 \, .
\ee
For $N=3$ $w$ and $M_6$ must vanish and hence we require $a=\tfrac49 \, b^2$ assuming  $d_2 =0$ and $d_4\ne 0$.
This gives $||d_4||^2 = \frac{10}{3}(1 -12\, d_0)^2$, and hence $d_0 = \frac{1}{15}, \, || d_4||^2 = \tfrac{2}{15}$  implying
$a_0 =1, \, S=\frac13$ or $d_0 = \frac{4}{45}, \, || d_4||^2 = \tfrac{2}{135}$ with $a_0=\tfrac43, \, S=\frac{10}{27}$. From
\eqref{wdef}
\be
\bigg( 1 -  \frac{12} {N(N+2)} \, a_0 \bigg )^{\! 2} \le \frac{9 (N-2)^2 }{ N(N-1)(N+2) } \, a_4\, , \quad a_2=0\, , \ a_4>0\, , \  \ N\le 16 \, .
\ee
Combining with \eqref{bounds} this gives
\be
\tfrac13 \, N \le a_0 \le \tfrac23 (N-1) \, ,  \qquad a_2=0\, , \  a_4>0\, , \ N\le 16 \, .
\ee
When this bound is saturated it is necessary that $||w|| =0$. For $N>16$ we may  combine \eqref{Mbound} with \eqref{S6bound}
to obtain the weaker bound
\be
b^2 \le \tfrac{(N-2)^2(N+8)^2}{6N(N+1)}\, a \, , \qquad N \ge 16 \, .
\ee

\section{Fixed Points  Valid for Any
\texorpdfstring{$N$}{N}}\label{sec:anyN}

For any $N$ the maximal symmetry is of course $O(N)$ with a potential  depending on a single coupling
\be
V_{O(N)}(\phi) = \tfrac{1}{24} \, \lambda\,  (\phi^2)^2 \, , \qquad \lambda_{ijkl} = \lambda ( \delta_{ij} \delta_{kl} +  \delta_{ik} \delta_{jl}
+ \delta_{il} \delta_{jk} ) \, ,
\ee
where  \eqref{RG1} reduces to just
\be
\lambda = \tfrac{1}{N+8} \, .
\ee
In this case
\be
S_{O(N)} = \tfrac{3N(N+2)}{(N+8)^2}\, , \quad a_0= \tfrac{N(N+2)}{N+8}\, ,\quad a_2=a_4 =0 \, .
\ee

More generally, fixed points  are obtained by considering a $p$-dimensional manifold of
couplings $\bG= \{ g_a \}$, $a=1,\dots,p$,  which is invariant under RG flow. Since the RG
equations are covariant under the action of $ O(N)$, then in general $\bG$
corresponds to all potentials $V(\phi)$  invariant under a  subgroup $H_{\rm sym} \subset O(N)$.  Invariably $\bG$
contains a one dimensional submanifold with $ H_{\rm sym} = O(N)$ but
there may also be higher dimensional submanifolds with enhanced symmetry
groups.  For theories determined by a  potential $V(\phi,g)$, $g\in \bG$,
there is a group action on $\bG$ given by   $V( h\,  \phi, g )  = V( \phi, g
\cdot {\hat h})$ where $ h \in H \subset  O(N)$.  The corresponding
symmetry group  is then the subgroup $H_{\rm sym} \subset  H$ defined by
$h \in H_{\rm sym}$, $g \cdot {\hat h} = g$ for arbitrary $g\in \bG$.  $H_{\rm sym} $ is a normal subgroup of $H$
and the action of the quotient group $Q_\bG= H/H_{\rm sym}$ on $\bG$ defines isomorphic theories and fixed
points related by the action of $Q_\bG$ are equivalent.
For an arbitrary potential depending on  the complete set of $\tfrac{1}{24} N(N+1)(N+2)(N+3)$
symmetric tensors $\lambda_{ijkl}$ then $H=O(N)$.

A standard procedure is to analyse the different subgroups $H_I  \subset
O(N)$ and then determine all possible quartic polynomials invariant under
each $H_I$.  However, differing $H_I$ may not have distinct quartic
polynomials so that $H_{\rm sym}$ may be  the union of various $H_I$.  An
alternative is to search for fixed points with a restricted symmetry where
the number of couplings is such that analytic and irrational numerical
solutions are possible for arbitrary $N$.  Inevitably this approach will
generate fixed points corresponding to decoupled theories but these can be
identified by looking at the stability matrix eigenvalues or checking the
additivity of $S,a_0,a_1$ for decoupled theories.

\subsection{Fixed Points with \texorpdfstring{$\S_N$}{SN} Symmetry}\label{sec:symm}

 For  an initial  search for fixed points for general $N$ we consider an ansatz with an overall
 $\S_N \times {\mathbb Z}_2$ symmetry which is obtained  by taking
 \begin{align}
&  \lambda_{iiii}  = x \, , \quad  \lambda_{iiij} =  y \, , \quad \lambda_{iijj} = z \, , \quad  \lambda_{iijk} = w \, , \quad
 \lambda_{ijkl} = u  \, , \nn \\
&  i \ne j \ne k \ne l, \qquad \mbox{no sums on}\ i,j,k,l\, .
\label{ans1}
\end{align}
Equivalently
\begin{align}
V_{\S_N}(\phi) = {}& \tfrac{1}{24}\, x\, {\ts \sum_i} \, \phi_i{\!}^4 + \tfrac{1}{6}\, y\, {\ts \sum_{i\ne j}} \, \phi_i{\!}^3 \phi_j  +
 \tfrac{1}{8}\, z\, {\ts \sum_{i\ne j}} \, \phi_i{\!}^2 \phi_j {\!}^2 + \tfrac{1}{4}\,  w \, {\ts \sum_{i\ne j\ne k} } \, \phi_i{\!}^2 \phi_j  \phi_k \nn \\
&{} +  \tfrac{1}{24} \, u\, {\ts \sum_{i\ne j\ne k \ne l} } \, \phi_i \phi_j  \phi_k \phi_l \, .
\end{align}
This ansatz is invariant under arbitrary $O(N)$ rotations if $x=3z, \, y=w = u = 0$ as then
$V_{\S_N}(\phi) = V_{O(N)}(\phi)$ with $\lambda=z$.
For  the $O(N-1)$ rotational subgroup, where infinitesimally
$ \delta \phi_i = \sum_j \omega_{ij} \phi_j$ with  $\omega_{ij}= - \omega_{ji}$ and $\sum_j \,\omega_{ij} =0$, then
\be
\delta  V_{\S_N}(\phi) =  \tfrac 16 ( x-y - 3 z + 3w ) \, {\ts \sum_{i\ne j}} \, \omega_{ij} \,  \phi_i{\!} ^3 \phi_j
+  \tfrac 12 ( y - 3w + 2u ) \, {\ts \sum_{i\ne j\ne k} } \, \omega_{ij} \,  \phi_i{\!} ^2\phi_j \phi_k \, .
\ee
Hence if $3z-x= 3w-y=2u$ there is a $O(N-1)$ rotation symmetry which leaves $\phi_i =\phi_j$ for all $i,j$ invariant.
 For $N=3$ it is only necessary that $x-3z =  y-3w$ for there to be a $O(2)$ symmetry.  When  $y,w,u$ are non zero there is just
the overall ${\mathbb Z}_2$ reflection symmetry.
For $y,w,u$ zero this extends to ${\mathbb Z}_2{\!}^N$ since the potential is invariant under $\phi_i \to -\phi_i$ for any $i$
and this then corresponds to a theory with hypercubic symmetry ${\mathbb Z}_2{\!}^N \rtimes \S_N$.
Otherwise theories
in which the couplings are changed in sign for an odd number of reflections belonging to  ${\mathbb Z}_2{\!}^N$ are equivalent.
There are two $\S_N$ singlet quadratic operators, ${\ts \sum_i} \, \phi_i{\!}^2 $ and $ {\ts \sum_{i\ne j}} \, \phi_i \phi_j $.
For $N=3$ the coupling $u$ is irrelevant, for $N=2$ both $w,u$ can be dropped.

There is one equivalence relation for the couplings in \eqref{ans1} given by
\begin{align}
(x,y,z,w,u) \sim{}& (x,y,z,w,u) \nn \\
& {}  - \tfrac{2}{N^3} \big ( 4(N-1)(N-2), \, (N-2)(N-4), \,  - 4(N-2), \, - 2(N-4) , \, 8 \big ) \nn \\
\noalign{\vskip -2pt}
& \qquad \times \big ( x - 3z + (N-4) ( y-3w) + 2 (N-3) u \big ) \, .
\label{equiv}
\end{align}
This corresponds to a reflection of $(\phi_1, \dots , \phi_N)$ through the hyperplane perpendicular to $(1,1,\dots,1)$.

The associated lowest order fixed point equations for the five couplings from \eqref{RG1} become
\begin{align}
x= {}& \beta_{\S_N,x} = 3x^2 + 3(N-1) (2 y^2 + z^2 + (N-2) w^2) \, ,  \nn \\
y = {}& \beta_{\S_N,y}  =  3(  xy + 3 yz) +3 (N-2)\big  ( 2 yw +  zw + 2w^2 + (N-3) wu \big )  \, , \nn \\
z= {}& \beta_{\S_N,z}  = 2xz + 6 y^2 +  (N+2) z^2 + (N-2)\big  ( 4yw + (N +7)  w^2  + 2(N -3) u^2 \big ) \, , \nn \\
 w = {}& \beta_{\S_N,w} =  2 y^2 + 2yz +  xw  + 8 yw + (N+7) zw + 2(5N-13)  w^2 + 2(N-3)yu \nn \\
 \noalign{\vskip -2pt}
&\hskip 1cm {} + (N-3)(N+4) wu + 2(N-3)(N-4) u^2  \, , \nn \\
 u = {}& \beta_{\S_N,u}  = 12 y w +12 z u+  3 (N+4) w^2 + 24(N-4) wu + 3(N-4)(N-5) u^2\, .
\label{redsol}
\end{align}
For $N=2$ just the first three equations with $w=u=0$ are relevant while for  $N\ge 3$ we need only keep the $x,y,z,w$
equations and set $u=0$.

For this ansatz
\begin{align}
a_{\S_N,0} ={}&  N\big  ( x + (N-1) \, z \big ) \, ,  \qquad \quad a_{\S_N,2} = N(N-1)\big  (2 \, y + (N-2) \,  w \big )^2 \, , \nn \\
a_{\S_N,4} = {}&N(N-1)\big ( \tfrac{1}{N+2}  \, (x-3z)^2 + \tfrac{4}{N+4} (N-2) \, (y-3w)^2 + (N-2)(N-3) \, u^2  \big )  \, , \nn \\
S_{\S_N} ={}&  N\big (x^2 + (N-1) (4\, y^2 + 3\, z^2 + 6(N-2)\, w^2 + (N-2)(N-3) \,u^2 ) \big ) \, ,
\end{align}
and in the decomposition \eqref{ldec}
\begin{align}
& (N+2) d_0 = x+(N-1)z \, , \qquad  d_{2,ii}=0\, \qquad(N+4) d_{2,ij} = 2y + (N-2) w \, , \nn \\
& (N+2)d_{4,iiii} = (N-1) (x-3z)\, , \quad (N+2) d_{4,iijj}= 3z-x \, , \nn \\
& (N+4) d_{4,iiij} = (N-2) ( y - 3w)\, , \quad (N+4) d_{4,iijk} = 2(3w-y) \, , \quad d_{4,ijkl} = u\, .
\end{align}

For $\Gamma_{ij}$ defined in \eqref{defGam} then with this ansatz
\be
\Gamma_{ii} = \tfrac{1}{N} \, S_{\S_N}  \quad  \mbox{no sum on}\  i \, , \qquad \Gamma_{ij} = v \, , \quad  i\ne j \,  ,
\ee
where
\begin{align}
v ={}&  2\, xy + 6\, yz + ( N-2) y^2 + 6(N-2) (y+z)w + 3(N-1)(N-2) w^2 \nn \\
&{} + 6(N-2)(N -3) uw + (N-2)(N-3)(N-4) u^2 \, .
\end{align}
For $v$ non zero there are two eigenvalues within this ansatz, namely
$\gamma = a_0/N + (N-1) v$ with degeneracy $1$, and
$\gamma= a_0/N-v$ with degeneracy $N-1$.

\subsubsection{\texorpdfstring{$O(N-1)$}{O(N-1)} Basis}

An alternative basis to that given in \eqref{ans1} is based on their transformation properties under the $SO(N-1)$
subgroup leaving  the $N$-vector $(1,1,\dots , 1)$ invariant. A basis of generators is given by
\be
(\omega_{rs})_{ij} = - (\omega_{sr})_{ij}  = \hdelta_{ri} \, \hdelta_{sj} -  \hdelta_{rj} \, \hdelta_{si}  \, , \quad  \hdelta_{ri}  = \delta_{ri} - \tfrac1N\, ,
\ee
which satisfies the commutation relation
\begin{align}
\big [ \omega_{rs} , \, \omega_{tu}\big ] {}_{ij} = {}& (\omega_{ru})_{ij} \, \hdelta_{st}  -  (\omega_{su})_{ij} \, \hdelta_{rt}
-  (\omega_{rt})_{ij} \, \hdelta_{su}  +  (\omega_{st})_{ij} \, \hdelta_{ru}  \, , \quad
\end{align}
and the completeness relation
\be
\tfrac12\,  {\ts \sum_{r\ne s} }\, (\omega_{rs})_{ij} \,(\omega_{rs})_{kl}
=    \hdelta_{ik} \, \hdelta_{jl} -  \hdelta_{il}\, \hdelta_{jk}  \, .
\ee

We may define
\be
(\delta_{rs} \phi)_i = {\ts \sum_j }\, (\omega_{rs})_{ij} \phi_j \, , \qquad \Delta = - \tfrac12\,  {\ts \sum_{r\ne s} }\, \delta_{rs} \delta_{rs} \,  ,
\ee
where $\Delta$ is essentially the Casimir for $SO(N-1)$. Then for
\begin{align}
X= {}&  {\ts \sum_i} \, \phi_i{\!}^4 \, , \qquad  Y = 4 \, {\ts \sum_{i\ne j}} \, \phi_i{\!}^3 \phi_j  \, , \qquad
Z=3 \, {\ts \sum_{i\ne j}} \, \phi_i{\!}^2 \phi_j {\!}^2 \, , \nn \\
W= {}& 6\,  {\ts \sum_{i\ne j\ne k} } \, \phi_i{\!}^2 \phi_j  \phi_k \, , \qquad
U= {\ts \sum_{i\ne j\ne k \ne l} } \, \phi_i \phi_j  \phi_k \phi_l \, ,
\end{align}
we have
\be
\Delta
\left(  \begin{smallmatrix} X \\ \noalign{\vskip 2pt}Y \\\noalign{\vskip 2pt} Z \\ \noalign{\vskip 2pt}W \\\noalign{\vskip 2pt} U \end{smallmatrix} \right )
= M
\left(  \begin{smallmatrix} X \\ \noalign{\vskip 2pt}Y \\\noalign{\vskip 2pt} Z \\ \noalign{\vskip 2pt}W \\\noalign{\vskip 2pt} U \end{smallmatrix} \right ) \, ,
\ee
where
\be
M = \begin{pmatrix} \frac{4}{N} (N-1)(N-2) & -\frac1N (N-2) & -\frac4N(N-2) &  \frac2N & 0 \\
 \noalign{\vskip 3pt}
- \frac{4}{N} (N-1)(N-2) & \frac1N (N-2)(3N-8)  &  \frac4N(N-2) & -\frac2N(3N-8)  &  \frac{24}{N} \\
 \noalign{\vskip 3pt}
-\frac{12}{N} (N-1)(N-2) & \frac3N (N-2) & \frac{12}{N}(N-2) &- \frac6N & 0 \\
 \noalign{\vskip 3pt}
\frac{12}{N} (N-1)(N-2) & -\frac3N (N-2)(3N-8)  &  -\frac{12}{N}(N-2) & \frac6N(3N-8)  & - \frac{72}{N} \\
 \noalign{\vskip 3pt}
0  & \frac6N (N-2)(N- 3) & 0 & - \frac{12}{N} (N-3)  &  \frac{48}{N}
\end{pmatrix}
\ee
The eigenvectors of $M$ determine the linear combinations of $x,y,z,w,u$ which correspond to different
representations of $O(N-1)$. There are three eigenvalues $0$, which is threefold degenerate, $3N$ and $4(N+1)$.
Correspondingly we define the couplings $\sigma, \, \rho, \, \tau_0, \, \tau_1, \, \tau_2$ where
\begin{align}
\sigma ={}& \tfrac{1}{N+2} \big ( x +(N-1) z\big ) \, , \quad  \rho = \tfrac{1}{N+4} \big ( 2y +(N-1) w \big ) \, , \nn \\
 \noalign{\vskip 3pt}
N \tau_0 ={}& \tfrac{1}{N+2}  (x-3z) + \tfrac{4}{N+4} (N-2) (y-3w)  -  \tfrac12 (N-2)(N-3) u\, , \nn \\
 \noalign{\vskip 3pt}
 N\tau_1 ={}& x  -y- 3 z +3w + (N-3) ( y-3w + 2u) \, , \nn \\
  \noalign{\vskip 3pt}
   N\tau_2 ={}& x  -y - 3 z +3w  - 3 ( y-3w + 2u) \, .
  \end{align}
  In the decomposition \eqref{ldec} $\sigma$ corresponds to $d_0$, $\rho$ to $d_{2,ij}$ and $\tau_0, \, \tau_1, \, \tau_2$ to
  $d_{4,ijkl}$.

  In this basis instead of \eqref{redsol}
  \begin{align}
  \sigma = {}& (N+8)\,  \sigma^2 + (N+16) \,  \tfrac{(N+4)(N-1)}{N+2}  \,\rho^2 \nn \\
 &{}  + \tfrac{8(N+4)(N-1)}{N(N+1)}\, \tau_0^2
+   \tfrac{8(N-1)(N-2)}{N(N+2)}  \,\tau_1^2 +   \tfrac{2(N-1)(N-2) (N-3) }{(N+1)(N+2)}\,\tau_2^2 \, , \nn \\
 \noalign{\vskip 3pt}
  \rho = {}& (N+16)\big (  \sigma   -  \tfrac{2N}{N+4}\,  \tau_0 \big ) \rho
  + (5 N+ 32)\,  \tfrac{2(N-2)}{N+4}   \,\rho^2 \nn \\
  &{}  + \tfrac{8(N+2)^2(N-2)}{N(N+1)(N+4) }\, \tau_0^2
+   \tfrac{2 (N-2)(N-4)}{N(N+4)} \,\tau_1^2 -    \tfrac{2(N-2)(N-3)}{(N+1)(N+4)}  \,\tau_2^2 \, , \nn \\
 \noalign{\vskip 3pt}
 \tau_0 = {}& 12 \, \sigma \tau_0  +  24 \tfrac{(N+2)(N-2)}{N+4} \, \rho \,  \tau_0
  -  ( N+ 16) \, \tfrac{3 N^2(N+1)}{2(N+2)(N+4)}\,\rho^2  \nn \\
  &{} - \tfrac{6 N(N+3)(N-2)}{(N+1)(N+4) } \, \tau_0^2
+   \tfrac{6 N(N-2)}{(N+2)(N+4)}  \,\tau_1^2 -  \tfrac{3 N(N-2) (N-3)}{(N+1)(N+2)(N+4)} \,\tau_2^2 \, , \nn \\
 \noalign{\vskip 3pt}
 \tau_1 = {}& 6 \big ( 2\,   \sigma + (N-4) \rho) \tau_1  +  \tfrac{3N}{N+1} \big ( 4 \, \tau_0 + (N-3) \tau_2 \big ) \tau_1\, , \nn \\
  \noalign{\vskip 3pt}
  \tau_2 = {}& 12 ( \sigma - 2 \rho) \tau_2  - \tfrac{24}{N+1} \, \tau_0 \tau_2
  + 6 \, \tau_1^2 + \tfrac{3}{N+1}( N^2-5N+2) \, \tau_2^2 \, .
  \end{align}
  When $N=3$  the $\tau_2$ equation is to be omitted and the remaining four equations are equivalent to those given in \cite{Safari}.
  These equations  correspond directly with the general form \eqref{RG4}.
  Clearly a consistent truncation is to set $\tau_1=0$  and for $O(N-1)$ symmetry $\tau_1=\tau_2=0$. The equivalence relation
  \eqref{equiv} becomes just
  \be
  \tau_1 \sim -\tau_1\, .
  \ee
 In terms of these couplings
  \begin{align}
  a_0 = {}& N(N+2) \sigma\, , \qquad a_2 = N(N-1)(N+4)^2 \rho^2 \, , \nn \\
  a_4 ={}& \tfrac{N-1}{N+1} \big (  4(N+2)(N+4) \, \tau_0^2 + 4(N+1)(N-2) \, \tau_1^2   +  N(N-2)(N-3)  \, \tau_2^2 \big ) \, .
  \end{align}

\subsubsection{Fixed Points}

Within this five coupling theory there are generically 18 real fixed points
of \eqref{redsol} including the trivial Gaussian theory.
Non trivial fixed points with rational $S_N$  at lowest order are given by
\begin{landscape}
\pagestyle{empty}
\setlength\LTcapwidth{\textwidth}
\setlength\LTleft{0pt}
\setlength\LTright{0pt}
\be
\hskip -1.7cm{
\begin{tabular}{l*{7}{c}}
\noalign{\vskip -3pt}
&  $\  x $ & $ y $ & $ z $ & $w $  & $u$&  \ $S_N $  \\
\noalign{\vskip 2pt}
\hline
\noalign{\vskip 5pt}
 I  &  $ \tfrac{1}{3}$ & $0 $ & $0 $ & $0 $ & $0$ & \ $ \tfrac19 N$   \\
\noalign{\vskip 8pt}
II  &  $\tfrac{N^3-8N^2+24N-16}{3N^3}$ & $-\tfrac{2(N-2)(N-4)}{3N^3}$ & $\tfrac{8(N-2)}{3N^3} $ & $ \tfrac{4(N-4)}{3N^3}$& $ -\tfrac{16}{3N^3}$ &\ $\tfrac19 N$   \\
\noalign{\vskip 8pt}
III &   $\tfrac{3}{N+8}$ & $ 0  $  & $\tfrac{1}{N+8} $  & $0 $  & $ 0 $ & $\tfrac{3N(N+2)}{ (N+8)^2} $ \\
\noalign{\vskip 5pt}
IV  &   $\tfrac{N-1}{3N}$ & $0$  & $\tfrac{1}{3N} $  & $0 $  & $ 0 $ &  $\tfrac{(N-1)(N+2)}{ 9N}  $ \\
\noalign{\vskip 8pt}
V &   $\tfrac{1}{N} + \tfrac{N-4}{N}\, x_{\rm{II}}$ &
 $\tfrac{N-4}{N}\, y_{\rm{ II}} $ & $\tfrac{1}{N}  + \tfrac{N-4}{N} \, z_{\rm{ II}} $ &
 $\tfrac{N-4}{N} \, w_{\rm{ II}} $ & $ \tfrac{N-4}{N}\, u_{\rm{ II}}$ &
 $\tfrac{(N-1)(N+2)}{ 9N} $ \\
\noalign{\vskip 8pt}
VI &   $\tfrac{N-1}{3(N+2)}$ & $-\tfrac{1}{3(N+2)}$  & $\tfrac{1}{3(N+2)} $  & $ 0 $  & $0$ &$ S_{T+}(N-1)$ \\
\noalign{\vskip 5pt}
VII  &   $\tfrac{1}{3N^2}$ & $\tfrac{1}{3N^2}$  & $\tfrac{1}{3N^2} $  & $ \tfrac{1}{3N^2} $   & $\tfrac{1}{3N^2} $   & $\tfrac{1}{ 9} $ \\
\noalign{\vskip 8pt}
VIII  &   $x_{\rm{VI}} + x_{\rm{VII}}$ & $y_{\rm{VI}} + y_{\rm{VII}}$   & $z_{\rm{VI}} + z_{\rm{VII}}$   &
$ w_{\rm{VII}}$   & $ u_{\rm{VII}}$ & $S_{\rm{VI}} + S_{\rm{VII}}$ \\
\noalign{\vskip 8pt}
IX   &   $\tfrac{3(N-1)^2}{N^2(N+7)}$ & $-\tfrac{3(N-1)}{N^2(N+7)}$  & $\tfrac{N^2-2N+3}{N^2(N+7)}$  & $ - \tfrac{N-3}{N^2(N+7)}$    &
$\tfrac{3}{N^2(N+7)}$  & $\tfrac{3(N^2-1)}{(N+7)^2}$   \\
\noalign{\vskip 8pt}
X  &   $x_{\rm{IX}} + x_{\rm{VII}}$ & $y_{\rm{IX}} + y_{\rm{VII}}$   & $z_{\rm{IX}} + z_{\rm{VII}}$   &
$w_{\rm{IX}} +  w_{\rm{VII}}$   & $ u_{\rm{IX}} + u_{\rm{VII}}$ & $S_{\rm{IX}} + S_{\rm{VII}}$ \\
\noalign{\vskip 8pt}
XI & $ \tfrac{(N-1)(N-2)(N-3)^2}{3N^2(N^2- 7N+ 14)}$ &$ - \tfrac{(N-2)(N-3)^2}{3N^2(N^2- 7N+ 14)}$ &  $\tfrac{(N-2)(N-3)}{N^2(N^2- 7N+ 14)}$ &
$ \tfrac{(N-3)(N-6)}{3N^2(N^2- 7N+ 14)}$ & $ -\tfrac{(N-6)}{N^2(N^2- 7N+ 14)}$ & $S_{T-}(N-1) $ \\
\noalign{\vskip 8pt}
XII  &   $x_{\rm{XI}} + x_{\rm{VII}}$ & $y_{\rm{XI}} + y_{\rm{VII}}$   & $z_{\rm{XI}} + z_{\rm{VII}}$   &
$w_{\rm{XI}} +  w_{\rm{VII}}$   & $ u_{\rm{XI}} + u_{\rm{VII}}$ & $S_{\rm{XI}} + S_{\rm{VII}}$ \\
\noalign{\vskip 8pt}
XIII${}_\pm$  &   $ \tfrac{   { \hskip -0.7cm { N^4 + 3N^2 +8N -8}  \atop  \pm 4(N-1)(N-2)\sqrt{N+1}}}{3N^3(N+3)}$ &
  $ -\scriptstyle{(N-2)} \tfrac{ {  \hskip-0.65cm {N^2  -N -4} \atop   \mp (N-4)\sqrt{N+1} } }{3N^3(N+3)}$
 &  $ \tfrac{ { \hskip-0.5cm{N^3 +  3N^2  - 8}  \atop   \mp 4  (N-2)\sqrt{N+1} } }{3N^3(N+3)}$   &
${\scriptstyle 2} \tfrac{ {  \hskip-0.6cm{N^2  -N -4}  \atop  \mp (N-4)\sqrt{N+1}}}{3N^3(N+3)}$   & $\tfrac{ {N^2 -  4N - 8 \atop   \pm 8\sqrt{N+1}}}{3N^3(N+3)}$  & $S_{T-}(N)$ \\
\noalign{\vskip 8pt}
XIV${}_\pm$  & \hskip -0.3cm
 $\scriptstyle{(N-1)(N-2)} \tfrac{   { \hskip 0.1cm { N^3 -4 N^2 +4N +16}  \atop  \pm 4(N-4)\sqrt{N+1}}}{3N^3(N^2-5N+8)}$ &
  $ -\scriptstyle{(N-2)(N-4)} \tfrac{ { \hskip-0.65cm {N^2  -N -4} \atop   \mp (N-4)\sqrt{N+1} }}{3N^3(N^2-5N+8)}$
 &  $\scriptstyle{4(N-2)} \tfrac{  {{ \hskip-0.6cm  N^2 -  N  - 4}  \atop   \mp 4  (N-2)\sqrt{N+1}} }{3N^3(N^2-5N+8)}$   &
${\scriptstyle 2(N-4)} \tfrac{ {  \hskip-0.6cm{N^2  -N -4}  \atop  \mp (N-4)\sqrt{N+1}}}{3N^3(N^2-5N+8)}$   &
$\scriptstyle{(N-4)}\tfrac{ {N^2 -  4N - 8 \atop   \pm 8\sqrt{N+1}}}{3N^3(N^2-5N+8)}$  & $S_{T+}(N)$ \\
\end{tabular}}
\label{tableN}
\ee

where
\be
S_{T-}(N) = \frac{N(N+1)(N+7)} {9 (N+3)^2} \, , \qquad S_{T+}(N) = \frac{N(N-1)(N-2)(N^2-6N+11)} {9 (N^2-5N + 8)^2} \, .
\ee
\end{landscape}
\clearpage

\newpage

 Cases I,II and IV,V as well as XIII${}_\pm$ and XIV${}_\pm$ are equivalent in that the couplings are related by $O(N)$
 rotations. Cases VIII, X and XII are decoupled theories.

At the fixed points given in \eqref{tableN} the results correspond essentially to theories with hypercubic, hypertetrahedral or $O(N)$
symmetry. For convenience we identify the hyperoctahedral group $B_n$ which is the symmetry group of a $n$-dimensional hypercube
expressible as a wreath product
\be
B_n =  {\mathbb Z}_2 \wr \S_n =  {\mathbb Z}_2{\!}^n\rtimes {\cal S}_n  \, , \qquad B_2 \simeq D_4 \, , \quad B_3 \simeq \S_4\times  {\mathbb Z}_2 \, ,
\ee
where $|B_n| = 2^n n!$ and we have used the notation for a wreath product.

Results for the lowest order anomalous dimensions for $\phi^4$ operators in the   hypercubic and hypertetrahedral cases were given in \cite{Seeking}
and extended in the hypercubic case in \cite{Antipin}.

\clearpage

\begin{landscape}
\pagestyle{empty}
\setlength\LTcapwidth{\textwidth}
\setlength\LTleft{0pt}
\setlength\LTright{0pt}
\be
\hskip - 1.5cm{
\begin{tabular}{l*{7}{c}}
\noalign{\vskip -3pt}
&  symmetry group &  $V_{\S_N}  $ & \quad  $ \ \ a_0 \ \ $ & \ \ $a_2  $ \ \  & $ \tfrac18 N - S_N -\tfrac 12 a_2 $ \\
\noalign{\vskip 2pt}
\hline
\noalign{\vskip 5pt}
 I,II &  $  B_N $  & $ \tfrac{1}{72} \,  \Phi_4, \   \tfrac{1}{72} \,  {\ts \sum_{j=1}^N} \, (v_j \!\cdot  \!\phi)^4  $  &
\quad $\tfrac13{\scriptstyle  N} $& $ 0$ & $\tfrac {N}{72} $ \\
\noalign{\vskip 8pt}
III &  $O(N)$    & $\tfrac{1} {8(N+8) } \, \Phi_2{}^2$ &\quad  $ \tfrac{N(N+2)}{N+8} $ & $ 0 $  & $\tfrac{ N(N-4)^2}{8(N+8)^2}$ \\
\noalign{\vskip 8pt}
IV,V  &   $  B_N $  & \raisebox{8pt}{ $ {\tfrac{1}{24N} \big (\Phi_2{\!}^2 + \tfrac13{{(N-4)}}\, \Phi_4 \big )} $,}
&   $ \tfrac23 {\scriptstyle (N-1) }$ & $0$ &  $ \tfrac{(N-4)^2}{72N} $ \\
\noalign{\vskip -8pt}
&  &  ${ \tfrac{1}{24N} \big (  \Phi_2{\!}^2  + \tfrac13 { (N-4)}\, {\ts \sum_{j=1}^N} \, (v_j  \! \cdot \! \phi)^4 \big ) } $  \\
\noalign{\vskip 8pt}
VI &  $ {\cal S}_N  \times {\mathbb Z}_2$  &  $\tfrac{1}{72(N+2)} \, {\ts \sum_{i<j}} \, (\phi_i -\phi_j )^4 $  & \ $\tfrac{2N(N-1) }{3(N+2)} $ &  \
$\tfrac{4N(N-1) }{9(N+2)^2} $    & $ \tfrac{N(N- 10)^2}{72(N+2)^2} $ \\
\noalign{\vskip 8pt}
VII & $O(N-1)\times {\mathbb Z}_2 $ &   $\tfrac{1} {72\, N^2} \, {\Phi}_1{\!}^4 $
&\ \ $ \tfrac13 $ & $\tfrac{N-1}{9N} $ & $ \tfrac{(3N-2)^2} {72N} $ \\
\noalign{\vskip 8pt}
VIII  & $O(N-1)\times {\mathbb Z}_2 $ &  $V_{\rm VI}(\phi)  +  V_{\rm VII}(\phi)  $ & $ \tfrac{2N^2 -N+2}{3(N+2)}$ &
 $ \tfrac{(N-1)(N-2)^2}{9N(N+2)^2} $ &  $ \tfrac{(N^2 - 8N +4)^2}{72N(N+2)^2}$ \\
\noalign{\vskip 8pt}
IX   & $O(N-1)\times {\mathbb Z}_2 $ & $\tfrac{1}{8N(N+7)} \, {\hat \Phi}_2{}^2 $ & $\tfrac{N^2-1}{N+7}$  &
$\tfrac{(N^2-1)(N+1)}{N(N+7)^2}$  & $ \tfrac{(N^2-7N -2)^2}{8N(N+7^2)}$ \\
\noalign{\vskip 8pt}
X  &   $ {\cal S}_N  \times {\mathbb Z}_2$ & $V_{\rm IX}(\phi)  +  V_{\rm VII}(\phi)  $  &   $\tfrac{3N^2+N+4}{3(N+7)}  $ &  $ \tfrac{4(N-1)(N-2)^2}{9N(N+7)^2}$ &$ \tfrac{(3N^2-19N+8)^2}{72N(N+7)^2}$  \\
\noalign{\vskip 8pt}
XI &  $ {\cal S}_N  \times {\mathbb Z}_2$ &  \hskip - 0.3cm $\tfrac{1}{72(N^2-7N +14)} \,
\Big (  { (N-5)} \,    {\ts \sum_{i<j}} \, (\phi_i {-\phi_j })^4  - \frac{3}{N}{(N-6)}{\hat  \Phi}_2 {}^2 \Big ) $
& $ \tfrac{(N-1)(N-2)(N-3)}{3(N^2-7N+14)} $ & $ \tfrac{(N-1)(N-2)^2(N-3)^2} {9N(N^2-7N+14)^2} $ &
$ \tfrac{(N^3 -9N^2 + 20N +12 )^2} {72N(N^2-7N+14)^2} $   \\
\noalign{\vskip 10pt}
XII  & $ {\cal S}_N  \times {\mathbb Z}_2$ & $V_{\rm XI}(\phi)  +  V_{\rm VII}(\phi)  $ &  $ \tfrac{N^3-5N^2 + 4N + 8}{3(N^2-7N+14)} $ &
$\tfrac{4(N-1)(N-4)^2}{9N(N^2-7N+14)^2}$ &  $ \tfrac{(N^3-11N^2 + 34 N-16)^2}{ 72N(N^2-7N+14)^2 }$  \\
\noalign{\vskip 8pt}
XIII${}_\pm$ & $\S_{N+1} \times  {\mathbb Z}_2$  &
$  \tfrac{1}{24(N+3)} \big ( \Phi_2{\!}^2  +  \frac13 {(N{+1})} \, {\ts \sum_{\alpha=1}^{N+1} }\, ( \phi_\pm{\hskip -0.8pt}^\alpha)^4 \big ) $
 & $ \tfrac{2N(N+1)}{3(N+3)} $ & $ 0$  &$  \tfrac{N(N-5)^2}{72(N+3)^2}$   \\
\noalign{\vskip 10pt}
XIV${}_\pm$  & $ {\cal S}_{N+1}  \times {\mathbb Z}_2 $&
$ \tfrac{1}{24(N^2- 5N +8)} \big ( \Phi_2{\!}^2  +  \frac13 {(N{+1})(N-{4}) }\, {\ts \sum_{\alpha=1}^{N+1} }\,
( \phi_\pm{\hskip -0.8pt}^\alpha)^4 \big ) $
&   $ \tfrac{N(N-1)(N-2)}{3(N^2-5N + 8)} $ & $ 0$
& $  \tfrac{N(N-4)^2(N-5)^2}{72(N^2-5N +8 )^2}$  \\
& \\
\end{tabular}}
\label{tableV}
\ee

\thispagestyle{empty}

\end{landscape}
\clearpage
In the above
\be
\Phi_n = {\ts \sum_{i=1}^N }\, \phi_i{\!}^n \, ,
\qquad (v_j)_i = \big (   \delta_{ji} - \tfrac{2}{N}  \big ) \, , \quad v_j \cdot v_k = \delta_{jk} \, ,
\ee
and for the cases with with hypertetrahedral symmetry
\begin{align}
{\hat \phi}_i = \phi_i - \tfrac1N\,  \Phi_1 \, , \ \  {\hat \Phi}_n = {\ts \sum_{i=1}^N }\, {\hat \phi}_i{\!}^n \,  , \ \ {\hat \Phi}_2 =
\Phi_2 - \tfrac1N\, \Phi_1{\!}^2 \, ,\nn\\
\quad V_{\rm VI} = \tfrac{1}{24(N+2)}
\big ( {\hat \Phi}_2{\!}^2  +  \tfrac13 N \, {\hat \Phi}_4 \big ) \, , \
\
V_{\rm XI} = \tfrac{1}{24(N^2-7N+14)} \big ( {\hat \Phi}_2{\!}^2  +  \tfrac13 N(N-5)\, {\hat \Phi}_4 \big )\, ,
\end{align}
and
\begin{align}
 \quad \phi_\pm{\!}^\alpha  =  \phi_i \, e_{\pm,i}{}^\alpha  =
 \begin{cases} \phi_\alpha + \tfrac1N \big ( - 1 \pm  \tfrac{1}{\sqrt{N+1}} \big ) \Phi_1 \, ,
\ \  &1\le  \alpha \le N \\ \mp  \tfrac{1}{\sqrt{N+1}} \, \Phi_1 \, , &
\alpha=N{+1}  \end{cases}\, ,
\end{align}
where $ \phi_\pm{\!}^\alpha $ satisfy the equations
\be
{\ts \sum_{\alpha=1}^{N+1} } \phi_\pm{\!}^\alpha =0, \  \quad  {\ts \sum_{\alpha=1}^{N+1} } \, ( \phi_\pm{\!}^\alpha)^2 = {\ts \sum_{i=1}^N }\, \phi_i{\!}^2 \, , \quad
 {\ts \sum_{i=1}^{N} } \, \pr_i \phi_\pm{\!}^\alpha \pr_i\phi_\pm{\!}^\beta = \delta^{\alpha\beta}- \tfrac{1}{N+1}  \, .
 \label{tetra}
\ee

  I,II correspond to $N$ decoupled Ising models, III to to the $O(N)$ invariant generalised Heisenberg fixed point $O_N$,  IV,V,
  where $y=w=u=0$, to the hypercubic
  symmetry fixed point $C_N$ and XIII${}_\pm$, XIV${}_\pm$ to the two hypertetrahedral fixed points $T_{N\pm}$. The cases with non zero $a_2$
  correspond to decoupled theories with an $N-1$ dimensional $O(N-1)$ or hypertetrahedral fixed point asoociated with a free
  Gaussian or Ising fixed point. The results for $a_2$ can all be obtained by applying  \eqref{a2}.  These decoupled theories
  are just those which maintain an overall $\mathcal{S}_N$ symmetry.

  The bound \eqref{Sbound} is saturated for $N=4$ by $O_4 = C_4$ and for $N=5$ by $T_{5+}=T_{5-}$.

As special cases
\begin{align}
N= {}& 3 \, , \ \  & &  V_{\rm IX}  = V_{\rm VI} \, , \  V_{\rm X}  = V_{\rm V} \, , \  V_{\rm XI}=0 \, , \ V_{\rm XII}  = V_{\rm VII} \, , \
 V_{\rm XIII_+}  = V_{\rm V} \, , \  V_{\rm XIII_-}  = V_{\rm IV} \, , \nn \\
& &   &V_{\rm XIV_+} = V_{\rm II} \, , \ V_{\rm XIV_-} = V_{\rm I} \, , \nn \\
N={}&  4\, , & &  V_{\rm III} =  V_{\rm IV} =V_{\rm V} = V_{\rm XIV_+} =  V_{\rm XIV_-} \, , \nn \\
N={}& 5 \, ,  & & V_{\rm XI}  = V_{\rm IX} \, , \ V_{\rm XII}  = V_{\rm X} \, , \  V_{\rm XIV_+}  = V_{\rm XIII_+} \, , \  V_{\rm XIV_-}  = V_{\rm XIII_-} \, ,\nn \\
N={}& 6 \, , & & V_{\rm XI}  = V_{\rm VI} \, , \ V_{\rm XII}  = V_{\rm VIII} \, .
\end{align}

Except for particular $N$ the remaining fixed point is irrational. The couplings satisfy $3z-x= 3w-y=2u$ and so there is a $O(N-1)$ symmetry.
The results for this case for low values of $N$ are
\be
\begin{tabular}{l*{5}{c}}
\noalign{\vskip -3pt}
&  $ N $  &  \ \  $ S $ \ \ & \ \ $a_0$  \ \  & \ \ $a_2$ \ \  & \ \ $ a_4 $ \\
\noalign{\vskip 2pt}
\hline
\noalign{\vskip 5pt}
& $3$   &  \ \ 0.370451 & 1.33713 & 0.000255 &  0.012651 \\
& $5$ &  \ \ 0.621937 &  2.67255 &  0.000171  & 0.009605 \\
& $6$ & \ \  0.738216 & 3.35878 & 0.002115 & 0.031859 \\
&  $7$ &  \ 0.848454 & 4.05973 & 0.008335 & 0.059079 \\
& $8$ & \  0.952091 & 4.77518 &  0.020705 & 0.086649 \\
& $9$ & \  1.048864 & 5.50436 & 0.040191 & 0.112193 \\
\end{tabular}
\label{tableB}
\ee
\ \

\subsection{Fixed Points with Continuous Symmetry}\label{sec:cont}

When $N$ factorises there are various fixed points which can be regarded as built from fixed points corresponding to the
factors of $N$. For $N=mn $ then for $\phi \to \vphi_{r a}$, $r=1, \dots , n, \ a= 1, \dots , m$, $n>1$ there are non trivial fixed points
obtained from the potential
\be
V_{M \hskip -0.8pt N}(\vphi)  = \tfrac18 \, \lambda \, (\vec \vphi {\,}^2)^2  + \tfrac{1}{24}\, g \, \textstyle{\sum_r}( {\vec \vphi}_{r}{\!}^2)^2 \, .
\ee
At the fixed point, after a suitable rescaling,  the necessary equations corresponds to \eqref{RG1} become just
\be
\lambda = \frac{4-m} { (m+8) N - 16(m-1)} \, , \qquad g =  \frac{3(N-4)} { (m+8) N - 16(m-1)} \, .
\ee
The symmetry group is $O(m)^n\rtimes S_n = O(m) \wr \S_n $ and this fixed point is denoted here by $M\! N_{m,n}$. $M\! N_{1,n}= C_n$,
and $M\! N_{2,2}=  O_4$. For these theories
\begin{align}
S= {}& \frac{3 m^2 n (n-1)(m^2 n + 2m (n-5) +16)}{  ((m+8) N - 16(m-1))^2}  \, , \quad
a_0 = \frac{6 m^2 n (n-1)}{  (m+8) N - 16(m-1)}  \, , \quad a_2 =0 \, ,\nn \\
a_4=  {}& \frac{3 m^2(m+2)  n (n-1)(N-4)^2 }{ (N+2) ((m+8) N - 16(m-1))^2}   \, , \quad
\tfrac18 m n - S = \frac{ m n (m-4)^2 ( N - 4)^2}{ ((m+8) N - 16(m-1))^2}  \, .
\end{align}
If $m=4$ and the $S$-bound is saturated this is just $n$ decoupled $O_4$ theories.

\subsection{Tetragonal Fixed Points}

In the condensed matter literature fixed points arising for systems with tetragonal symmetry are of interest
\cite{3coupling1,MudrovV1,MudrovV2,RGrev}.
These can be modelled by considering the potential
\be
V(\vphi,\psi) = \tfrac18 \,\lambda \big  ( \vphi^2 + \psi^2 \big )^2 + \tfrac{1}{24} \, g \,  {\ts \sum_{a=1}^n} \big ( \vphi_a{\!}^4 + \vphi_a{\!}^4 \big)
+ \tfrac{1}{4}\,  h\,  {\ts \sum_{a=1}^n} \,  \vphi_a{\!}^2 \vphi_a{\!}^2 \, ,
\ee
where $N=2n$. This has the symmetry $D_4{\!}^n \rtimes \S_n$. The potential is invariant under
equivalence relation $g\sim (g+3h)/2, \, h \sim  (g-h)/2$.
The fixed point equations for the three couplings are just
\be
\lambda= 2(n+4) \lambda^2 + 2(g+h)\lambda \, , \quad g = 12\,  \lambda g + 3(g^2+h^2) \, , \quad
h= 12\,  \lambda h + 2(g+2h)h \, ,
\ee
which have 8 solutions but two pairs related the equivalence relation. The results all have $a_2=0$ since there is a unique
quadratic invariant and the non trivial ones can be summarised by
\be
 \begin{tabular}{ l c c c c c  c  c c }
\noalign{\vskip 5pt}
Fixed Point & $S$ & $a_0$ & $a_4$  & $ \{ \kappa \}$  \\
\hline
\noalign{\vskip 5pt}
$O_{2n}$ &  $\tfrac{3n(n+1)}{(n+4)^2}$ & $\tfrac{2n(n+1)}{n+4}$ & 0 & $ - \tfrac{n-2}{n+4}(2), \, 1$\\
\noalign{\vskip 5pt}
$I^{2n} $ & $ \tfrac{2}{9} n$ & $\tfrac{2}{3} n $ & $\tfrac{2n(2n-1)}{9(n+1)} $ & $-\tfrac13(2), \, 1$\\
\noalign{\vskip 5pt}
$O_2{\!}^{n}$ & $ \tfrac{6}{25} n$ & $ \tfrac{4}{25} n$ & $ \tfrac{6n(n-1)}{25(n+1)} $ & $-\tfrac15, \, \tfrac15, \, 1$\\
\noalign{\vskip 5pt}
$C_{2n} $ & $\tfrac{(n+1)(2n-1)}{9n }$ & $\tfrac23(2n-1) $ & $\tfrac{(n-2)^2 (2n-1)}{9n(n+1)} $ & $ - \tfrac{n-2}{3n}, \, \tfrac{n-2}{3n}, \, 1$\\
\noalign{\vskip 5pt}
$M\! N_{2,n} $ & $\tfrac{3n (2n^2-3n+1)}{(5n-4)^2 }$ & $\tfrac{6n(n-1)}{5n-4} $ & $\tfrac{6n(n-1)(n-2)^2}{(n+1)(5n-4)^2} $ &
$ \tfrac{n-2}{5n-4}(2), \, 1$
\end{tabular}
\nn
\ee
The tetragonal symmetry is enhanced at each fixed point and within this three coupling theory $M\!N_{2,n}$ is stable for $n>2$. At the
$M\!N_{2,n}$ fixed point $g=3h$ which is invariant under the equivalence             on on the couplings. This agrees with Michel's theorem
requiring that a stable fixed point is unique.

\section{`Double Trace' Perturbations}\label{sec:doubletrace}

A wide range of  fixed points can be obtained by perturbations of decoupled theories. Assuming $\phi_i = (\vphi_a, \psi_r)$,
$ a=1, \dots, m,  r=1,\dots, n$,  $N=m+n$, the starting point is
\be
V(\phi ) = V_1(\vphi) + V_2(\psi)  \, ,
\label{V12}
\ee
where  the potentials $V_1,V_2$ are invariant  under subgroups $H_1 \subset O(m) , \, H_2\subset O(n)$.
After imposing  \eqref{RG2}  for $V_1,\, V_2$  generate  fixed points $FP1, \, FP2$,
 with symmetry groups $H_1,\, H_2$  so that \eqref{V12} then corresponds to $FP1\cup FP2$.
We assume  that at each fixed point
there are unique quadratic polynomials which are invariant under $H_1, \, H_2$ respectively
and that these have anomalous dimensions $\mu_1, \, \mu_2$ at the fixed points defined by $V_1,\, V_2$.
In the examples of relevance here these polynomials are just $\vphi^2, \, \psi^2$ and the eigenvalue equation  \eqref{eigmn}
requires
\be
\mu_1 \, \delta_{ab} = \lambda_{1,abcc} \, , \quad  \mu_2 \, \delta_{rs} = \lambda_{2,rs\hskip 0.5 pt tt} \quad \Rightarrow \quad
m \, \mu_1 = a_{0,1} \, , \quad n \, \mu_2 = a_{0,2} \, .
\label{mumu}
\ee
For the decoupled theory then
\be
S_N = \tfrac12 m \, \mu_1(1-\mu_1) + \tfrac12 n \, \mu_2(1-\mu_2) \, , \quad
a_0= m\, \mu_1 + n\, \mu_2 \, , \quad a_2 = \frac{m n}{m+n} \, (\mu_1 - \mu_2)^2 \, .
\ee

We then consider a perturbed theory obtained by
\be
V(\phi ) = V_1(\vphi) + \delta V_1(\vphi) + V_2(\psi) + \delta V_2(\vphi)  + U(\vphi,\psi)   \, ,
\label{VP}
\ee
where for just single quadratic invariants $\vphi^2, \psi^2$,
\be
U(\vphi,\psi) = h \, \tfrac14 \, \vphi^2 \, \psi^2 \, ,
\label{defU}
\ee
which preserves a $H_1 \times H_2$ symmetry.  With $\vep \ne 1$ at zeroth order in $h$, $d- \Delta_{\vphi^2\psi^2}
= (1-\mu_1 - \mu_2) \vep$ so this perturbation  is relevant for
\be \epsilon= 1 - \mu_1 -\mu_2 >0 \, .
\ee
The additional term may be regarded as a double trace perturbation and the symmetry
ensures the form \eqref{VP} is preserved under any RG flow generated by the coupling $h$ and we may require
$\delta \mu_1 \, \delta_{ab} = \delta\lambda_{1,abcc} \, , \  \delta \mu_2 \, \delta_{rs} =
\delta\lambda_{2,rs\hskip 0.5 pt tt}$.
At any fixed point with $m\ne n$ there are then two distinct invariant quadratic operators.

 At lowest order the fixed point equation \eqref{RG2} requires
\be
h ( 1-\mu_1 - \delta \mu_1 - \mu_2 - \delta \mu_1 ) = 4  \, h^2 \, .
 \ee
 Hence at this order we may take
 \be
 h = \tfrac{1}{4} \, \epsilon\,,
 \ee
 so that for  a relevant perturbation $h>0$.
 If the decoupled fixed points are such that $\epsilon$ is small we may set up a perturbation expansion in $\epsilon$. At order $h^2$  a non trivial  fixed point must satisfy
 \be
 \delta V_1(\vphi) - V_{1,ab}(\vphi) \delta V_{1,ab} ( \vphi) = \tfrac18 n  \, h^2 (\vphi^2)^2 \, , \ \
 \delta V_2(\psi) - V_{2,rs}(\psi) \delta V_{2,rs} ( \psi) = \tfrac18 m \, h^2  (\psi^2)^2 \, .
 \label{Vpert}
 \ee
 To the extent that these equations can be inverted to define $\delta V_1, \,\delta V_2$ it is possible to set up a series
 expansion in $\epsilon$ which should converge in the neighbourhood of $\epsilon =0$ so that there is a new non decoupled
 fixed point with symmetry $H_1 \times H_2$. In general the theory defined by \eqref{VP} may have several fixed
 points some of which have an enhanced  symmetry but for $h$ non zero and small the fixed point is here denoted as
 $B_{FP1*FP2}$.

As a special case we may impose $m=n, \ V_1 = V_2$ and then $H_1 = H_2=H $.  There is an additional
 ${\mathbb Z}_2$ symmetry under $\vphi\leftrightarrow \psi$ so that  $\phi_i = (\vphi_a, \psi_a)$,
$ a=1, \dots, m$ and there is a single $H^2 \rtimes {\mathbb Z}_2$ invariant quadratic operator $\vphi^2 + \psi^2$.
Fixed points obtained with the additional  ${\mathbb Z}_2$ symmetry  are generally rational.

 These results can be used to determine the change in $S_N$ as in \eqref{Sbound}. From \eqref{VP} to lowest order
 \be
 \delta S_N = 2 \, \lambda_{1,abcd}\, \delta \lambda_{1,abcd} + 2 \, \lambda_{2,rstu}\, \delta \lambda_{2,rstu} + 6 \, mn \, h^2\, .
 \label{varS}
 \ee
 \eqref{Vpert} is equivalent to
 \be
 \delta \lambda_{1,abcd} - S_{6,abcd}\, \lambda_{1,abef} \delta \lambda_{1,cdef} = n\,  h^2 ( \delta_{ab}\delta_{cd} +
 \delta_{ac}\delta_{bd}  + \delta_{ad}\delta_{bc} ) \, ,
 \label{lpert}
 \ee
 and equivalently for $\delta \lambda_{2,rstu}$. Crucially the right hand side is orthogonal to perturbations
 of the form \eqref{vomega} which form a null space for the stability matrix.   Hence \eqref{lpert} and its partner for
 $\delta \lambda_{2,rstu}$ immediately give
 \be
 -  \lambda_{1,abcd}\, \delta \lambda_{1,abcd}  = 3 \, n \, h ^2 \, \lambda_{1,aa bb} = 3\,  m n \, h^2 \, \mu_1 \, , \quad
-  \lambda_{2,rstu}\, \delta \lambda_{2,rstu}  = 3\,  m n \, h^2 \, \mu_2 \, ,
 \ee
 using \eqref{mumu}. Then \eqref{varS} becomes
 \be
 \delta S_N = 6\, m n \, h^2 (1-\mu_1 - \mu_2) =  \tfrac 38 \, m n \, \epsilon^3 \, .
 \ee
 For theories in which $S_N$ is close to the bound \eqref{Sbound} it is necessary from \eqref{bounds} that $a_0$ is close
 to $\tfrac12 N$. By virtue of \eqref{mumu} $\mu_1, \, \mu_2$ are close to $\tfrac 12$ when the initial theories defining
 $FP_1, \, FP_2$ are close to saturating their respective $S$-bounds. If $S_{FP1} = \tfrac18 m , \, S_{FP2} = \tfrac18 n$ then
 necessarily $\mu_1 = \mu_2= \tfrac12$ and $\epsilon =0$. When
$\epsilon > 0$, as for a relevant perturbation,  $ \delta S_N >0$.

For $a_0$
\be
\delta a_0 = \delta \lambda_{1,aabb} +  \delta \lambda_{2,rrss} + 2\,  m n \, h=
m \, \delta \mu_1 + n\, \delta \mu_2 +  2\,  m n \, h \,  .
\ee
With $ \delta \mu_1, \, \delta \mu_2$ of ${\rm O} (\epsilon^2)$ this gives to ${\rm O}(\epsilon)$
\be
\delta a_0= \tfrac12 \, mn \, \epsilon  \, , \qquad \delta a_2 =- \tfrac{m\, n}{2(m+n)} \, (m-n)(\mu_1 - \mu_2) \, \epsilon\,  .
\ee

\subsection{Perturbed Cubic Theories}

For the  ${\mathbb Z}_2$ symmetric case  we assume $V_1=V_2$ each  correspond to a theory which generates the $C_m$ fixed point so that
\be
V_1(\vphi) =  \tfrac18 \, \lambda \, (\vphi^2)^2 + \tfrac{1}{24} \, g \, {\ts \sum_a}\, \vphi_a{\!}^4 \, .
\label{V1}
\ee
For the potential $V_1(\vphi)+V_1(\psi) + \tfrac14 \, h \, \vphi^2 \psi^2 $,
after rescaling, the fixed point equations reduce to
\begin{align}
\lambda ={}&  ( m+8) \, \lambda^2+ 2\, \lambda g + m\, h^2 \,  , \quad g = 3\, g^2 + 12 \, \lambda g\, , \nn \\
h = {}& 2(m+2)\, \lambda h  + 2 \, g h + 4\, h^2\, .
\end{align}
There are seven non trivial real solutions which include  the $O_{2m}, \, C_{2m}, \, M \! N_{m,2}$ fixed points as well as two decoupled
$C_m$ and $O_m$  fixed points in addition to $2m$ decoupled Ising fixed points. There is one new fixed point, which we denote
by $CC_m= B_{C_m*C_m} $, with symmetry  $(B_m \times B_m)\rtimes {\mathbb Z}_2$,  and which gives for the $O(N)$ invariants
\begin{align}
S_{2m} = {}& \frac{m(m^2-3m+8)(2m^2-9m+16)}{9(m^2-4m+8)^2} \, , \quad a_0 =  \frac{2m(m^2-3m+8)}{3(m^2-4m+8)} \, , \quad a_2=0\, , \nn \\
a_4 = {}& \frac{2m(m-1)(m-2)^2(m^2-3m+8)}{9(m+1)(m^2-4m+8)^2}  \, ,  \quad \tfrac14 m - S_{2m} = \frac{m(m-2)^2(m-4)^2}{36 ( m^2-4m+8)^2}\,  .
\end{align}
For this example $CC_1= O_2, \, CC_2 = O_4$ but  novel rational fixed points arise for $m\ge 3$  although $CC_4= O_4{\!}^2$,
which is a decoupled theory as in this case $h=g=0$.

For $m\ne n$ there is no longer the $ {\mathbb Z}_2$ symmetry relating $V_1, \, V_2$ and the couplings extend to
$\lambda_1, \, \lambda_2, \, g_1, \, g_2$ as well as $h$. The lowest order fixed point equations become
\begin{align}
\lambda_1 ={}&  ( m+8) \, \lambda_1{\!}^2+ 2\, \lambda_1 g_1 + n\, h^2 \,  , \quad g _1= 3\, g_1{\!}^2 + 12 \, \lambda_1 g_1\, , \nn \\
\lambda_2 ={}&  ( n+8) \, \lambda_2{\!}^2+ 2\, \lambda_2 g_2 + m\, h^2 \,  , \quad g _2 = 3\, g_2{\!}^2 + 12 \, \lambda_2 g_2\, , \nn \\
h = {}& \big ( (m+2)\, \lambda_1 + (n+2)\lambda_2 \big )  h  +  (g_1+g_2)  h + 4\, h^2\, .
\end{align}
The symmetry group is $B_n \times B_m$.
For $g_1=g_2=0$, corresponding to a perturbed theory with  $O(m)\times O(n)$ symmetry, the non trivial irrational fixed points are referred to as
biconical. For this case when $m=n, \, \lambda_1 = \lambda_2$  the symmetry group is $ O(m)^2 \rtimes {\mathbb Z}_2 $ and the
fixed point is identical to $M\!N_{m,2}$.

\subsection{Perturbed Tetrahedral Theories}

In a similar fashion to the cubic case we take $V_1=V_2$ to be given by a potential which generates the tetrahedral fixed points
\be
V_1(\vphi) =  \tfrac18 \, \lambda \, (\vphi^2)^2 + \tfrac{1}{24} \, g \, {\ts \sum_\alpha}\, (\vphi_a e_a{\!}^\alpha )^4 \, .
\ee
where $e_a{\!}^\alpha$ here define the $m{+1}$ vertices of a hypertetrahedron in $m$-dimensions,
$\sum_\alpha e_a{\!}^\alpha = 0 $,   $\sum_\alpha e_a{\!}^\alpha e_b{\!}^\alpha
= \delta_{ab}$. After rescaling the fixed point equations become
\begin{align}
\lambda ={}&  ( m+8) \, \lambda^2+ 2\, \tfrac{m}{m+1}\,  \lambda g +  \tfrac{1}{(m+1)^2} \, g^2+ m\, h^2 \,  , \quad
g = 3\,\tfrac{m-1}{m+1} \,  g^2 + 12 \, \lambda g\, , \nn \\
h = {}& 2(m+2) \, \lambda h  + 2 \, \tfrac{m}{m+1}\, g h + 4\, h^2\, .
\end{align}
The associated fixed points reproduce hitherto known ones but two new ones, with symmetry  $(\S_{m+1} \times \S_{m+1})\rtimes {\mathbb Z}_2$,
 which may be denoted as  $TT_{1,m}$,
\begin{align}
S_{2m} = {}& \frac{m(m+1)^2(2m^2- 5m+11)}{9(m^2-m+4)^2} \, , \quad a_0 =  \frac{2m(m+1)^2}{3(m^2-m+4)} \, , \quad a_2=0\, , \nn \\
a_4 = {}& \frac{2m(m+1)^2(m-2)^2}{9(m^2-m+4)^2}  \, ,  \quad \quad \tfrac14 m - S_{2m} = \frac{m(m-2)^2(m-5)^2}{36 ( m^2-m+4)^2}\,  ,
\end{align}
and $TT_{2,m}$,
\begin{align}
S_{2m} = {}& \frac{m(m^3 - 2m^2 -19m+56)(2m^3- 13m^2 +19m+16)}{9(m^3- 5 m^2 + 24)^2} \, , \nn \\
a_0 = {}&  \frac{2m(2m^3- 13m^2 +19m+16)}{3(m^3- 5 m^2 + 24)} \, , \qquad a_2=0\, , \nn \\
a_4 = {}& \frac{m(m+1)^2(m-2)^3(2m^3- 13m^2 +19m+16)}{9(m+1)(m^3- 5 m^2 + 24)^2}  \, ,  \nn \\
\tfrac14 m - S_{2m} =  {}& \frac{m(m-2)^2(m-4)^2(m-5)^2}{36 ( m^3 - 5 m^2+ 24)^2}\,  .
\end{align}
Here, for low $m$,  $TT_{1,2}=TT_{2,2} = O_4, \ TT_{1,3} = CC_3, \ TT_{2,3} = C_6, \ TT_{2,4} = O_4{}^2$ while $TT_{1,5} =TT_{2,5} = T_{5+}{\!}^2$.

For $m\ne n$ the equations extend to
\begin{align}
\lambda_1 ={}&  ( m+8) \, \lambda_1{\!}^2+ 2\tfrac{m}{m+1}\, \lambda_1 g_1 + \tfrac{1}{(m+1)^2}\, g_1{\!}^2 + n\, h^2 \,  ,
\quad g _1= 3\, \tfrac{m-1}{m+1} \, g_1{\!}^2 + 12 \, \lambda_1 g_1\, , \nn \\
\lambda_2 ={}&  ( n+8) \, \lambda_2{\!}^2+ 2\tfrac{n}{n+1}\, \lambda_2 g_2 +  \tfrac{1}{(n+1)^2} \, g_2{\!}^2 + m\, h^2 \,  , \quad \ \
g _2= 3\, \tfrac{n-1}{n+1} \, g_2{\!}^2 + 12 \, \lambda_2 g_2\, , \nn \\
h = {}& \big ( (m+2)\, \lambda_1 + (n+2)\lambda_2)  h  + \big  ( \tfrac{m}{m+1}\, g_1+ \tfrac{n}{n+1}\, g_2 \big )  h + 4\, h^2\, .
\end{align}
Setting $g_1=g_2=0$ reduces this to the biconical case as before.

\subsection{Multi-conical Theories}

To obtain fixed points with more than two quadratic invariants the biconical case is naturally extended to potentials
\be
V(\vphi_1,\dots, \vphi_n ) = \tfrac18 \,  {\ts \sum_{r=1}^{n}} \lambda_r  \, (\vphi_r{\!}^2)^2
+ \tfrac18\,   {\ts \sum_{r\ne s}} \, h_{rs }  \, \vphi_r {\!}^2 \vphi_s{\!}^2 \, , \quad h_{rs}= h_{sr} \, ,
\ee
where each $\vphi_r$ has $m_r$ components so that $N= \sum_{r=1}^n \, m_r$. There are $\frac12 n(n+1)$ couplings
and generically  the symmetry is
$ O(m_1)\times \dots \times O(m_n)$. If $m_r = m_{r'}, \, \lambda_{r} =  \lambda_{r'}, \, h_{rs} = h_{r's}$ for $r,r' \in S$ and all $s$
there is an additional $\S_{\dim S}$ symmetry. If $m_r = m, \, \lambda_r= \lambda, \, h_{rs}=h$ for all $r,s$ then this is equivalent
to the $M\!N_{m,n}$ model in  subsection \ref{sec:cont}. Of course if $h_{rs}= 0$ for $r\in S, \, s\in S'$, $ S \cap S'= \emptyset$ then this is a decoupled
theory.
The lowest order fixed point equations reduce to
\begin{align}
\lambda_r = {}& (m_r +8) \lambda_r{\!}^2 +  {\ts \sum_{s\ne r}} \, m_s \, h_{rs}{\!}^2 \, ,  \nn \\
h_{rs}  = {}& \big ( (m_r+2)\lambda_r + (m_s +2) \lambda_s \big ) h_{rs}  + 4 \, h_{rs} {\!}^2 +   {\ts \sum_{t\ne r,s}} \, m_t \, h_{rt} h_{st}  \, .
\end{align}
Such fixed points in the triconical case, $n=3$, were consider by Eichhorn {\it et al} \cite{Eichhorn}.
Fixed points which are not reducible to decoupled or biconical theories
with additional symmetry are here denoted by $B_{O_{m_1}*O_{m_2}*\dots *O_{m_n}}$ with $O_1=I$. Extensions where the initial theories have
cubic or tetrahedral symmetry are easily obtained.

\subsection{Fixed Points for Theories containing
\texorpdfstring{$\S_n$}{Sn} Symmetries}

A wider range of fixed points can be obtained using the results for $\S_N$ symmetry obtained in 3.1 with additional fields.
Many examples are encompassed by taking $N=n+m$ and imposing $\S_n \times \S_m \times {\mathbb Z}_2{\!}^2$
where theories with $\S_n , \, \S_m$  symmetry, as discussed in subsection
\ref{sec:symm},
are linked by products of quadratic operators  which are singlets under $\S_n$ and $\S_m$. The potential
becomes
\begin{align}
V(\vphi,\psi) = {}& V_{\S_n} (\vphi) + V'{\!} _{\S_m}(\psi) + U(\vphi,\psi)\,  ,\nn \\
 U(\vphi,\psi)= {}& \tfrac14 \big ( s \, {\ts \sum_a}\,  \vphi_a{\!}^2 \, {\ts \sum_r}\,  \psi_r{\!}^2+ t \, {\ts \sum_{a\ne b}}\,   \vphi_a \vphi_b \,
  {\ts \sum_r}\,  \psi_r{\!}^2\nn \\
  \noalign{\vskip -1pt}
&\quad {} +  p \, {\ts \sum_a}\,  \vphi_a{\!}^2 \,  {\ts \sum_{r\ne s}}\,   \psi_r \psi_s +q\,   {\ts \sum_{a\ne b}}\,   \vphi_a \vphi_b \,
 {\ts \sum_{r\ne s}}\,   \psi_r \psi_s  \big ) \, ,
 \label{VSNM}
\end{align}
with couplings $x,y,z,w,u,x',y',z',w',u',s,t,p,q$. The resulting fixed points have up to four invariant quadratic operators.
An $\S_n \times O(m)$ symmetric theory is obtained  for  $V'{\!} _{\S_m}(\psi) \to V_{O(m)}(\psi) $ and $p=q=0$.

For the $\S_n \times \S_m$ invariant theory
\begin{align}
S_{n+m}= {}&S_{\S_n} + S'{\!}_{\S_m} + 6\,  n m \big ( s^2+ (n-1)t^2 + (m-1) p^2 + (n-1)(m-1) q^2 \big ) \, , \nn \\
a_0 = {}& a_{\S_n,0} + a'{\!}_{\S_m,0} +  2\, n m \, s \, , \nn \\
a_2 = {}& \tfrac{mn}{m+n}\big ( x -x'  + (n-1)z - (m-1)z' - (n-m)s \big )^2 \nn \\
\noalign{\vskip -2pt}
&{}+ n(n-1) \big ( 2y + (n-2)w + m\, t \big )^2 + m(m-1) \big ( 2y' + (m-2)w '+ n \, p \big )^2\, .
\end{align}

For the potential \eqref{VSNM} the fixed point equations reduce to
\begin{align}
&x = \beta_{\S_n,x} + 3m \big (  s^2 +(m-1)p^2 \big )\, , \qquad y = \beta_{\S_n,y} + 3m \big (  s\, t + (m-1) p\,q\big )  \, , \nn \\
& z = \beta_{\S_n,z} + m ( s^2 + 2 \, t^2 ) + m(m - 1)(p^2 + 2\, q^2 )  \, , \nn \\
& w= \beta_{\S_n,w} + m( s\,t + 2  \, t^2 ) + m(m-1) ( p\,q+2\, q^2)  \, , \qquad
 u = \beta_{\S_n,u} + 3m( t^2  + (m-1) q^2 ) \, , \nn \\
& s = \big ( x+x'+ (n-1) z + (m-1) z' \big ) s + (n-1) \big ( 2\, y +  (n-2) w \big ) t + (m-1) \big ( 2\, y' +  (m-2) w' \big ) p\nn \\
\noalign{\vskip -2pt}
&\quad {}+   4\big ( s^2 +  (n-1)\, t^2+ (m-1) p^2 + (n-1)(m-1) q^2 \big )  \, ,\nn \\
&  t =  \big ( 2y + (n-2) w \big ) s + \big  ( 2z +  4(n-2)w +(n-2)(n-3) u + x' + (m-1)z'  \big )t \nn \\
\noalign{\vskip -2pt}
& \quad {}  + (m-1) \big ( 2\, y'+(m-2)w' \big ) q  + 4\big (2\, s  +  (n-2)\, t \big ) t + 4(m-1) \big (2\, p + (n-2)q\big ) q  \, , \nn \\
& q=\big ( 2\, y + (n-2) w \big ) p + \big ( 2\,y'+( m-2) w' \big ) t \nn \\
& \quad {}+  \big (2\, z + 4(n-2) w + (n-2)(n-3) u + 2\, z' + 4(m-2)w' + (m-2) (m-3)u'\big ) q\nn \\
& \quad {}+ 8\, s \, q + 8 \, t \, p + 8 \big ( (n-2) t +  ( m-2) p \big ) q + 4( m-2)(n-2) q^2 \, ,
\end{align}
together with those obtained by $(x,y,z,w,u,t,n) \leftrightarrow (x',y',x',w',u',p,m)$
and where the $\beta$-functions are given in  \eqref{redsol}. For $m=n$ we may reduce the couplings by
 imposing the symmetry condition $(x',y',z',w',u',p)=(x,y,z,w,u,t)$.

Solving these equations for various $n,m$ generates  of the
order $50-200$ fixed points, though many are decoupled theories. The non trivial fixed points may be denoted by $B_{\S_n*\S_m}$.
These include all the fixed points found in the perturbed cubic and tetrahedral  theories considered in subsections 4.1 and 4.2.
There are further special cases when just  $S_m \to O_m, \, C_m, \, T_{m}$ and  there is an additional $O(m), \, B_m, \, \S_{m+1}$ symmetry.

\section{Fixed Points for Low \texorpdfstring{$N$}{N}}\label{sec:lowN}

As described in the introduction we have looked for fixed points  numerically where the bound \eqref{Sbound} is close to being
saturated. In each case we have determined the eigenvalues  $\{ \gamma \}$ for the anomalous dimension matrix $\Gamma$
and the stability eigenvalues  $\{ \kappa \}$. These are are too lengthy  to add in here but are available on request. The number
of  different eigenvalues $\gamma$ in general correspond to the number of independent invariant quadratic forms.
The eigenvalues $\{\kappa\}$ serve to identify decoupled theories which we have checked through the additivity of $S_N, \, a_0$.
For fully interacting cases $\kappa=1$ is non degenerate, the eigenvector being the coupling at the fixed point.
The number of zero eigenvalues for $\kappa$   is equal to or greater than  $ \dim \mathfrak {so}(N) -  \dim \mathfrak {H}$, where
 $\dim \mathfrak {H} $ is the dimension of the Lie algebra of the symmetry group at the fixed point $H_{\rm fp} < O(N)$ and
 $  \dim \mathfrak {so}(N)= \tfrac12 N (N-1) $. Additional zero eigenvalues arise if there is a bifurcation point.
 For the hypercubic symmetry case $C_N$ there are a further  $(N{-1})$  zero $\kappa$ at this lowest order but these become
 non zero at the next order in the $\vep$ expansion. If $H_{\rm fp}$ is discrete then there will be at a minimum  $\tfrac12 N (N-1) $
 zero $\kappa$.

 The eigenvalues $\{\gamma \}$ and $\{\kappa\}$ fall into degenerate groups which correspond to irreducible representations of the
 symmetry group $H_{\rm fp}$ for the particular fixed point. Where possible we have identified the symmetry group in terms
 the analysis undertaken in the previous sections. Of course these are all subgroups of $O(N)$ and commonly are formed in terms
 of direct or semi-direct products of simpler groups. For a non abelian symmetry group $H_{\rm fp}$ the larger it is then
 larger the dimensions of possible representations.

 The fixed points may be realised in a restricted theory  $F_V$ corresponding to a potential $V$ with $p$ couplings and
  symmetry group  $H_{\rm sym} \subseteq H_{\rm fp}$ since the symmetry may be  enhanced at the fixed point.
 The eigenvalues $\{\gamma \}$ remain of dimension $N$ but $\dim \{\kappa\}_{F_V} = p$, the $p$  eigenvalues $\{ \kappa\}_{F_V}$
 are a subset of the complete set of $\tfrac{1}{24}N(N+1)(N+2)(N+3)$ eigenvalues $\{\kappa\}$ obtained by solving
 \eqref{keig}.

For small $N$ results for fixed points which can not be reduced to decoupled products of fixed point theories (including free theories)
with lower $N$ are enumerated  in various tables below. Analytic results have been added to complement the numerical ones. As already
said our searches may not find all fixed points but we hope this to be the case for $N=3,4,5,6$. We use the terminology
 where  $I$ is the usual Ising fixed point, $O_N, \ C_N, \ T_{N\pm}$
denote the  fixed points which are present for general $N$ with $O(N)$, cubic  and tetrahedral symmetry.
Of course $O_1 = I$, $O_2$ is the XY model and   $O_3$ is the Heisenberg model.   $C_2=I^2= I\cup I$, $T_{3-}= C_3, \
 T_{3+}=I^3=  I \cup I \cup I$ while $T_{4+}= O_4$ and $T_{5+} = T_{5-}$.

In the tables the degeneracies for $\gamma$ are listed in order for decreasing value of $\gamma$.

\subsection{\texorpdfstring{$N=1,2,3$}{N=1,2,3}}
\begin{tabular}{ l c c c c c  c  c c }
$N=1$ & $S_1$ & $a_0$ & $a_2$ & $a_4$ & Symmetry  & \makecell{\# different $\gamma$\\[-2pt] and degeneracies} & $ \# {\kappa<0}, \, {=0}$  \\
\hline
\noalign{\vskip 5pt}
$I$ &   $\tfrac19$ & $ \tfrac13$   &0  &0  & ${\mathbb Z}_2$  & 1(1)  & $0, \, 0$
\end{tabular}\\[6pt]
$\kappa = \nu =1, \ \mu = \tfrac13$\\
\begin{tabular}{ l c c c c c  c  c c }
$N=2$ & $S_2$ & $a_0$ & $a_2$ & $a_4$ & Symmetry  & \makecell{\# different $\gamma$\\[-2pt] and degeneracies} & $ \# {\kappa<0}, \, {=0}$  \\
\hline
\noalign{\vskip 5pt}
$O_2$ &   $\tfrac{6}{25}$ & $ \tfrac45$   & 0  & 0  & $O(2) $   & 1(2) & $0, \, 0$
\end{tabular}\\[6pt]
$\big \{\kappa \big \} {}_{O_2} = \big \{ 1(1), \, \tfrac45(2), \, \tfrac15(2) \big  \} , \
 \big \{ \nu \big \} {}_{O_2} = \big \{ 1(2), \, \tfrac35(2)\big  \}, \
 \big \{ \mu \big \} {}_{O_2} = \big \{\tfrac25(1), \, \tfrac15(2) \big
 \}$\\[16pt]
\begin{adjustbox}{max width=\textwidth}
\begin{tabular}{ l c c c c c  c  c c }
$N=3$ & $S_3$ & $a_0$ & $a_2$ & $a_4$ & Symmetry&   \makecell{\# different $\gamma$\\[-2pt] and degeneracies}&   $ \# {\kappa<0}, \, {=0}$   \\
\hline
\noalign{\vskip 5pt}
$O_3$  &   $\tfrac{45}{121}$ & $ \tfrac{15}{11}$    &0  &0  & $O(3) $   & 1(3)   & $0, \, 0$ \\
 \noalign{\vskip 5pt}
$C_3$  &   $\tfrac{10}{27}$ & $ \tfrac43$   & 0  &   $\tfrac {2}{135} $ &   $ B_3 $    & 1(3) &  $1, \, 5$ \\
 \noalign{\vskip 5pt}
$B_{O_2*I}$  &   $0.370451$ & $ 1.33713 $   & 0.000255  &   $ 0.01265$ &   $ O(2) \times  {\mathbb Z}_2 $ & 2(2,1)   & $1, \, 2$ \\
\end{tabular}
\end{adjustbox}\\[6pt]
$\big \{\kappa \big \} {}_{O_3} = \big \{ 1(1), \, \tfrac{8}{11}(5), \, \tfrac{1}{11}(9) \big  \} , \
 \big \{ \nu \big \} {}_{O_3} = \big \{ 1(3), \, \tfrac{6}{11}(7)\big  \},\
 \big \{ \mu \big \} {}_{O_3} = \big \{\tfrac{5}{11}(1), \,
 \tfrac{2}{11}(5) \big  \}$, \\[2pt]
$\big \{ \kappa \big \} {}_{C_3} = \big \{ 1(1), \, \tfrac{5}{9}(2), \,  \tfrac{1}{18}(9 \pm \sqrt{33})(3,3), \, 0(5), \, -\tfrac{1}{9}(1) \big  \}, \
 \big \{ \nu \big \} {}_{C_3} = \big \{ 1(3), \,  \tfrac23(1), \,
 \tfrac{5}{9}(3), \,  \tfrac49(3) \big \}$,\\[2pt]
$\big \{ \mu \big \} {}_{C_3} = \big \{ \tfrac49(1), \, \tfrac{2}{9}(3),
\,  \tfrac19(2) \big \}$

Perturbing two decoupled Ising fixed points leads to just  $O_2$  as the
unique non decoupled theory. However, for $N=3$ there is the irrational
biconical fixed point  $B_{O_2*I }$  as well as $O_3$.

\subsection{\texorpdfstring{$N=4$}{N=4}}
\begin{align}
\noalign{\vskip 4pt}
&\begin{adjustbox}{max width=\textwidth}
\begin{tabular}{ l c c c c c  c  c c }
$N=4$ & $S_4$ & $a_0$ & $a_2$ & $a_4$ & Symmetry &  \makecell{\# different $\gamma$\\[-4pt] and degeneracies} &   $ \# {\kappa<0}, \, {=0}$    \\
\hline
\noalign{\vskip 5pt}
$O_4$   &   $\tfrac12 $ & $ 2 $   &0  &0  & $O(4) $  &  1(4)  & $0, \, 25 $ \\
 \noalign{\vskip 5pt}
$T_{4-}$ &   $\tfrac{220}{441}$ & $ \tfrac{40}{21}$   & 0  &   $\tfrac {20}{441} $ &   $\S_5 \times  {\mathbb Z}_2 $  &  1(4) & $15, \, 6$  \\
 \noalign{\vskip 5pt}
$B_{\S_3*I}$ & 0.499115 & 1.92406 & 0.000328& 0.036117&   $ \S_3 \times  {\mathbb Z}_2 \times  {\mathbb Z}_2$  & 3(1,2,1) & $14,\,  6$\\
 \noalign{\vskip 5pt}
 ${\hat B}_{O_2*I*I}$ & 0.499144 & 1.92641 & 0.000359& 0.034994 &  $ D_4 \times {\mathbb Z}_2$  & 3(1,2,1)  & $13, \, 6$  \\
  \noalign{\vskip 5pt}
$O_2\circ O_2 $  & 0.499606 & 1.95458 & 0.000273& 0.021851& $ O(2) $ & 2(2,2) & $12, \, 5$ \\
\end{tabular}
\end{adjustbox}\nn \\
& \big \{ \kappa \big \} {}_{O_4} = \big \{ 1(1), \, \tfrac{2}{3}(9), \, 0(25) \big  \} , \
\big \{ \nu \big \} {}_{O_4} = \big \{ 1(4), \, \tfrac{1}{2}(16)\big  \}, \ \big \{ \mu \big \} {}_{O_4} = \big \{\tfrac{1}{2}(1), \, \tfrac{1}{2}(9) \big  \},
\nn \\
&  \big \{ \kappa \big \} {}_{T_{4-}} = \big \{ 1(1), \, \tfrac{1}{42} (21 \pm \sqrt{201})(4,4), \, \tfrac{1}{42} (9 \pm \sqrt{129})(5,5), \, 0(6), \,
 -\tfrac{1}{21} (1),  \, - \tfrac{4}{21}(5), \, - \tfrac27 (4)\big  \}, \nn \\
&  \big \{ \nu \big \} {}_{T_{4-}} = \big \{ 1(4), \, \tfrac{5}{7}(1), \, \tfrac{11}{21} (6), \, \tfrac{10}{21}(4), \,  \tfrac{1}{3} (5)\big  \} , \,
 \big \{ \mu \big \} {}_{T_{4-}} = \big \{\tfrac{10}{21}(1), \, \tfrac{5}{21} (4), \, \tfrac{2}{21}(5) \big  \} \nn
\end{align}
Note that $\tfrac{220}{441}= 0.498866$ and the dimension of space of slightly marginal quartic deformations is  35.
Of course assuming a particular symmetry group
restricts the space of couplings and hence the dimension of the stability matrix is thereby restricted although
the fixed points found with a single quadratic invariant are identical to the $O_4$ or $T_{4-}$ cases above or correspond to decoupled theories.

The 25 zero modes for $\kappa$ for the $O_4$ fixed point correspond to the $O(4)$ representation given by symmetric traceless $4$-index tensors and are accidental in that they are removed at higher orders in the $\vep$ expansion.  At next order and reinstating
$\vep$ explicitly  the 25 dimensional representation has
$\kappa = - \frac16 \vep^2$. At lowest order in $\vep$ the $O_4$ fixed point is degenerate with $C_4$ and $T_{4+}$. At ${\rm O}(\vep^2)$
for these fixed points  the different cubic and tetrahedral symmetry  representations develop different eigenvalues. For $C_4$ we have for
$\{\kappa\}$ an expansion  to ${\rm O}(\vep^2)$
\begin{align}
C_4: \quad & \vep-\tfrac{13}{24}\vep^2\ (1) \, , \quad \tfrac 23 \vep - \tfrac{5}{36} \vep^2\ (3)  \, ,  \quad \tfrac 23 \vep - \tfrac{5}{9} \vep^2 \ (6)  \, ,  \nn \\
&  \tfrac16 \vep^2 (1)\, ,
 \quad 0 \ (6) \, , \quad 0 \ (3) \, ,  \   - \tfrac16 \vep^2 \  (2) \, ,  \quad   - \tfrac14 \vep^2 \ (6) \, ,  \quad  - \tfrac12 \vep^2 \ (1) \, .
 \label{Cep2}
\end{align}
For the $T_{4+}$ fixed point the $O_4$ eigenvalues for $\kappa$ also split at ${\rm O}(\vep^2)$ giving
\begin{align}
T_{4+}: \quad & \vep-\tfrac{13}{24}\vep^2 \  (1) \, , \quad  \tfrac 23 \vep + \tfrac{5}{18} \vep^2 \ (4)  \, ,  \quad
\tfrac 23 \vep - \tfrac{35}{36} \vep^2 \ (5)  \, ,  \nn \\ &   \tfrac16 \vep^2  \ (1)\, ,
\quad  \tfrac18(\sqrt{41}-1)\vep^2 \ (4) \, , \quad  0\ (6) \, , \quad 0 \ (5) \, , \quad
   - \tfrac23 \vep^2 \  (5) \, ,  \quad   -  \tfrac18(\sqrt{41}+1)\vep^2 \ (4)  \, .
  \label{Tep2}
\end{align}
The degeneracies of course correspond to dimensions of representations of the respective symmetry groups $B_4$ and
$\S_5 \times  {\mathbb Z}_2 $. In both cases there remains a representation with $\kappa=0$ which should become non zero
at higher order in the $\vep$ expansion.

The three irrational fixed points which were found in our numerical search were also obtained  in \cite{Codello4}.
One of these appears within the biconical framework described here, the other two appear to be special to $N=4$. For the case with largest $S_4$ we follow \cite{Codello4} and consider the potential
\begin{align}
V_1(\phi) = {}& \tfrac18 \, \lambda_1 ( \phi_1{\!}^2 + \phi_2{\!}^2)^2 +  \tfrac18 \, \lambda_2 ( \phi_3{\!}^2 + \phi_4{\!}^2)^2  + \tfrac14\, h
 ( \phi_1{\!}^2 + \phi_2{\!}^2)( \phi_3{\!}^2 + \phi_4{\!}^2)\nn \\
&{}  + \tfrac16 \, {\hat h} \big ( \phi_1{\!}^3 - 3 \, \phi_1 \phi_2{\!}^2 ,\,  \phi_2{\!}^3 - 3 \, \phi_1{\!}^2  \phi_2 \big ) \cdot \big ( \phi_3, \, \phi_4 \big ) \, ,
\label{VC}
\end{align}
which has four couplings.
For ${\hat h} =0$ this would correspond to a biconical theory with $O(2)\times O(2)$ symmetry which has no non trivial fixed points
other than $O_4$. With ${\hat h}$ non zero, the symmetry is reduced to $O(2)$ corresponding to $\delta \phi_1 = \phi_2 , \, \delta \phi_2 = -\phi_1 , \,
\delta \phi_3 = - 3 \, \phi_4 , \, \delta \phi_4 = 3\,  \phi_3$. This has  a ${\mathbb Z}_3$ subgroup
 generated by $2\pi/3$ rotations of $\phi_1, \, \phi_2$ which leave $\phi_3, \, \phi_4$ invariant.
The ${\mathbb Z}_3$ symmetry   ensures that the potential \eqref{VC} is preserved under RG flow.
Theories related by ${\hat h}\to - {\hat h}$ are equivalent. With this symmetry there are two quadratic invariants. The fixed point equations
for this case give
\be
\lambda_1 = 2 (5 \lambda_1{\!}^2 + h ^2 + 2\, {\hat h}{}^2 ) \, , \ \  \lambda_2 = 2 (5 \lambda_2{\!}^2 + h ^2  )\, , \ \
h= 4 ( ( \lambda_1+\lambda_2 ) h + h^2 + {\hat h}{}^2 )\, , \ \ {\hat h}=  6 (\lambda_1 + h) {\hat h}\, ,
\ee
and
\begin{align}
S_4 ={}&  24 (\lambda_1{\!}^2 + \lambda_2{\!}^2 + h^2 ) + 32 \, {\hat h}^2 \, , \qquad a_0 = 8 (\lambda_1 + \lambda_2 + h) \, ,\nn \\
a_2= {}& 16 (\lambda_1-\lambda_2)^2 \, , \qquad \quad  a_4 = 4 ( \lambda_1 + \lambda_2 - 2 h)^2 +32\, {\hat h}^2 \, .
\end{align}
Apart from $O_4$ the only non trivial fixed point is the irrational one labelled by $O_2\circ O_2$  above.

For the other case 8 couplings are necessary
\begin{align}
V_2(\phi) = {}& \tfrac18 \, \lambda  ( \phi_1{\!}^2 + \phi_2{\!}^2)^2 +   \tfrac{1}{24} \, g ( \phi_1{\!}^4 + \phi_2{\!}^4) +
\tfrac{1}{24}  \, x_1 \,  \phi_3{\!}^4 + \tfrac{1}{24} \, x_2 \,\phi_4{\!}^4   + \tfrac14\, z \,  \phi_3{\!}^2 \phi_4{\!}^2 \nn \\
&{}  + \tfrac14\, h_1 ( \phi_1{\!}^2 + \phi_2{\!}^2)\phi_3{\!}^2  + \tfrac14\, h_2 ( \phi_1{\!}^2 + \phi_2{\!}^2)\phi_4{\!}^2
 +  h\,  \phi_1 \phi_2 \phi_3  \phi_4 \, .
\label{VC2}
\end{align}
For $h=0$ this corresponds to a triconical type theory with symmetry group of order 32,
$({\mathbb Z}_2{\!}^2 \rtimes {\mathbb Z}_2) \times {\mathbb Z}_2 \times {\mathbb Z}_2$ where ${\mathbb Z}_2{\!}^2 \rtimes {\mathbb Z}_2
\simeq D_4$ the two dimensional cubic symmetry group. For $h$ non zero the symmetry is reduced to $D_4 \times {\mathbb Z}_2$ since
a $\pi/2$ rotation of $(\phi_1 , \phi_2)$ then requires also a reflection $\phi_3\to -\phi_3$. There are clearly three quadratic invariants.
Theories related by $h\to -h$ and also $x_1 \leftrightarrow x_2, \, h_1 \leftrightarrow h_2$ are equivalent.
The fixed point equations require
\begin{align}
\lambda={}&  10 \lambda^2 + 2 g \lambda + h_1{\!}^2 +h_2{\!}^2 + 4 h^2 \, , \qquad g= 3g^2 +12 \lambda g - 12 h^2 \, ,\nn \\
z= {}& 4 z^2 + (x_1+x_2) z + 2h_1 h_2 + 4  h^2\, , \quad
x_1 =  3 x_1{\!}^2 + 3 z^2 + 6 h_1{\!}^2 \, , \quad x_2 =  3 x_2{\!}^2 + 3 z^2 + 6 h_2{\!}^2 \, , \nn \\
h_1={}& 4 h_1{\!}^2 + ( 4 \lambda + g +x_1) h_1 + z h_2 + 4 h^2 \, ,\quad h_2= 4 h_2{\!}^2 + ( 4 \lambda + g +x_2) h_2 + z h_1 + 4 h^2 \, ,\nn \\
h = {}& 2(\lambda + z)h + 4 (h_1+h_2) h \, .
\end{align}
Solutions with $h$ non zero give the tetrahedral fixed point $T_{4-}$ and also the case listed as ${\hat B}_{O_2*I*I}$ above.

Rational fixed points
for $N=4$ were obtained  in \cite{Brezin2} by looking for fixed points arising from quartic potentials invariant under all
possible subgroups of $O(4)$ subject to there being a single invariant quadratic form. The subgroups of $O(4)$ are non trivial
\cite{duval,Conway,Farrill}.
Here we list their symmetry types discussed in  \cite{Brezin2} for which  the group acting on the couplings defining equivalent theories
is discrete,
the number of couplings corresponding to the number of independent quartic potentials necessary to realise the required symmetry
and the associated fixed points in our notation. These always involve the $O(4)$ symmetric case but include those corresponding to
decoupled theories which are omitted above.
\begin{align}
\hskip -1cm
\begin{tabular}{ l c c c c } symmetry(dimension) & number of couplings & fixed points ${\rm O}(\vep)$ &  fixed points ${\rm O}(\vep^2)$ \\
\hline \noalign{\vskip 5pt}
 di-icosahedral (240) &   2 &  $T_{4-}, \, O_4$ & $T_{4+}$  \\
\noalign{\vskip 3pt}
dipentagonal (40)& 3 & $ O_2{\!}^2, \, T_{4-}, \, O_4$  &$T_{4+}, \, M\!N_{2,2}$  \\
\noalign{\vskip 3pt}
trigonal-cubic (48)& 3 &  $ I^4, \, O_4 $ & $ C_4$ \\
\noalign{\vskip 3pt}
orthotetragonal (32) & 4 & $I^4, \, O_2{\!}^2, \, O_4$ &$ C_4, \, M\!N_{2,2} $ \\
\noalign{\vskip 3pt}
diorthorhombic (32) & 5 & $I^4, \, O_2{\!}^2, \, O_4$   & $ C_4, \, M\!N_{2,2} $ \\
\end{tabular} \nn
\end{align}
Each example is intended  to correspond to a different chain of symmetry breaking of $O(4)$ and therefore to be independent.
The different symmetry groups in each case are described in Appendix \ref{appV4}.
However for a special restriction of the couplings the dipentagonal case reduces to the  di-icosahedral  which  is identical with the hypertetrahedral theory
 restricted to $N=4$.
Each of the fixed points arising from solving the RG equations for the various potentials can be obtained starting from theories
with just two couplings and also  can in principle be extended to arbitrary $N$.
The $O_4$ fixed point splits at ${\rm O}(\vep^2)$. An $O(4)$ symmetric fixed point remains but other fixed points with
lesser symmetry can emerge and are given above  for each example. The $\kappa$ eigenvalues for $C_4$ and $T_{4+}$ are
listed in \eqref{Cep2} and \eqref{Tep2}. For $M\!N_{2,2}$, which has a $O(2)^2 \rtimes {\mathbb Z}_2$ symmetry, the eigenvalues are
\begin{align}
M\! N_{2,2}: \quad & \vep-\tfrac{13}{24}\vep^2\ (1) \, , \quad \tfrac 23 \vep + \tfrac{5}{36} \vep^2\ (1) \, , \quad \tfrac 23 \vep - \tfrac{5}{18} \vep^2\ (4)  \, ,  \quad \tfrac 23 \vep - \tfrac{25}{36} \vep^2 \ (4)  \, ,  \nn \\
&  \tfrac16 \vep^2\ (5)\, ,
 \quad 0 \ (4) \, , \quad    - \tfrac{1}{12} \vep^2 \  (4) \, ,  \quad   - \tfrac13 \vep^2 \ (8) \, ,  \quad  - \tfrac12 \vep^2 \ (4) \, .
 \label{Oep2}
\end{align}

The lowest order invariants $S_N, \,a_0, \,a_4$ do not distinguish these new fixed points but they can be
identified in terms  of their stability matrix eigenvalues $\{ \kappa \}$ when calculated to ${\rm O}(\vep^2)$ as was given in
two cases with cubic and tetrahedral symmetry in \eqref{Cep2} and \eqref{Tep2} above.
To identify the
split fixed points we consider the $\beta$-functions for $N=4$ to two loops in a canonical form
\begin{align}
\beta_\lambda =  {}& 12\, \lambda^2 - 78\,  \lambda^3 + ( 1- 17 \, \lambda ) \, a_{rs} \,g^r g^s  -  \tfrac23 \, b_{rst} \, g^r g^s g^t \, , \nn \\
\beta_g{\!}^r =  {}& 12\, \lambda \, g^r  - 102\,  \lambda^2 g^r  + ( 1-  12\, \lambda )\,  b_{st}{}^r \, g^s g^t + {\rm O} ( g^3 ) \, ,   \quad
b_{rst} = b_{rs}{}^u a_{ut} \, ,
\end{align}
where $g^r$ are couplings for $\phi^4$  operators  which are symmetric traceless tensors and $\lambda$ is,
as previously, the coupling for the $O(4)$ invariant $(\phi^2)^2$. Split solutions are obtained
by requiring at lowest order in a $\vep$ expansion that $\lambda= {\rm O}(\vep), \ g^r = {\rm O}(\vep^2)$ and then
\be
\vep \, \lambda = \beta_\lambda  +  {\rm O}(\vep^4) \, , \qquad  \vep \, g^r = \beta_g{}^r +  {\rm O}(\vep^5) \, ,
\ee
requires
\be
\lambda = \tfrac{1}{12} \, \vep + \tfrac{13}{288}\, \vep^2  +  {\rm O}(\vep^3)  \, , \qquad
b_{st}{}^r \, g^s g^t =  \tfrac16 \, g^r \, \vep^2  +  {\rm O}(\vep^5) \, .
\ee
These results are not affected by three loop contributions and it is straightforward to determine $b_{st}{}^r$ in each case.\footnote{For
the cubic, tetrahedral and di-icosahedral cases there is one coupling $g$ and $(a,b) = (\tfrac16,1 ),  \ (\tfrac{1}{15}, \tfrac15),\ ( \tfrac{20}{3}, -2)$
respectively. For the pentagonal and trigonal-cubic cases there are two couplings with the non zero $a_{rs}, \, b_{rs}{\!}^t$ given by
$a_{11}=\frac43, \,  a_{22}=\frac13, \,b_{11}{\!}^1 =2, \, b_{22}{\!}^1=-\frac14, \, b_{12}{\!}^2= b_{21}{\!}^2 =-1$
and  $a_{11}=\frac16, \, a_{22}=\frac{1}{12}, \, b_{11}{\!}^1 =1, \, b_{22}{\!}^1=-\frac14, \, b_{12}{\!}^2=  b_{21}{\!}^2 =-\frac12, \, b_{22}{\!}^2 =\frac12$.}

\subsection{\texorpdfstring{$N=5$}{N=5}}
\begin{align}
&\begin{adjustbox}{max width=\textwidth}
\begin{tabular}{ l c c c c c c c c } $N=5$ & $S_5$ & $a_0$ & $a_2$ & $a_4$ & Symmetry &$ \hskip - 1cm$
  \makecell{\# different $\gamma$\\[-4pt] and degeneracies} & $ \# {\kappa<0}, \, {=0}$ \\
\hline \noalign{\vskip 5pt}
$O_5$ &   $\tfrac{105}{169} $ & $ \tfrac{35}{13} $   &0  &0  & $O(5) $  &  1(5)  & $55, \, 0 \  \, $ \\
\noalign{\vskip 5pt}
$C_5$ & $\tfrac{28}{45}$ & $\tfrac83$ & 0 & $\tfrac{4}{315}$&
$B_5$ & 1(5) & $40, \, 14$ \\
\noalign{\vskip 5pt} $T_{5\pm}$ & $\tfrac{5}{8}$ & $ \tfrac52$ & 0 &
$\tfrac {5}{56} $ & $\S_6 \times {\mathbb Z}_2 $ & 1(5) & $39, \, 11$ \\
\noalign{\vskip 5pt}
$B_{I* O_4}$ & 0.621937 & 2.67255 & 0.000170& 0.009605& $ {\mathbb Z}_2  \times O(4)  $ & 2(4,1) & $50,\,  4$\\
\noalign{\vskip 5pt}
$B_{C_2*O_3}$ & 0.622163 & 2.66667 & 0.000118 & 0.012561 &  $B_2 \times O(3) $  & 2(3,2)  & $46, \, 7$  \\
\noalign{\vskip 5pt}
$B_{C_3*O_2}$ & 0.622230 & 2.66560 & 0.000056 & 0.013157 & $ B_3   \times O(2) $ & 2(2,3)  & $41, \, 9$  \\
\noalign{\vskip 5pt}
$B_{I*O_2*O_2}$& 0.623037 & 2.63897 & 0.000064 & 0.026068 & ${\mathbb Z}_2  \times  O(2) \times O(2) $ & 3(2,1,2)  & $40, \, 8$  \\
\noalign{\vskip 5pt}
$B_{C_3*O_2}$ & 0.623040 & 2.63881 & 0.000066 & 0.026139 & $ B_3 \times O(2) $  & 2(3,2)  & $38, \, 9$  \\
\noalign{\vskip 5pt}
$B_{O_2*O_3}$ & 0.623053 & 2.63808 & 0.000082 & 0.026474 & $O(2)\times O(3)$ & 2(3,2) & $37,\,  6$\\
\end{tabular}
\end{adjustbox}\nn\\
& \big \{ \kappa \big \} {}_{O_5} = \big \{ 1(1), \,
  \tfrac{8}{13} (14), \, - \tfrac{1}{13} (55) \big  \}, \ \big \{ \nu \big \} {}_{O_5} = \big \{ 1(5), \, \tfrac{6}{13}(30)\big  \}, \
  \big \{ \mu \big \} {}_{O_5} = \big \{\tfrac{7}{13}(1), \, \tfrac{2}{13}(14) \big  \},\nn\\
& \big \{ \kappa \big \} {}_{T_{5+}} =  \big \{ \kappa \big \} {}_{T_{5-}} =\big \{ 1(1), \,
  \tfrac{1}{4} (2 \pm \sqrt{2})(5,5), \, \tfrac{1}{12} (2 \pm
  \sqrt{10})(9,9), \, 0(11), \,
-\tfrac{1}{4} (21), \, - \tfrac{5}{12} (9) \big  \}, \nn\\
&  \big \{ \nu \big \} {}_{T_{5+}} = \big \{ \nu \big \} {}_{T_{5-}} = \big \{ 1(5), \, \tfrac{3}{4}(1), \, \tfrac{1}{2} (15), \, \tfrac{1}{3}(9), \,  \tfrac{1}{4} (5)\big  \} , \,
 \big \{ \mu \big \} {}_{T_{5+}} =  \big \{ \mu \big \} {}_{T_{5-}} = \big \{\tfrac{1}{2}(1), \, \tfrac{1}{4} (5), \, \tfrac{1}{12}(9) \big  \} , \nn \\
& \big \{ \kappa \big \} {}_{C_5} = \big \{ 1(1), \,
  \tfrac{11}{15}(4),\, \tfrac{1}{15}(1),\, \tfrac{1}{30} (7 \pm
  \sqrt{97})(10,10), \, 0(14), \,  -\tfrac{1}{15} (5), \,
-\tfrac{2}{15} (20), \, - \tfrac{1}{5} (5) \big  \}, \nn \\
&  \big \{ \nu \big \} {}_{C_5} = \big \{ 1(5), \,  \tfrac{8}{15}(5), \, \tfrac{2}{5}(10), \,  \tfrac{7}{21}(15) \big \} ,\
 \big \{ \mu \big \} {}_{C_5} = \big \{ \tfrac{8}{15}(1), \, \tfrac{1}{5}(4), \,  \tfrac{2}{15}(10) \big \} \nn
\end{align}
For comparison for the first three rational cases $S_5 =\{0.6213, \, 0.6222, \, 0.625\}$.
For $N=5$ the total dimension of the space  of slightly marginal deformations is 70. For $C_5$ at ${\rm O}(\vep^2)$ the $\kappa=0$
stability eigenmodes split, 4 have $\kappa= - \frac{323}{7425}\vep$ while 10 remain zero as expected since this fixed point
has no continuous symmetries. The tetrahedral fixed point also has no continuous symmetries but
there is an additional zero eigenvalue $\kappa$ for $T_{5\pm}$
 since this is a bifurcation point where the two tetrahedral fixed points coincide. At higher orders the tetrahedral fixed points are
 become complex in a region $N_{{\rm crit}-}< N < N_{{\rm crit}+}$ \cite{Seeking}.

\subsection{\texorpdfstring{$N=6$}{N=6}}

There are more fixed points since 6 is non prime, rational fixed points are given by
\begin{align}
&\begin{tabular}{ l c c c c c c c c } $N=6$ & $S_6$ & $a_0$ & $a_2$ & $a_4$ & Symmetry & \# different $\gamma$ & $ \# {\kappa<0}, \, {=0}$ \\
\noalign{\vskip -2 pt}
& & & & & &  and degeneracies \\
\hline \noalign{\vskip 5pt}
$O_6$ &   $\tfrac{36}{49} $ & $ \tfrac{24}{7} $   &0  &0  & $O(6) $  &  1(6)  & $105, \, 0 \  \, $ \\
\noalign{\vskip 5pt}
$C_6$ & $\tfrac{20}{27}$ & $\tfrac{10}{3}$ & 0 & $\tfrac{5}{108}$&
$B_6 $ & 1(6) & $84, \, 20$  \\
\noalign{\vskip 5pt}
$M \! N_{2,3}$  & $\tfrac{90}{121}$ & $\tfrac{36}{11}$ & 0 & $\tfrac{9}{121}$&
$O(2)^3\rtimes S_3$ & 1(6) & $86, \, 12 $  \\
\noalign{\vskip 5pt}
$CC_3 $ & $\tfrac{56}{75}$ & $\tfrac{16}{5}$ & 0 & $\tfrac{8}{75}$& $(\S_4 \times  {\mathbb Z}_2)^2 \rtimes {\mathbb Z}_2$ & 1(6) & $79, \, 15 $  \\
\noalign{\vskip 5pt}
$M\! N_{3,2} $ & $\tfrac{216}{289}$ & $\tfrac{54}{17}$ & 0 & $\tfrac{135}{1156}$& $O(3)^2\rtimes {\mathbb Z}_2$ & 1(6) & $77, \, ~ 9 $  \\
\noalign{\vskip 5pt}
$T_{6+}$ & $\tfrac{110}{147}$ & $ \tfrac{20}{7}$ & 0 &
$\tfrac {5}{21} $ & $\S_7 \times {\mathbb Z}_2 $ & 1(6) & $84, \, 15$ \\
\noalign{\vskip 5pt} $T_{6-}$ & $\tfrac{182}{243}$ & $ \tfrac{28}{9}$ & 0 &
$\tfrac {35}{243} $ & $\S_7 \times {\mathbb Z}_2 $ & 1(6) & $83, \, 15$ \\
\end{tabular} \nn
\end{align}
Numerically for these theories $S_6 =\{0.7347, \,  0.7407, \, 0.7438,\, 0.7467, \,  0.7474, \, 0.7483, \, 0.7490\}$.
These results demonstrate that the value of $S$ does
not uniquely characterise a fixed point, thus $S_6= \frac{20}{27}$ corresponds to $C_3{\!}^2=C_3 \cup C_3$ as well as $C_6$, although
$a_0 = \frac83, \, \frac{10}{3}$ respectively,   and  $S_6=\frac{90}{121}$ corresponds to $O_3{\!}^2=O_3\cup O_3$ as well as $M\!N_{2,3}$
while $a_0 = \frac{30}{11}, \, \frac{36}{11}$.

For the $C_6$ fixed point five zero eigenvalues $\kappa$ become
$-\frac{215}{3402} \vep^2$ at ${\rm O}(\vep^2)$.

Irrational cases for $N=6$ based on a numerical search with $S>0.74$ and using results for fixed points obtained for
various special cases are
\begin{align}
&\begin{adjustbox}{max width=\textwidth}
\begin{tabular}{ l c c c c c c c c } $N=6$ & $S_6$ & $a_0$ & $a_2$ & $a_4$ & Symmetry & \makecell{\# different $\gamma$\\[-2pt] and degeneracies} & $ \# {\kappa<0}, \, {=0}$ \\
 \hline \noalign{\vskip 5pt}
 $B_{I*O_5}$ & 0.738216& 3.35878& 0.002115 &
0.031859& $  {\mathbb Z}_2 \times O(5) $    & 2(5,1) & $99,\,  5$ \\
\noalign{\vskip 5pt}
 $B_{C_2*O_4}$ & 0.739865& 3.33333 & 0.001752 &
0.044369& $  B_2 \times O(4) $    & 2(3,3) & $94,\,  9$ \\
\noalign{\vskip 5pt}
$B_{C_3*O_3}$ & 0.740572 & 3.32649 & 0.001091 &
0.048323& $  B_3 \times O(3) $    & 2(3,3) & $90,\,  12$\\
\noalign{\vskip 5pt}
$B_{C_4*O_2}$  & 0.740798 & 3.32758 & 0.000520 & 0.048438 &$ B_4 \times O(2) $    & 2(2,4) & $85,\,  14$\\
\noalign{\vskip 5pt}
$B_{O_2*O_4}$  & 0.744334 & 3.23615 & 0.002037 & 0.088569 & $O(2)\times O(4)$ & 2(4,2) & $94,\,  8$\\
\noalign{\vskip 5pt}
$B_{I*O_2*O_3}$& 0.744373 & 3.23709 & 0.001886 & 0.088318 &  ${\mathbb Z}_2 \times O(2)\times O(3) $ &
3(3,1,2) & $90, \, 11$  \\
\noalign{\vskip 5pt}
$B_{C_4*O_2}$ & 0.7443770 & 3.23720 & 0.001868 & 0.088288 & $ B_4 \times O(2) $   &
2(4,2)  & $87, \, 14$  \\
\noalign{\vskip 5pt}
 $B_{\S_4*O_2}$ & 0.7443773 & 3.23721 & 0.001867 & 0.088286 &  $\S_4\times {\mathbb Z}_2\times O(2)$  &
3(3,1,2)  & $86, \, 14$  \\
\noalign{\vskip 5pt}
$B_{C_2*O_2*O_2}$& 0.744379 & 3.23726 & 0.001860 & 0.088272 & $ B_2 \times O(2) \times O(2)$  &
3(2,2,2)  & $85, \, 13$  \\
\noalign{\vskip 5pt}
$B_{O_2*O_2*O_2}$ & 0.744437 & 3.23901 & 0.001605 & 0.087776 &  $ ( O(2)^2 \rtimes {\mathbb Z}_2)  \times O(2) $ &
2(4,2)  & $85, \, 12$  \\
\noalign{\vskip 5pt}
$B_{I*I*O_2*O_2}$ & 0.746610 & 3.19983 & 0.000125 & 0.106603 & $ ( {\mathbb Z}_2 \times O(2) )^2 \rtimes {\mathbb Z}_2  $
& 2(4,2) & $83, \, 13$  \\
\noalign{\vskip 5pt}
$B_{\S_4*O_2}$& 0.746638 & 3.19991& 0.000063 & 0.106637 &  $\S_4\times {\mathbb Z}_2\times O(2)$ & 3(2,3,1)  & $81, \, 14$  \\
\noalign{\vskip 5pt}
$B_{I*O_2*O_3}$& 0.746962 & 3.18917 & 0.000112 & 0.111220 &  ${\mathbb Z}_2 \times O(2)\times O(3) $ &
3(2,3,1) & $80, \, 11$  \nn \\
\noalign{\vskip 5pt}
$B_{C_3*O_3}$ & 0.746991 & 3.18955 & 0.000030 &
0.111147&$ B_3 \times O(3) $  & 2(3,3) & $78,\,  12$\\
\noalign{\vskip 5pt}
\end{tabular} \nn
\end{adjustbox}
\end{align}
For $N=6$ the total dimension of the space  of slightly marginal deformations is 126. The dimension of $\mathfrak {so}(6) $ is of
 course $15$, for $14$ zero modes there remains an $O(2)$ continuous symmetry.

 \subsubsection{Mukamel and Krinsky Model}

 An example of an $N=6$ theory intended to be related to phase transitions in antiferromagnets was considered by Mukamel and Krinsky
 long ago \cite{3coupling1}. Taking $\phi_i = (\vphi_a ,\psi_a)$, for $a=1,2,3$, this is based on the potential, with slight changes of notation from
  \cite{3coupling1},
 \begin{align}
 V(\vphi,\psi) = {}&  \tfrac{1}{24}\, g_1 \,  {\ts \sum_{a=1}^{3}} \big ( \vphi_a{\!}^4 +  \psi_a{\!}^4 \big )
 +   \tfrac{1}{4}\, g_2 \,  {\ts \sum_{1\le a < b\le 3}} \big ( \vphi_a{\!}^2  \vphi_b{\!}^2  + \psi_a{\!}^2  \psi_b{\!}^2 \big )
+   \tfrac{1}{4}\, g_3 \,  {\ts \sum_{a=1}^{3}} \, \vphi_a{\!}^2  \psi_a{\!}^2 \nn \\
&{} +   \tfrac{1}{4}\, g_4\,  ( \vphi_2{\!}^2 \, \psi_1{\!}^2  + \vphi_3{\!}^2  \, \psi_2{\!}^2 + \vphi_1{\!}^2 \, \psi_3{\!}^2   \big )
+   \tfrac{1}{4}\, g_5\,  ( \vphi_1{\!}^2 \, \psi_2{\!}^2  + \vphi_2{\!}^2  \, \psi_3{\!}^2 + \vphi_3{\!}^2 \,  \psi_1{\!}^2   \big ) \,  .
\label{VMK}
 \end{align}
 This has a ${\mathbb Z}_2{\!}^6$ symmetry resulting from $\vphi_a\to -\vphi_a$ or $\psi_a \to - \psi_a$ for any individual $a$.
 There are also  three  ${\mathbb Z}_2$ symmetries $\vphi_a \leftrightarrow \psi_b, \, \vphi_b \leftrightarrow \psi_a, \, \vphi_c \leftrightarrow \psi_c$ for
 $ a \ne b \ne c\ne a$ and further the  ${\mathbb Z}_3$ cyclic symmetry generated by $ \phi_a \to \phi_{a+1}, \, \psi_a \to \psi_{a+1}  \mod 3$.
 These form ${\mathbb Z}_2{\!}^3 \rtimes {\mathbb Z}_3 \simeq \S_4$, the symmetry group of a tetrahedron.
 By considering just  $ \psi_a \to \psi_{a+1}  \mod 3$ and also
 $\vphi_a \leftrightarrow \psi_a$ theories related by permutations of $g_3,g_4,g_5$ are equivalent.

 For this theory
 \be
 \Gamma_{ij} = \big ( g_1{\!}^2 +  6\,  g_2{\!}^2 + 3(  g_3{\!}^2 + g_4{\!}^2  + g_5{\!}^2 )\big ) \delta_{ij} \, .
 \ee

There are 20 fixed points,  the non trivial fixed points realised by this theory can be summarised, with previous notation
 and  excluding equivalent fixed points related by permutations of $g_3,g_4,g_5$, by
 \begin{align}
&\begin{tabular}{ l c c c c c c c c }
Fixed Point & $g_1$ & $g_2$ & $g_3$ & $g_4$  & $g_5$ & $ \{\kappa \}$ \\
\hline \noalign{\vskip 5pt}
$I^6$ & $ \tfrac13$ & $0$  & $0$  & $0$  & $0$ &  $- \tfrac13(4), \, 1$ \\
 \noalign{\vskip 5pt}
 $I^6$ & $ \tfrac16$ & $0$  & $\tfrac16$  & $0$  & $0$ &    $-1, \, -\tfrac23, \, - \tfrac13(2), \, 1$ \\
 \noalign{\vskip 5pt}
$O_2{\!}^3$ &   $\tfrac{3}{10} $ & $0 $   & $\tfrac{1}{10} $  &0  & $0 $  &   $-\tfrac35, \, -\tfrac24, \, - \tfrac15, \, -\tfrac15, \, 1$ \\
\noalign{\vskip 5pt}
$O_6$ & $\tfrac{3}{14} $ &  $\tfrac{1}{14} $  &  $\tfrac{1}{14} $ &  $\tfrac{1}{14} $  &  $\tfrac{1}{14} $  &$- \tfrac17(4), \, 1$   \\
\noalign{\vskip 5pt}
$C_3{\!}^2$ &   $\tfrac{2}{9} $ & $ \tfrac19  $   & $0  $  &0  & $0 $  &    $ -\tfrac79(2), \, - \tfrac19(2), \, 1$ \\
\noalign{\vskip 5pt}
$C_6$ &   $\tfrac{5}{18} $ & $\tfrac{1}{18} $   & $\tfrac{1}{18} $   & $\tfrac{1}{18} $  & $\tfrac{1}{18} $ &
$ -\tfrac19(3), \,  \tfrac19, \, 1$  \\
\noalign{\vskip 5pt}
$C_6$ &   $\tfrac{2}{9} $ & $\tfrac{1}{18} $   & $\tfrac{1}{9} $   & $\tfrac{1}{18} $  & $\tfrac{1}{18} $ &
$-\tfrac13, \, -\tfrac29, \, - \tfrac19, \, \tfrac19, \, 1$  & \\
\noalign{\vskip 5pt}
$O_3{\!}^2$ &   $\tfrac{3}{11} $ & $ \tfrac{1}{11}  $   & $0  $  &0  & $0 $  &  $-\tfrac{7}{11}(2), \, -\tfrac{1}{11}, \, \tfrac1{11}, \, 1$ \ & \\
\noalign{\vskip 5pt}
$M \! N_{2,3}$  & $\tfrac{3}{11}$ & $\tfrac{1}{22}$ & $\tfrac{1}{11}$  & $\tfrac{1}{22}$&
$\tfrac{1}{22} $ &$-\tfrac3{11}, \, -\tfrac2{11}, \, \tfrac1{11}(2), \, 1$   & \\
\noalign{\vskip 5pt}
$CC_3 $ & $\tfrac{7}{30}$ & $\tfrac{1}{10}$ &  $\tfrac{1}{30}$ &  $\tfrac{1}{30}$&$\tfrac{1}{30}$ &
 $-\tfrac{7}{15}(2), \, -\tfrac{1}{15}, \, \tfrac1{15}, \, 1$  \\
\noalign{\vskip 5pt}
$M\! N_{3,2} $ & $\tfrac{9}{34}$ & $\tfrac{3}{34}$ &  $\tfrac{1}{34}$ &  $\tfrac{1}{34}$  & $\tfrac{1}{34}$  &
 $-\tfrac{7}{17}(2), \, \tfrac{1}{17}(2), \, \, 1$ \ & \\
\end{tabular} \nn
\end{align}
These are ordered in terms of increasing $S_6$. Even restricting to the five couplings in \eqref{VMK}
all fixed points are unstable, as was realised by Mukamel and Krinsky.  At non decoupled fixed points the symmetry is enhanced.
For fixed points linked by a RG flow arising from a relevant perturbation the change in $S_6$ between the two fixed points
must be positive.  Hence  there cannot be any RG flow from
$C_3{\!}^2$  to $C_6$ or from $O_3{\!}^2$ to $M \! N_{2,3}$  although there  may be a flow to $M\! N_{3,2} $.

If we impose $g_5=g_4 = g_3$ the potential reduces to that for perturbed cubic theories given by \eqref{VP}, \eqref{defU} and $V_1=V_2$
as in \eqref{V1} where $g_1= g+3\lambda, \, g_2 = \lambda, \, g_3 = h$ and $m=3$. The symmetry group is enhanced to
$B_3{\!}^2 \rtimes {\mathbb Z}_2$, note $| B_2{\!}^2 \rtimes {\mathbb Z}_2| \big /
| {\mathbb Z}_2 {\!}^6 \times \S_4 | =3 $.
With just three couplings the table becomes
  \begin{align}
&\begin{tabular}{ l c c c c c c c c }
Fixed Point & $g_1$ & $g_2$ & $g_3$ & $ \{\kappa \}$ \\
\hline \noalign{\vskip 5pt}
$I^6$ & $ \tfrac13$ & $0$  & $0$  &  $- \tfrac13(2), \, 1$ \\
 \noalign{\vskip 5pt}
$O_6$ & $\tfrac{3}{14} $ &  $\tfrac{1}{14} $  &  $\tfrac{1}{14} $  & $- \tfrac17(2), \, 1$   \\
\noalign{\vskip 5pt}
$C_3{\!}^2$ &   $\tfrac{2}{9} $ & $ \tfrac19  $   & $0 $ &   $- \tfrac19(2), \, 1$ \\
\noalign{\vskip 5pt}
$C_6$ &   $\tfrac{5}{18} $ & $\tfrac{1}{18} $   & $\tfrac{1}{18} $   &
$ -\tfrac19, \,  \tfrac19, \, 1$  \\
\noalign{\vskip 5pt}
$O_3{\!}^2$ &   $\tfrac{3}{11} $ & $ \tfrac{1}{11}  $   & $0  $  &  $ -\tfrac{1}{11}, \, \tfrac1{11}, \, 1$ \ & \\
\noalign{\vskip 5pt}
$CC_3 $ & $\tfrac{7}{30}$ & $\tfrac{1}{10}$ &  $\tfrac{1}{30}$ &
 $-\tfrac{1}{15}, \, \tfrac1{15}, \, 1$  \\
\noalign{\vskip 5pt}
$M\! N_{3,2} $ & $\tfrac{9}{34}$ & $\tfrac{3}{34}$ &  $\tfrac{1}{34}$ &
 $\tfrac{1}{17}(2), \, \, 1$ \ & \\
\end{tabular} \nn
\end{align}
In this restricted case there is a stable fixed point $M\! N_{3,2} $ which provides a potential endpoint for the RG flow.

\subsection{\texorpdfstring{$N=7$}{N=7}}

The rational fixed points are more limited in this case
\begin{align}
&\begin{tabular}{ l c c c c c c c c } $N=7$ & $S_7$ & $a_0$ & $a_2$ & $a_4$ & Symmetry & \makecell{\# different $\gamma$\\[-2pt] and degeneracies} & $ \# {\kappa<0}, \, {=0}$ \\
\hline \noalign{\vskip 5pt}
$O_7$ &   $\tfrac{21}{25} $ & $ \tfrac{21}{5} $   &0  &0  & $O(7) $  &  1(7)  & $182, \, 0 \  \, $ \\
\noalign{\vskip 5pt}
$C_7$ & $\tfrac{6}{7}$ & $4 $ & 0 & $\tfrac{2}{21}$&
$B_7 $ & 1(7) & $154, \, 27$  \\
\noalign{\vskip 5pt}
$T_{7+}$ & $\tfrac{105}{121}$ & $ \tfrac{35}{11}$ & 0 &
$\tfrac {5}{21} $ & $\S_8 \times {\mathbb Z}_2 $ & 1(7) & $154, \, 21$ \\
\noalign{\vskip 5pt} $T_{7-}$ & $\tfrac{196}{225}$ & $ \tfrac{56}{15}$ & 0 &
$\tfrac {28}{135} $ & $\S_8 \times {\mathbb Z}_2 $ & 1(7) & $153, \, 21$ \\
\end{tabular} \nn
\end{align}
Numerically for these theories $S_7 =\{0.84, \,  0.85714, \, 0.86777,\, 0.87111\}$.

The number of irrational fixed points proliferate. By a combination of numerical methods searching for solutions with $S_7 >0.865$  and using
results for a variety of cases we obtained

\begin{landscape}
\pagestyle{empty}
\setlength\LTcapwidth{\textwidth}
\setlength\LTleft{0pt}
\setlength\LTright{0pt}
\begin{longtable}{l c c c c c c c c } $N=7$ & $S_7$ & $a_0$ & $a_2$ & $a_4$ &
  Symmetry & \makecell{\# different $\gamma$\\[-4pt] and degeneracies} & $ \# {\kappa<0},\,{=0}$ \\ \hline \noalign{\vskip 5pt}
{$B_{I*O_6}$} & 0.848454 & 4.05973 & 0.008335 & 0.059079
& $\mathbb{Z}_2\times O(6)$ & 2(6,1) & $175,\,6$\\
\noalign{\vskip 5pt}
{$B_{C_3*O_4}$} & 0.855735 & 3.97989 & 0.005630 & 0.098402
& $B_3\times O(4)$ & 2(4,3) & $164,\,15$\\
\noalign{\vskip 5pt}
{$B_{C_3*C_4}$} & 0.857146 & 3.99516 & 0.000681 &
0.098711 &$B_3\times B_4$ & 2(3,4) & $156,\,21$\\
\noalign{\vskip 5pt}
{$B_{C_5*O_2}$} & 0.857297 & 3.98590 & 0.001676 & 0.099839
&$O(2)\times B_5$ & 2(2,5) & $155,\,20$\\
\noalign{\vskip 5pt}
{$B_{O_2*O_5}$} & 0.862416 & 3.82034 & 0.010508 & 0.161683
& $O(2)\times O(5)$ & 2(5,2) & $169,\,10$\\
\noalign{\vskip 5pt}
{$B_{I*O_2*O_4}$} & 0.863351 & 3.82328 & 0.008369 &
0.162715 & $\mathbb{Z}_2\times O(2)\times O(4)$& 3(2,4,1) & $164,\,14$\\
\noalign{\vskip 5pt}
{$B_{C_2*O_2*O_3}$} & 0.863688 & 3.82583 & 0.007459 &
0.162621 & $B_2\times O(2)\times O(3)$ & 3(3,2,2) & $160,\,17$\\
\noalign{\vskip 5pt}
{$B_{C_5*O_2}$} & 0.863748 & 3.82704 & 0.007224 &
0.162369 &$O(2)\times B_5$ & 2(5,2) & $158,\,20$\\
\noalign{\vskip 5pt}
{$B_{C_2*C_3*O_2}$} & 0.863750 & 3.82693 & 0.007230 &
0.162405 &$O(2)\times B_2\times B_3$ & 3(3,2,2) & $156,\,20$\\
\noalign{\vskip 5pt}
{$B_{O_2*O_2*C_3}$} & 0.863776 & 3.82689 & 0.007183 &
0.162473 & $O(2)\times O(2)\times B_3$& 3(2,3,2) & $155,\,19$\\
\noalign{\vskip 5pt}
$B_{I*O_2*O_2*O_2}$ & 0.865351 & 3.85371 & 0.001426 & 0.157379 &
$\mathbb{Z}_2\times O(2)\times (O(2)^2\rtimes \mathbb{Z}_2)$ & 3(2,1,4) &
$155,\,18$\\
\noalign{\vskip 5pt}
$B_{I*O_2*O_2*O_2}$ & 0.865360 & 3.84698 & 0.002082 & 0.159497 &
$\mathbb{Z}_2\times O(2)\times O(2)\times O(2)$& 4(2,1,2,2) & $154,\,18$\\
\noalign{\vskip 5pt}
$B_{O_2*O_2*C_3}$& 0.865363 & 3.85323 & 0.001450 & 0.157553 &
$B_3\times
(O(2)^2\rtimes\mathbb{Z}_2)$ & 2(3,4) & $153,\,19$\\
\noalign{\vskip 5pt}
$B_{O_2*O_2*C_3}$& 0.865370 & 3.84723 & 0.002036 & 0.159439 &
$O(2)\times O(2)\times B_3$ & 3(3,2,2) & $152,\,19$\\
\noalign{\vskip 5pt}
{{$B_{O_2*O_2*O_3}$}} & 0.865427558 & 3.84923 & 0.001721 &
0.158937 & $O(3)\times(O(2)^2\rtimes\mathbb{Z}_2)$ & 2(3,4) & $152,\,16$\\
\noalign{\vskip 5pt}
$B_{O_2*O_2*O_3}$ & 0.865427563 & 3.84907 & 0.001738 & 0.158988 &
$O(2)\times O(2)\times O(3)$ & 3(3,2,2) & $151,\,16$\\
\noalign{\vskip 5pt}
{$B_{I*C_2*O_4}$} & 0.8712962 & 3.68437 & 0.002552 &
0.223496 & $\mathbb{Z}_2\times B_2
\times O(4)$ & 3(4,2,1) & $162,\,15$\\
\noalign{\vskip 5pt}
$B_{I*C_2*C_4}$& 0.87129773 & 3.684606 & 0.002536 & 0.223423 &
$\mathbb{Z}_2\times B_2\times O(4)$& 3(4,2,1) & $155,\,21$\\
\noalign{\vskip 5pt}
& 0.87129775 & 3.684611 & 0.002536 & 0.223421 & & 4(2,2,2,1) & $153,\,20$\\
\noalign{\vskip 5pt}
{$B_{C_3*O_4}$} & 0.8712983 & 3.68496 & 0.002516 &
0.223311 &$B_3\times O(4)$ & 2(4,3) & $161,\,15$\\
\noalign{\vskip 5pt}
& 0.8712989 & 3.70402 & 0.001456 & 0.217183 & & 3(4,2,1) & $152,\,21$\\
\noalign{\vskip 5pt}
& 0.8712994 & 3.68487 & 0.002519 & 0.223342 & & 3(4,2,1) & $153,\,19$\\
\noalign{\vskip 5pt}
{$B_{I*O_3*C_3}$} & 0.8712996 & 3.68516 & 0.002503 &
0.223247 & $\mathbb{Z}_2\times B_3 \times O(3)$ & 3(3,1,3) & $157,\,18$\\
\noalign{\vskip 5pt}
{$B_{C_3*C_4}$}& 0.871299832 & 3.6852003 & 0.00250046 &
0.2232359 & $B_3\times O(4)$ & 2(4,3) & $154,\,21$\\
\noalign{\vskip 5pt}
$B_{I*C_3*C_3}$ & 0.871299833 & 3.6852004 & 0.00250045 & 0.2232358 &
$\mathbb{Z}_2\times B_3\times B_3$ & 3(3,1,3) & $153,\,21$\\
\noalign{\vskip 5pt}
$B_{O_2*C_2*C_3}$ & 0.87129986 & 3.68521 & 0.002500 & 0.223234 &
$O(2)\times B_2\times B_3$ & 3(2,2,3) & $152,\,20$\\
\noalign{\vskip 5pt}
$B_{O_2*O_2*C_3}$& 0.871301 & 3.68547 & 0.002483 & 0.223153 &
$B_3\times
(O(2)^2\rtimes\mathbb{Z}_2)$ & 2(4,3) & $152,\,19$\\
\noalign{\vskip 5pt}
$B_{C_3*T_4}$& 0.871304 & 3.70466 & 0.001409 & 0.216987 &
$B_3\times
\mathcal{S}_5\times\mathbb{Z}_2$ & 2(4,3) & $151,\,21$\\
\noalign{\vskip 5pt}
& 0.871305 & 3.70164 & 0.001581 & 0.217961 & & 5(1,2,1,2,1) & $151,\,21$\\
\noalign{\vskip 5pt}
& 0.87130606 & 3.70132 & 0.001598 & 0.218064 & & 5(1,1,2,2,1) & $150,\,21$\\
\noalign{\vskip 5pt}
$B_{\mathcal{S}_5*O_2}$& 0.871306 & 3.70264 & 0.00152144 & 0.217639 &
$\mathcal{S}_5\times\mathbb{Z}_2\times O(2)$ & 3(2,4,1) & $151,\,20$\\
\noalign{\vskip 5pt}
& 0.871310 & 3.70227 & 0.001536 & 0.217767 & & 4(1,2,1,3) & $150,\,21$\\
\noalign{\vskip 5pt}
& 0.871311 & 3.70195 & 0.001553 & 0.217871 & & 4(1,1,2,3) & $149,\,21$\\
\noalign{\vskip 5pt}
& 0.871314 & 3.70006 & 0.001655 & 0.218486 & & 5(1,2,2,1,1) & $150,\,20$\\
\noalign{\vskip 5pt}
& 0.8713147 & 3.69972 & 0.0016724 & 0.218597 & & 5(1,2,1,2,1) & $149,\,20$\\
\noalign{\vskip 5pt}
{$B_{I*O_2*O_4}$} & 0.8713152 & 3.68073 & 0.002703 &
0.224709 & $\mathbb{Z}_2\times O(2)\times O(4)$& 3(2,4,1) & $161,\,14$\\
\noalign{\vskip 5pt}
{$B_{I*I*O_2*O_3}$} & 0.871316 & 3.68092 & 0.002691 &
0.224648 & $\mathbb{Z}_2\times \mathbb{Z}_2\times O(2)\times O(3)$ &
4(3,2,1,1) & $157,\,17$\\
\noalign{\vskip 5pt}
{$B_{I*O_2*C_4}$}& 0.87131659 & 3.68096 & 0.002689 & 0.224637 &
$\mathbb{Z}_2 \times O(2)\times B_4$&
3(2,4,1) & $154,\,20$\\
\noalign{\vskip 5pt}
& 0.87131661 & 3.68097 & 0.002688 & 0.224636 & & 4(2,2,2,1) & $152,\,19$\\
\noalign{\vskip 5pt}
{{$B_{I*O_2*O_2*O_2}$}} & 0.871318 & 3.68121 & 0.002673 &
0.224559 & $\mathbb{Z}_2\times O(2)\times (O(2)^2\times\mathbb{Z}_2)$&
3(2,4,1) & $152,\,18$\\
\noalign{\vskip 5pt}
& 0.8713206 & 3.68941 & 0.002233 & 0.221922 & & 3(4,2,1) & $151,\,21$\\
\noalign{\vskip 5pt}
& 0.87132074 & 3.68949 & 0.002229 & 0.221899 & & 5(1,2,1,2,1) & $150,\,21$\\
\noalign{\vskip 5pt}
& 0.87132076 & 3.6895 & 0.002228 & 0.221894 & & 5(1,1,2,2,1) & $149,\,21$\\
\noalign{\vskip 5pt}
$B_{C_3*T_4}$& 0.8713233 & 3.69025 & 0.002183 & 0.221659 &
$B_3\times\mathcal{S}_5\times \mathbb{Z}_2$ & 2(4,3) & $150,\,21$\\
\noalign{\vskip 5pt}
& 0.87132340 & 3.69033 & 0.002178 & 0.221632 & & 4(1,2,1,3) & $149,\,21$\\
\noalign{\vskip 5pt}
& 0.87132342 & 3.69035 & 0.002177 & 0.221627 & & 4(1,1,2,3) & $148,\,21$\\
\noalign{\vskip 5pt}
$B_{\mathcal{S}_5*O_2}$& 0.871337 & 3.68539 & 0.00241684 & 0.223251 &
$\mathcal{S}_5\times\mathbb{Z}_2\times O(2)$ & 3(2,4,1) & $150,\,20$\\
\noalign{\vskip 5pt}
& 0.87133668 & 3.68544 & 0.0024141 & 0.223236 & & 5(1,2,2,1,1) & $149,\,20$\\
\noalign{\vskip 5pt}
& 0.87133669 & 3.68545 & 0.0024136 & 0.223233 & & 5(1,2,1,2,1) & $148,\,20$\\
\noalign{\vskip 5pt}
$B_{T_4*O_3}$& 0.872241 & 3.68634 & 0.000557 & 0.224839 &
$O(3)\times\mathcal{S}_5\times\mathbb{Z}_2$ & 2(4,3) & $150,\,18$\\
\noalign{\vskip 5pt}
& 0.872269 & 3.68187 & 0.000737 & 0.226337 & & 4(1,2,3,1) & $149,\,18$\\
\noalign{\vskip 5pt}
& 0.872273 & 3.68132 & 0.000758 & 0.226521 & & 4(1,1,2,3) & $148,\,18$\\
\noalign{\vskip 5pt}
$\hat{B}_{(O_2\circ O_2)*O_3}$& 0.872388 & 3.66736 & 0.001223 & 0.231267 & $O(2)\times O(3)$ & 3(2,2,3) & $147,\,17$\\
\noalign{\vskip 5pt}
{$B_{O_3*O_4}$} & 0.8724124 & 3.65263 & 0.001847 & 0.236084 &
$O(3)\times O(4)$& 2(4,3) & $160,\,12$\\
\noalign{\vskip 5pt}
{$B_{I*O_3*O_3}$} & 0.8724128 & 3.65273 & 0.001842 & 0.236054 &
$\mathbb{Z}_2\times O(3)\times O(3)$ & 3(3,1,3) & $156,\,15$\\
\noalign{\vskip 5pt}
{$B_{C_4*O_3}$} & 0.8724129 & 3.65275 & 0.001841 & 0.236049 &
$O(3)\times B_4$& 2(4,3) & $153,\,18$\\
\noalign{\vskip 5pt}
{$B_{O_2*O_2*O_3}$} & 0.872413 & 3.65286 & 0.001835 &
0.236012 & $O(3)\times(O(2)^2\rtimes\mathbb{Z}_2)$& 2(4,3) & $151,\,16$\\
\noalign{\vskip 5pt}
{$B_{T_4*O_3}$}& 0.872418318 & 3.654206 & 0.0017663 & 0.235587 &
$O(3)\times\mathcal{S}_5\times\mathbb{Z}_2$ & 2(4,3)& $149,\,18$\\
\noalign{\vskip 5pt}
& 0.872418321 & 3.654208 & 0.0017662 & 0.235586 & & 4(1,1,2,3) & $147,\,18$\\
\noalign{\vskip 5pt}
$\hat{B}_{(O_2\circ O_2)*O_3}$& 0.872419 & 3.65456 & 0.001749 &
0.235474 & $O(2)\times O(3)$ & 3(2,2,3) & $146,\,17$\\
\noalign{\vskip 5pt}
\end{longtable}
For this case  the total dimension of the space  of slightly marginal deformations is 210. We have not identified all fixed points as
previously, in particular those with four and five quadratic invariants.  There are no fixed points which have the maximal possible
seven quadratic invariants, and also 21 zero $\kappa$, as would be expected
if there were just ${\mathbb Z}_2$ symmetry.
Thus our numerical search fails to gain the prize promulgated in \cite{RychkovS}.
\end{landscape}
\pagestyle{plain}

\begin{figure}[H]
  \centering
  \includegraphics[scale=0.99]{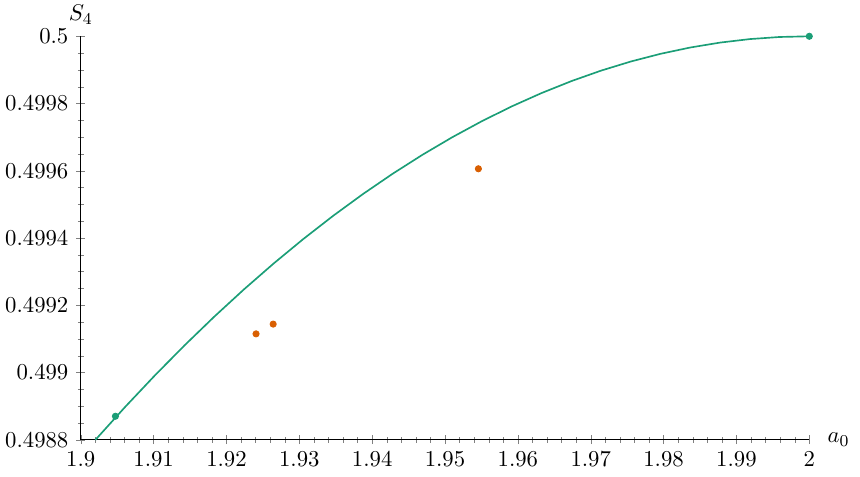}
  \caption{Fully-interacting fixed points for $N=4$.}
  \label{fig:N4}
\end{figure}

\begin{figure}[H]
  \centering
  \includegraphics[scale=0.99]{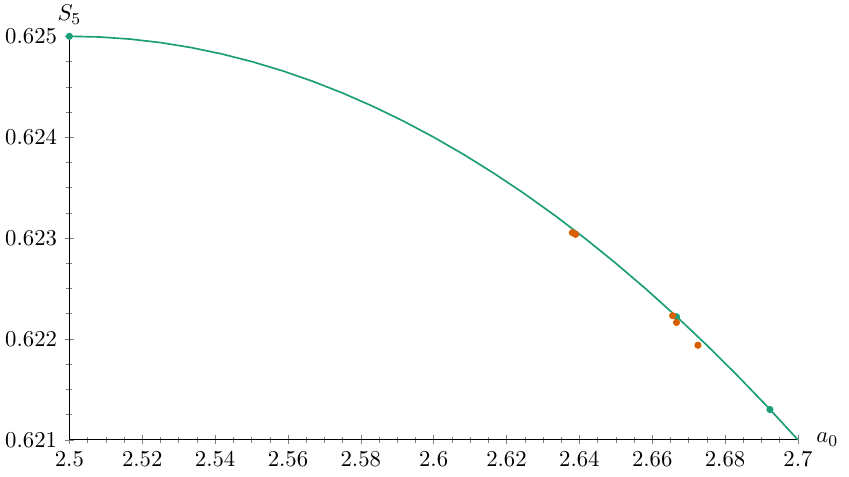}
  \caption{Fully-interacting fixed points for $N=5$. The leftmost point
    corresponds to $T_{5\pm}$ and the rightmost to $O(5)$.}
  \label{fig:N5}
\end{figure}

\begin{figure}[H]
  \centering
  \includegraphics{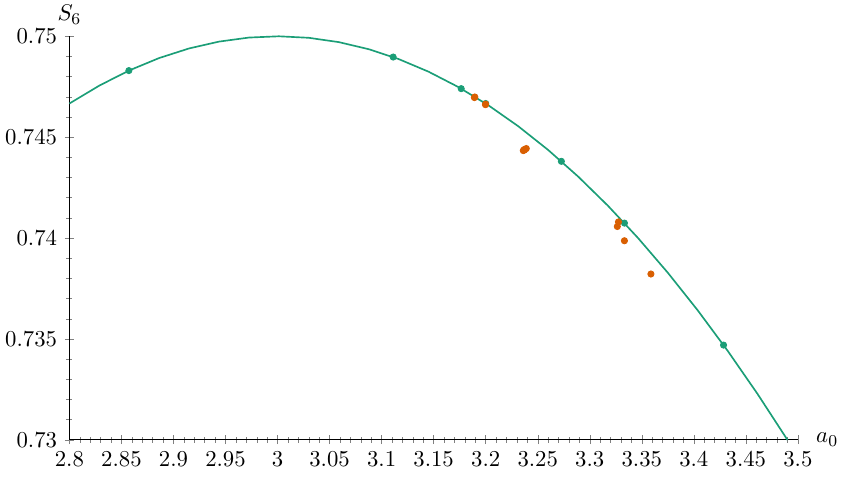}
  \caption{Fully-interacting fixed points for $N=6$. The leftmost point
    corresponds to $T_{6+}$ and the rightmost to $O(6)$.}
  \label{fig:N6}
\end{figure}

\begin{figure}[H]
  \centering
  \includegraphics{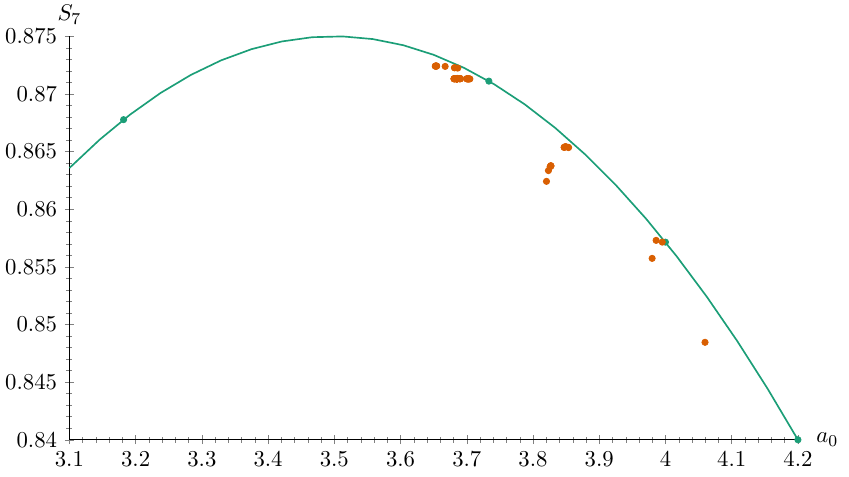}
  \caption{Fully-interacting fixed points for $N=7$. The leftmost point
    corresponds to $T_{7+}$ and the rightmost to $O(7)$.}
  \label{fig:N7}
\end{figure}

Following~\cite{HogervorstToldo} we present a pictorial representation of
the $N=4,5,6,7$ fully interacting fixed points in Figures
\ref{fig:N4}--\ref{fig:N7}. Rational fixed poins are shown in green and
irrational in orange. The line $S_N-\tfrac18
N+\tfrac{1}{2N}(a_0-\tfrac12N)^2=0$ is shown in green. The reasons behind the apparent clustering of fixed
points as well as the qualitative change in their distribution between
$N=4$ and $N=5$ are unclear.\footnote{The $N=3$ fully-interacting fixed
points are distributed similarly to the $N=4$ ones.}

\section{Six Index Case}\label{sec:sixindex}

In $3-\vep$ dimensions there may be fixed points starting from the renormalisable  interaction $V(\phi)=\frac{1}{6!} \lambda_{ijklmn} \phi_i \phi_j \phi_k
\phi_l \phi_m \phi_n$. At lowest order possible fixed points in the $\vep$ expansion
are determined, with a suitable rescaling, by finding  solutions of
\be
\lambda_{ijklmn} = S_{10,ijklmn} \, \lambda_{ijkpqr} \lambda_{lmnpqr} \, ,
\label{fixp6}
\ee
where $S_{n,ijklmn}$ here denotes the sum over the $n$ permutations, with unit weight, necessary to ensure the sum is fully
symmetric in $ijklmn$.  This is equivalent to
\be
V(\phi) = \tfrac12 \, V_{ijk}(\phi) V_{ijk}(\phi) \, .
\label{V6eq}
\ee

As before in \eqref{ldec} the coupling can be decomposed, with a similar notation for symmetrisation, as
\be
\lambda_{ijklmn} =d_0 \,  S_{15,ijklmn} \, \delta_{ij} \delta_{kl} \delta_{mn} + S_{45,ijklmn} \, \delta_{ij} \delta_{kl} d_{2,mn}
+  S_{15,ijklmn} \, \delta_{ij} d_{4,kl mn}  + d_{6,ijklmn}\, ,
\label{RG6}
\ee
for $d_2, \, d_4, d_6$ symmetric traceless tensors.
With this expansion
\begin{align}
a_0={}&    N(N+2)(N+4) d_0= \lambda_{iijjkk}   \, ,  \nn \\
a_2 = {}& (N+4)^2(N+6)^2\,  || d_2 ||^2 =  \lambda_{ijkkmm} \lambda_{ijllnn} - \tfrac{1}{N} \, a_0{\!}^2 \, , \nn \\
a_4 = {}& (N+8)^2\,  || d_4 ||^2 =  \lambda_{ijklmm} \lambda_{ijklnn} - \tfrac{6}{N+4} \, a_2 - \tfrac{3}{N(N+2)} \, a_0{\!}^2 \, , \nn \\
|| \lambda ||^2 = {}& \lambda_{ijklmn} \lambda_{ijklmn} = 15 N(N+2) (N+4) \, d_0{}^2 + 45(N+4) ( N+6) \, ||d_2||^2 \nn \\
&\hskip 2.8cm {}  +  15 (N+8 ) \,  ||d_4||^2 +  ||d_6||^2 \, .
\label{exp6}
\end{align}
The fixed point equation \eqref{RG6}, by contracting indices, requires
\begin{align}
N(N+2)(N+4) d_0 = {}& 6 N (N+2)(N+4) ( 3N+22) d_0{}^2 + 36 (N+4) (N+6) (N+11) || d_2||^2 \nn \\
&{}  + 6 (N+8)(N+18) || d_4||^2 + 4 \, || d_6||^2\, .
\label{fp6}
\end{align}
This requires
\be
d_0 \le \tfrac{1}{6(3N+22)} \,  .
\label{d0}
\ee
Combining \eqref{exp6} and \eqref{fp6}
\begin{align}
|| \lambda ||^2 & + \tfrac32 (N+8)^2 \, || d_4||^2 + 9 (N+4)(N+6)^2\,  || d_2||^2\nn \\
 ={}&  \tfrac{1}{288} \, N(N+2) - \tfrac12 N (N+2) \big ( 3(N+4)d_0  -\tfrac{1}{12} \big )^2  \le \tfrac{1}{288} \, N(N+2)  \, .
 \label{S6bound2}
\end{align}
The constraint \eqref{d0} does not here lead to any modification for low $N$.

For the $O(N)$ invariant theory, $\lambda_{ijklmn}= \lambda\,S_{15,ijklmn}\,  \delta_{ij}\delta_{kl} \delta_{mn}$ or
$V(\phi) =  \tfrac{1}{48} \lambda (\phi^2)^3$.
At the fixed point solving \eqref{fixp6} $\lambda =  1/6(3N+22)$ and  then \eqref{d0} is saturated
\be
d_0 = \frac{1}{6(3N+22)} \, , \quad a_2 = a_4 = || d_6||^2 = 0 \, , \qquad   || \lambda ||^2 = \frac{5N(N+2)(N+4)}{12(3N+22)^2}\, .
\label{6dON}
\ee
For $N=1$, $ || \lambda ||^2 = \tfrac{1}{100}$ so that without any decoupled free theories we expect $ || \lambda ||^2 > \tfrac{1}{100} N$.
For $N$ decoupled non free theories
\be
 || \lambda ||^2 = \tfrac{N}{100}\, , \ \ a_0 = \tfrac{N}{10} \, , \ \ a_2 = 0 \, , \ \  a_4  =\tfrac{N(N-1)}{100(N+2)}\, ,
 \ \ ||d_6||^2 = \tfrac{N(N-1)(N-2)}{100 (N+4)(N+8)} \, .
 \ee
 For $N\to \infty$ in the $O(N)$ case $||\lambda||^2 \sim \tfrac{5}{108} N, \, a_0 \sim \frac{1}{18} N^2$.

Assuming cubic symmetry $B_N$ there are  just three couplings,
\be
V_C(\phi) = \tfrac{1}{6!}\,  g_1 {\ts \sum_i} \, \phi_i{\!}^6  +
 \tfrac{1}{48}\, g_2\, {\ts \sum_{i\ne j}} \, \phi_i{\!}^4 \phi_j {\!}^2 + \tfrac{1}{48}\,  g_3 \, {\ts \sum_{i\ne j\ne k} } \, \phi_i{\!}^2 \phi_j{\!}^2 \phi_k{\!}^2 \, ,
 \label{V6cubic}
\ee
and
\begin{align}
|| \lambda ||^2 ={}&  N\big ( g_1{\!}^2 + 15(N-1) \, g_2{\!}^2 + 15(N-1)(N-2) \, g_3{\!}^2\big )  \, , \nn \\
 a_0 = {}&  N \big ( g_1+ 3(N-1) g_2 + (N-1)(N-2) g_3 \big ) \, , \qquad a_2=0 \, , \nn \\
 a_4 = {}& \tfrac{N(N-1)}{N+2} \big ( g_1 + (N-7) g_2 - 3(N-2)g_3 \big )^2 \, , \quad
  a_6 =  \tfrac{N(N-1)(N-2) }{(N+4)(N+8))} \big ( g_1 -15  g_2  + 30 \big )^2 \, ,
\end{align}
The lowest order equations are just
\begin{align}
g_1={}&  10\, g_1{\!}^2 +30 (N-1) g_2{\!}^2 \, , \quad g_2 = 4\,  g_1 g_2 + 36\, g_2{\!}^2 + 12 (N-2) g_2 g_3 + 18 (N-2) g_3{\!}^2 \, , \nn \\
g_3 ={}&  6\, g_2{\!}^2 +36\,  g_2 g_3 + 6(3N-5) g_3{\!}^2\, .
\end{align}
This example was considered in \cite{Seeking}, for $g_1/15 = g_2/3 = g_3 =\lambda$ it reduces to the $O(N)$ case. For $N$ not very large there are three non trivial real solutions. Apart from the decoupled theory with  $ || \lambda ||^2 = \tfrac{1}{100} N$ there is the $O(N)$ symmetric fixed point and
an irrational one with cubic symmetry. For $N=14$ these coincide so  $N=14$ plays a similar role to $N=4$ in the four index case. For $N>14$
the cubic fixed point has higher $ || \lambda ||^2 $. For $14034<N<14035$ there is a bifurcation point and for higher $N$  two new
real fixed points, one  of which is stable within this three coupling theory. In each case $ || \lambda ||^2  \propto N$ for large $N$ where
the different solutions for $g_1  = {\rm O}(1), \,  g_2  = {\rm O}(N^{-\frac12}), \,  g_3  = {\rm O}(N^{-1})$ give
\be
|| \lambda ||^2  \sim \big ( \tfrac{1}{100}, \, \tfrac{5}{108}, \, \tfrac{25}{448}, \,  \tfrac{25}{448}, \, \tfrac{38}{675} \big ) N\, .
\ee
Although not obviously required by \eqref{S6bound2} it appears in general that $|| \lambda ||^2 \propto N$ for large $N$.

For the case of tetrahedral symmetry
\be
V_T(\phi) =  \tfrac{1}{48} \lambda (\phi^2)^3 +  \tfrac{1}{48} g_1 \, \phi^2 {\ts \sum_\alpha} \, (\phi^\alpha)^4 +
 \tfrac{1}{72} g_2 \, \big ( {\ts \sum_\alpha} \, (\phi^\alpha)^3 \big )^2 + \tfrac{1}{6!} g_3 \,  {\ts \sum_\alpha} \, (\phi^\alpha)^6\, ,
 \ee
 for
 \be
 \phi^\alpha = \phi_i e_i{\!}^\alpha \, \ \alpha =1,\dots, N+1 \, , \quad e_i{\!}^\alpha e_i{\!}^\beta = \delta^{\alpha\beta}- \tfrac{1}{N+1} \, , \
 {\ts \sum_\alpha} e_i{\!}^\alpha = 0 \, , \  { \ts \sum_\alpha} e_i{\!}^\alpha e_j{\!}^\alpha = \delta_{ij}\, .
 \ee
 This has the symmetry group $\S_{N+1} \times {\mathbb Z}_2$.
 For $N=2$ only two couplings are necessary since the potential is invariant for
 $(\lambda ,g_1,g_2,g_3)\sim (\lambda - \mu -\frac12 \nu,g_1 + 2 \mu ,g_2-\nu ,g_3 + 30\nu)$
 while for $N=3,4$,  $(\lambda ,g_1,g_2,g_3)\sim (\lambda -\frac12 \mu,g_1 + 3 \mu ,g_2 + 2\mu ,g_3 -  60\mu)$ so the couplings reduce
 to three. When $N=3$  $V_T(\phi)$ is equivalent to $V_C(\phi)$.

 The $O(N)$ invariants reduce to
\begin{align}
|| \lambda ||^2 ={}&  15N(N+2)(N+4)  \lambda^2  + \tfrac{30N}{N+1} \big ( 3 N(N+4) g_1 + 4(N-1) g_2 + \tfrac{N^2}{N+1} \,  g_3) \lambda , \nn \\
&{}+  15 N  \tfrac{N^3+1}{(N+1)^3} \big ( (N+8) g_1 +\tfrac{2N}{N+1} g_3  \big ) g_1 + \tfrac{90N^3}{(N+1)^2}  g_1{\!}^2
 \nn \\
&{} +10N  \tfrac{ (N-1)^2 } {(N+1)^2} \big ( 18 g_1+ ( N+9) g_2 + \tfrac{2N}{N+1}g_3 \big ) g_2 + N \tfrac{N^5+1}{(N+1)^5} \, g_3{\!}^2 \, ,
\nn \\
 a_0 = {}&  N(N+2)(N+4) \lambda + \tfrac{N}{N+1}  \big ( 3 N (N+4) g_1+ 4(N-1) g_2 \big )  +  \tfrac{N^3}{(N+1)^2} \, g_3  \, , \qquad a_2=0 \, , \nn \\
 a_4 = {}& \tfrac{N(N-1)(N-2)}{(N+1)^3(N+2)} \big ( (N+1)(N+9) g_1 + 6N  g_2  + N g_3 \big )^2 \, .
\end{align}

The fixed point equations from \eqref{V6eq} become
\begin{align}
\lambda ={}& 6 (3N+22) \lambda^2 +  6 g_1{\!}^2  + \tfrac{72}{N+1}( N g_1 - g_2 )\lambda - \tfrac{18(N-3)}{(N+1)^2} ( g_1 + g_2)^2 \, , \nn \\
g_1 = {}& ( 78 g_1 + 36 g_2 + 4 g_3) g_1 + 12 \big ( (N+18) g_1 + 6 g_2 + \tfrac{N}{N+1} g_3 \big ) \lambda - 12 \tfrac{N-1}{(N+1)^2} ( g_1+g_3 ) g_2\nn \\
&{}+ \tfrac{6}{N+1} \big ( 3(N-5) g_2{\!}^2 - 22 g_1{\!}^2 - 40 g_1 g_2 \big ) \, , \nn \\
g_2 ={}& (N+16) g_2{\!}^2 + 72 g_1 g_2 + 120 g_2 \lambda - \tfrac{1}{N+1} \big ( 3(N+32)g_1{\!}^2 + 88 g_2{\!}^2+180 g_1 g_2 \big ) \nn \\
&{}+ \tfrac{2N}{(N+1)^2} \big ( (N-1) g_2  - 3 g_1 \big )  g_3 - \tfrac{1}{(N+1)^3}\,  g_3{\!}^2 \, ,  \nn \\
g_3={}& 30(N+32) g_1{\!}^2 + 1440 g_1 g_2 + 540 g_2{\!}^2 + 10 g_3{\!}^2 + 120 g_3 \lambda \nn \\
&{} + \tfrac{60}{N+1} \big ( (4N-3) g_1 + 3 (N-1) g_2 \big ) g_3 - \tfrac{30N}{(N+1)^2} \, g_3{\!}^2 \, .
\end{align}
For $N=2,3$ the fixed points are identical to those  for cubic symmetry.   Apart from the $O(N)$ invariant fixed point at
 $N=4$ there is one tetrahedral fixed point and two tetrahedral fixed points  are present for $N>4$.
One of these collides with the  $O(N)$ fixed point when $N=14$ and for higher $N$ has a larger $|| \lambda ||^2 $. There are also
bifurcation points for $319<N<320$, $507<N<508$,  $14035<N<14036$ and $15695<N<15696$, where two new fixed points emerge.

The cubic and tetrahedral  solutions all have $a_2=0$ reflecting the fact that there is just one quadratic invariant.
More generally possibilities arise with just  $\S_N$ symmetry which encompasses the cubic and tetrahedral cases but for which there
are two quadratic invariants. Extending \eqref{ans1} in this case
\begin{align}
&  \lambda_{iiiiii}  = x_1 \, , \quad  \lambda_{iiiiij} =  x_2 \, , \quad \lambda_{iiiijj} = x_3 \, , \quad  \lambda_{iiiijk} = x_4 \, , \quad
 \lambda_{iiijjj} = x_5   \, , \quad \lambda_{iiijjk} = x_6  \nn \\
 & \lambda_{iiijkl} = x_7 \, , \quad \lambda_{iijjkk} = x_8\, , \quad \lambda_{iijjkl} = x_9\, , \quad \lambda_{iijklm} = x_{10}\, , \quad
 \lambda_{ijklmn} = x_{11} \, ,
\label{ans2}
\end{align}
with $i,j,k,l,m,n$ all different.  The 11 couplings  reduce to 9  for $N=4$ and 7 for $N=3$. The resulting potential with just $x_1=g_1, \,
x_3=g_2, \, x_8 = g_3$ non zero is identical to \eqref{V6cubic}. As was the case with the four index case the $\beta$-function equations
can be simplified by imposing $O(N-1)$ symmetry. This requires
\begin{align}
&5(x_3- x_4) = x_1 -x_2\, , \quad 2(x_5-x_6 ) = 4(x_6-x_7) = x_2-x_4\, , \quad 3(x_6- x_9) = 2(x_4-x_7) \, , \nn \\
 & 3(x_8- x_9) = x_3 -x_6\, , \quad 3 x_9 = 2 x_{10} + x_6\, , \quad 3 x_{10} = 2 x_{11}+ x_7\, ,
 \label{ONcon}
\end{align}
leaving just four couplings.  Defining
\begin{align}
\sigma = {}& x_1 + 3 (N-1) x_3 + (N-1)(N-2) x_8 \, , \nn \\
 \rho = {}&.  x_2 +  (N-2) (\tfrac12 x_4 + 2 x_6) + \tfrac12 (N-2)(N-3)  x_9 \, ,  \nn \\
 \tau = {}& x_1 + (N-7) x_3 - 3 (N-2)  x_8 \, ,  \qquad \upsilon = x_1 - 15 x_3 + 30 x_8 \, ,
\end{align}
then along with \eqref{ONcon}
\begin{align}
a_0 = N  \sigma \, , \quad a_2 =4 (N-1) \rho^2 \, , \quad a_4 = \tfrac{ N^4 (N^2-1)}{4(N+2)(N+4)}\, \tau^2\, , \quad
a_6 = \tfrac{ N^6 (N^2-1)(N+3)}{256(N+4)(N+6)(N+8)}\, \upsilon^2\, ,
\end{align}
the fixed point equations become
\begin{align}
\sigma = {}& \tfrac{6(3N+22)}{(N+2)(N+4)} \, \sigma^2 + \tfrac{144(N-1)(N+11)}{(N+4)(N+6)} \, \rho^2 +
 \tfrac{3N^3(N^2-1)(N+18)}{2(N+2)(N+4)(N+8)} \,  \tau^2 + \tfrac{N^5(N^2-1)(N+3)}{64(N+4)(N+6)(N+8)} \,  \upsilon^2 \, ,\nn \\
 \rho={}& \big ( \tfrac{24(N+11)}{(N+2)(N+4)} \, \sigma- \tfrac{18N^3(N+1)(N+14)}{(N+2)(N+4)^2(N+6)} \, \tau ) \rho
 - \tfrac{N^5(N+1)(N+3)(N+18)}{16(N+4)(N+6)(N+8)^2}\, \tau \, \upsilon \nn \\
 &{}+ \tfrac{12(N-2)(N^2 + 40 N+264)}{(N+4)^2(N+6)}\, \rho^2 +  \tfrac{N^3 (N-2)(N+1)(N^2 + 52  N+ 432)}{4(N+4)^2(N+8)^2}\, \tau^2
 + \tfrac{N^5(N-2)(N+2)(N+3)}{128(N+6)(N+6)^2}\, \upsilon^2 \, ,
 \nn \\
 \tau ={}& \big ( \tfrac{12(N+18) }{(N+2)(N+4)}\, \sigma+ \tfrac{24(N^2-4)(N^2+52N+432)}{(N+4)^2(N+6)(N+8) }\, \rho \big ) \tau
 - \tfrac{3N^2 (N+2)(N+3) (N+18)}{(N+4)(N+6)^2(N+8)} \, \rho \, \upsilon \nn \\
 &{} +\tfrac{N^3(N^2-4) (N+3)(N+32)}{4(N+4)(N+6)(N+8)^2}\, \tau\,\upsilon \nn \\
 &{} -\tfrac{432(N+8)(N+14)}{(N+4)^2(N+6)^2}\, \rho^2
 - \tfrac{36N^3(N-2) (N+3) (N+14)}{(N+2)(N+4)^2(N+8)^2}\, \tau^2 - \tfrac{3N^5(N-2)(N+3)(N+5)}{64(N+6)^2(N+8)^2} \, \upsilon^2 \, ,
 \nn \\
\upsilon = {}&  \big ( \tfrac{120}{(N+2)(N+4)}\, \sigma + \tfrac{720(N-2)}{(N+6)(N+8)}\, \rho - \tfrac{90N^3(N-2)(N+5)}{(N+2)(N+6)(N+8)^2} \, \tau
\big ) \upsilon\nn \\
&{}- \tfrac{2880(N+18)}{(N+4)(N+6)(N+8)}\, \rho \,\tau  + \tfrac{120N(N-2)(N+32)}{(N+4)(N+8)^2} \, \tau^2
+ \tfrac{5N^4(N^2-4)(N+5)}{8(N+4)(N+6)(N+8)^2}  \, \upsilon^2 \, .
\end{align}
For low $N$ there is one non trivial solution in addition to the $O(N)$ symmetric one, this has higher $||\lambda||^2$ for $N>14$. For
$206<N<207$ there is a bifurcation. For very large $N$ the solutions converge to the $O(N)$ symmetric fixed point.

\subsection{Fixed Points for Low \texorpdfstring{$N$}{N}}

In a similar fashion to previously we have looked numerically snd analytically for fixed points for low $N$. For the
values of $N$ considered here the fixed point with maximal $||\lambda||^2$ is always that with $O(N)$ symmetry as in \eqref{6dON}.

\begin{flushleft}
\begin{tabular}{ l c c c c c  c  c c }
$N=1$ & $||\lambda||^2$ & $a_0$ & $a_2$ & $a_4$ & $a_6$ &Symmetry  &
\makecell{\# different $\gamma$\\[-2pt] and degeneracies} & $ \# {\kappa<0}, \, {=0}$  \\
\hline
\noalign{\vskip 5pt}
$O_1$ &   $\tfrac{1}{100}$ & $ \tfrac{1}{10}$   & 0  & 0  & 0 &  ${\mathbb Z}_2$   & 1(1) & $0, \, 0$ \\
\end{tabular}

\begin{tabular}{ l c c c c c  c  c c  c }
$N=2$ & $||\lambda||^2$ & $a_0$ & $a_2$ & $a_4$ & $a_6$ &
Symmetry  & \makecell{\# different $\gamma$\\[-2pt] and degeneracies} & $ \# {\kappa<0}, \, {=0}$  \\
\hline
\noalign{\vskip 5pt}
$O_2$ & $\tfrac{5}{196}$ & $ \tfrac{2}{7}$   & 0  & 0  & 0 &  $O(2)$   & 1(2) & $0, \, 0$ \\
\end{tabular}
\end{flushleft}
\begin{align}
&\begin{adjustbox}{max width=\textwidth}
\begin{tabular}{ l c c c c c  c  c c }
$N=3$ & $||\lambda||^2$ & $a_0$ & $a_2$ & $a_4$ & $a_6$ &
Symmetry&   \makecell{\# different $\gamma$\\[-2pt] and degeneracies}&   $ \# {\kappa<0}, \, {=0}$   \\
\hline
\noalign{\vskip 5pt}
Cubic  &   $0.03923$ & $ 0.41911$    &0  & 0.085681 &  $ 0.002453 $   & $B_3$   & 1(3)    & $16, \, 3$ \\
 \noalign{\vskip 5pt}
 & 0.04148 & 0.48136 & 0.00107 & 0.005309 & 0.000378 &  & 2(1,2) & 15,\, 3
  \\
 \noalign{\vskip 5pt}
$S_3$  &   $0.04218$ & $ 0.49592$   & 0.00201  &   $0.003627$ &  0.000668
&  $ O(2)  $    & 2(2,1) &  $14, \, 2$ \\
 \noalign{\vskip 5pt}
$O_3$  &   $\tfrac{175}{3844}$ & $ \tfrac{35}{62}$   & 0  &   0  &  0&  $ O(3)   $ & 1(3)   & $13, \, 0$ \\
\end{tabular}
\end{adjustbox}
\nn
\end{align}
For $N=4$
\begin{align}
&\begin{adjustbox}{max width=\textwidth}
\begin{tabular}{l c c c c c c c c } $N=4$ & $||\lambda||^2$ & $a_0$ & $a_2$ & $a_4$ &
$a_6$ & Symmetry & \makecell{\# different $\gamma$\\[-4pt] and degeneracies} & $ \# {\kappa<0},\,{=0}$ \\ \hline \noalign{\vskip 5pt}
Tetrahedral & 0.049805 &  0.51113 & 0 & 0.019328 & 0.005234 &$\S_5 \times {\mathbb Z}_2 $  & 1(4) & $64, \, 6 $ \\
\noalign{\vskip 5pt}
$S_4$ & 0.057895 & 0.65408 & 0.002867 & 0.014789 & 0.004372 & &  2(1,3)  & $60, \, 6$\\
\noalign{\vskip 5pt}
$S_4$ & 0.058085 & 0.68190 & 0.004683 & 0.013291 & 0.002510 & &  2(1,3)  & $59, \, 6$\\
\noalign{\vskip 5pt}
$S_4$ & 0.061224 & 0.74130& 0 & 0.014043 & 0.001042 & & 3(1,2,1) & $58, \, 6$\\
\noalign{\vskip 5pt}
& 0.061568 & 0.74663 & 0.000438& 0.01338& 0.000738 & &  1(4)  & $52, \, 6$\\
\noalign{\vskip 5pt}
& 0.061631 & 0.74646 & 0.000538 & 0.01327 & 0.001211  & & 3(2,1,1) & $57, \, 6$  \\
\noalign{\vskip 5pt}
& 0.063151 & 0.78865 & 0.001131 & 0.01075 & 0.000490 & & 2(2,2) & $58, \, 5$\\
\noalign{\vskip 5pt}
& 0.063167& 0.79036 & 0.001857 & 0.01014& 0.000647 & & 4(1,1,1,1) & $56, \, 6$\\
\noalign{\vskip 5pt}
Cubic & 0.064443 & 0.79957 & 0 & 0.01039 & 0.001515 & $B_4$ & 1(4) & $55, \, 6$\\
\noalign{\vskip 5pt}
& $\frac{29}{450}$ & $\frac{4}{5}$ & 0 & $\frac{7}{675}$ &
$\frac{1}{675}$ & & 1(4) & $54, \, 6$\\
\noalign{\vskip 5pt}
& 0.06487& 0.82359 & 0.0002390 & 0.009052 & 0.000427 & & 2(2,2) & $55, \, 5$\\
\noalign{\vskip 5pt}
& 0.064911  & 0.82528 & 0.001028& 0.008363 & 0.000671 & & 3(2,1,1) & $54, \, 5$\\
\noalign{\vskip 5pt}
$B_{O_2*O_2}$ & $\frac{365}{5618}$ & $\frac{44}{53}$ & 0 &
$\frac{25}{2809}$ & 0 & $O(2)^2\rtimes\mathbb{Z}_2$ & 1(4) & $54, \, 4$\\
\noalign{\vskip 5pt}
$ S_4$ & 0.065297& 0.82785 & 0.001050 & 0.007989 & 0.001179 & & 2(1,3) & $52, \, 6$\\
\noalign{\vskip 5pt}
& 0.066461 & 0.86738& 0.001292 & 0.005244 & 0.000402 & & 2(2,2) & $51, \, 5$\\
\noalign{\vskip 5pt}
$S_4$& 0.067240 & 0.88811 & 0.001475 & 0.003493 & 0.000425 &
$O(3)$ & 2(3,1) & $50, \, 3$\\
\noalign{\vskip 5pt}
$O_4$ & $\frac{20}{289}$ & $\frac{16}{17}$ & 0 & 0 & 0 & $O(4)$ & 1(4) & $49, \, 0$\\
\noalign{\vskip 5pt}
\end{tabular} \nn
\end{adjustbox}
\end{align}

\begin{figure}[H]
  \centering
  \includegraphics{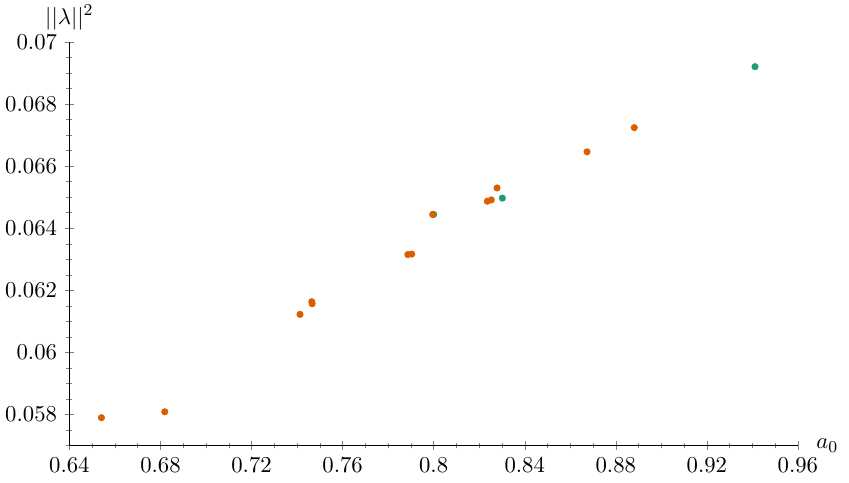}
  \caption{Fully-interacting fixed points for $N=4$ in $d=3-\vep$.}
  \label{fig:N4_d3}
\end{figure}

A pictorial representation of these fixed points (excluding the first one)
is given in Figure \ref{fig:N4_d3}. The distribution of these fixed points
is similar to that of Figure \ref{fig:N4} of the $d=4-\vep$ case.

For $N=5$ the number of solutions explode; those found by us are given in
Appendix \ref{appFPs}.

\section{Conclusion}

The large number of potential fixed points which appear close together in the lowest order
$\vep$ expansion equations for $N=6$ and larger are presumably quite
fragile. Reinstating $\vep $, for two fixed points where $||\lambda_1-\lambda_2|| = \xi \, \vep $ at lowest order
then what happens at higher orders in the $\vep$ expansion for $\vep \sim \xi$  is
far from clear. In many of the examples obtained numerically $\xi  \sim 10^{-3}, 10^{-4}$.
Which fixed points survive in the interesting case when $\vep=1$
is not at all apparent. In most cases,  though perhaps not all, they can be
understood in terms of perturbations of combinations of fixed point
solutions for  lower $N$. None of the solutions obtained here numerically
saturate the bound \eqref{Sbound} when the requirement that the fixed point
is not a combination of decoupled fixed points is imposed.

Nevertheless there exist sporadic non trivial fixed point solutions which saturate the
bound for higher $N$.  These arise for  $N=m n$, $m,n>1$, and there is a
$O(m) \times O(n)/{\mathbb Z}_2$ symmetry with the scalar fields belonging
to the bivector representation. These theories have a large $n$ limit \cite{Pelissetto}.
There are just two couplings. Apart from $m=n=2$, where the fixed
point is only $O_4$, the first real fixed point solution arises  for $m=22, \, n=2$ where there are two fixed points
having $S_+ = \tfrac{6831}{1250}, \, S_- = \tfrac{20328}{3721}$.
When $m,n$ satisfy a certain diophantine equation  the two fixed points present in general  coincide and the bound is saturated.
For this the integer solutions are obtained by $(m_{i+1}, \, n_{i+1})= ( 10 m_i - n_i +4, \,
m_i)$ and $m_1=7, \, n_1=1$ \cite{RychkovS}. For $i  \geq 4$, $m_i \approx
\tfrac34 \,\alpha^i, \ n_i \approx \tfrac34 \,\alpha^{i-1} $, $\alpha = 5 +
\sqrt{24}$.  Whether there are any other similar non trivial solutions
satisfying the bound is unknown.

Unless strong symmetry conditions are imposed then for any $N\ge 6$ the stability matrix will have negative eigenvalues. This
suggests according to standard lore that any phase transition is first order so that the apparent fixed point will not realise a CFT.
Nevertheless the large numbers of solutions suggest that any classification of CFTs when $d=3$ for instance is likely to be very non trivial.

The discussion here is also incomplete in that an implicit assumption made is that the expansion involves only integer
powers of $\vep$ so that the lowest order contributions to the $\beta$-functions determine the leading contribution to the $\vep$ expansion.
This assumption breaks down near a bifurcation point \cite{Seeking}.
However, the analysis  may be   potentially more complicated away from bifurcation points in that in some theories
the $\beta$-function corresponding to one, or more, particular  coupling is zero to lowest order in the loop expansion but is non
zero at the next order.  This situation  can arise in large $N$ limits
where couplings are rescaled by fractional powers of $N$.\footnote{Another
example was recently observed for a $\phi^3$ theory in $6-\vep$ dimensions where the one loop contribution to the
$\beta$-function was zero in~\cite{Gracey}.}
The expansion of the equations for a fixed point then typically involves
$\sqrt{\vep}$. As an illustration, for couplings $h,\,g_a$, $a=1,\dots ,n$,
then if to second order in a loop expansion the $\beta$-functions can be truncated to the form
\begin{align}
\beta_h(h,g) ={}&  k_a g_a \, h  + k\, h^3 +{\rm O}(  g^2  h)  \,,   \nn \\
 \beta_a(h,g) ={}&  b_a(g) + \tfrac12 \, k_a \, h^2 +  c_a (g)h^2   + {\rm O} (g^3 , g^2 h^2)\, ,  \quad b_a(g) = {\rm O}(g^2) , \,  c_a(g) = {\rm O}  (g)  \, ,
\end{align}
assuming that at lowest order $\beta_ h = \pr_h A, \, \beta_a = \pr_a A$ and also $\beta_h(-h,g) =- \beta_h(h,g)$, $\beta_a(-h,g)=  \beta_a(h,g)$.
 Solving $\vep \,h = \beta_h(h,g) , \ \vep\, g_a =  \beta_a(h,g)$ neglecting cubic terms requires $h=0$ unless $g_a$ are
 constrained by $k_a g_a  = \vep$ and then, subject to this constraint, the $n$ equations
  $\vep \, g_a =   b_a(g) + \tfrac12 k_a \, h^2 $  potentially determine $g_a, \,  h={\rm O}(\vep)$,
 with the higher order terms generating the usual perturbative expansion in $\vep$.
 However, if $k_a$ are zero, or can be scaled away in a large $N$ limit, we may take  $h \sim h_0 \, \vep^\frac12, \,
 g_a =  {\rm O} ( \vep)$. The higher order terms then involve an expansion in powers of $\sqrt{\vep}$.
 Similar  scenarios arise in melonic theories \cite{Giombi,Benedetti,Benedetti2}. The graphs relevant for the lowest order $\beta$-function in $\phi^4$ and $\phi^6$
theories may not be sufficient to generate the melonic interactions for fields with three or more indices.

The discussion of fixed points for the  six index coupling undertaken
in section \ref{sec:sixindex} is less complete. All cases considered are such that
 $||\lambda||^2 $ depends linearly on $N$ for large $N$ although there is
 no bound analogous to \eqref{Sbound}.  Perhaps such  a bound might follow
 from a more detailed analysis of the fixed point equations. The $O(N)$
 symmetric  fixed point is no longer the one with maximal  $||\lambda||^2 $ for $N>14$. However, the
 relevance to physical theories is tenuous.

\ack{This investigation started as part of a part III project in Natural Sciences in
Cambridge with Igor Timofeev. His initial
contributions involved  the analysis of $\S_N$ symmetry and finding bounds in the
six index case. We are grateful to the authors of \cite{Codello4} for sending us an
early copy of their work. AS  would like to thank Matthijs
Hogervorst for valuable discussions and also sending a preliminary version of \cite{HogervorstToldo}.

We are very grateful to Slava Rychkov who stimulated much of this work and who
read this paper making many valuable suggestions.

This work has been partially supported by STFC consolidated grants ST/P000681/1, ST/T000694/1.

This research used resources provided by the Los Alamos
National Laboratory Institutional Computing Program, which is supported by
the U.S.\ Department of Energy National Nuclear Security Administration
under Contract No.\ 89233218CNA000001. Research presented in this article
was supported by the Laboratory Directed Research and Development program
of Los Alamos National Laboratory under project number 20180709PRD1.}

\begin{appendices}

\section{Alternative Formulation}

An index free notation due to Michel \cite{Michel3} is convenient in many cases for analysing  the fixed point equation \eqref{RG1}.
 For symmetric four index tensors $u_{ijkl}, \, v_{ijkl}$, scalar products and symmetric tensor products defined by
\be
u\cdot v = u_{ijkl} v_{ijkl} \, , \qquad ( u v )_{ijkl} =  u_{ijmn} v_{mnkl} \, , \quad  (u\vee v)_{ijkl} = ( P_4 \cdot ( u v ))_{ijkl}  \, ,
\ee
for $P_4$ the projector onto symmetric four index tensors. Clearly $|| v||^2 = v\cdot v$ and the triple product
 $(u\vee v)\cdot w = u\cdot (v\vee w)$ is completely symmetric in $u,\, v , \, w$. $(v\vee v)_{iijj} = \tfrac13
 (2   ||v||^2  + v_{ijkk} v_{ijll})$ so that $v\vee v= 0$ implies $v=0$.
 This product is commutative but not associative.
 Note that $(v\vee v)\vee (v\vee v) \neq v \vee (v\vee(v\vee v))$.

The fixed point and eigenvalue equations are then
\be
3\, \lambda \vee \lambda = \lambda \,, \qquad 6\, \lambda \vee v = ( \kappa+1) v \, .
\label{fpk}
\ee
The solutions $\{\kappa_r, \, v_r\}$ can be chosen to form an orthonormal basis so that
\be
v_r \cdot v_s = \delta_{rs}\, , \qquad \kappa_0 = 1 \, , \quad v_0 = \tfrac{1}{|| \lambda ||} \lambda\, .
\ee
In terms of this basis
\be
3\, v_r \vee v_s = {\ts \sum_t} \, C_{rst} \, v_t \, , \quad C_{rst}  = 3\,  (v_r \vee v_s )\cdot v_t = C_{(rst)}\, , \quad
C_{rs0} = \tfrac{1}{2|| \lambda ||} (\kappa_r+1) \delta_{rs} \, .
\ee

Directly from \eqref{fpk}
\be
\kappa + 1 = \tfrac{18}{||v||^2}\, (\lambda\vee \lambda ) \cdot (v\vee v)=  \tfrac{18}{||v||^2}\, (\lambda \lambda) \cdot P_4 \cdot (v v) \ge 0 \, .
\ee
Hence $\kappa \ge -1$.

A proof that $\kappa \le 1$ in general  is not  immediately evident but should follow from bounds on the  product. Using \eqref{fpk} twice
\begin{align}
&\tfrac{1}{36}(1- \kappa^2 ) \, || v||^2 =  ( \lambda \wedge \lambda)\cdot (v \wedge v)- (\lambda \wedge v) \cdot (\lambda\wedge v)  \nn \\
&{} = \tfrac23 \big ( (\lambda \lambda)_{ijkl} (v v)_{ikjl} - (\lambda v )_{ijkl} (\lambda v)_{ikjl} \big ) +
 \tfrac16 \big ( (\lambda \lambda)_{ijkl} (v v)_{klij} - (\lambda v )_{ijkl} (\lambda v)_{klij} \big ) \, .
\label{relK}
\end{align}
For any symmetric matrices $A,B$,  $\tr (A^2 B^2 - ABAB) = \tfrac12 \tr ([A,B][B,A]) \ge 0$ so that the second term is positive.

For $n$ decoupled theories where
\be
 3\, \lambda_r \vee \lambda_s = \delta_{rs}\,   \lambda_r \, , \quad
 \lambda_r \cdot \lambda_s = 0 \, , \ \ r\neq s\, , \qquad r,s=1, \dots n \, ,.
 \label{lrel}
 \ee
 then with
\be
{\ts \sum_r} \, \lambda_r = \lambda , \, \quad v_i  =  {\ts \sum_r} \, x_{i,r}  \lambda_r  \, , \quad
 {\ts \sum_r} \, x_{i,r} \, || \lambda_r ||^2 =0  \, \quad   {\ts \sum_r} \, x_{i,r}x_{j,r}\, || \lambda_r ||^2 = \delta_{ij} \, ,
 \ee
 there are $n-1$ unit eigenvectors $v_i$, $i=1,\dots, n-1$, with $\kappa_i=1$, such that
 \be
3 \, \lambda \vee v_i = v_i \, , \ \ \lambda \cdot v_i = 0 \, , \ \ v_i \cdot v_j = \delta_{ij} \, ,
\quad 3\,  v_i \vee v_j  =  \delta_{ij} \, \tfrac{1}{ ||\lambda ||^2 } \, \lambda
+  {\ts \sum_k} \, C_{ijk}  \, v_k  \, ,
\label{veig}
\ee
where the symmetric tensor  $C_{ijk} = \sum_r \, x_{i,r} x_{j,r} x_{k,r} ||\lambda_r ||^2 $.
Conversely finding  eigenvectors satisfying \eqref{veig} implies the presence of two or more decoupled theories.
For $v_0 = \lambda/||\lambda||$, and extending the index range for $i$ to $0,1,\dots,n-1$ with $x_{0,r}= 1/ ||\lambda ||$,
   then we may define $\lambda_r/||\lambda_r|| = \sum_i \, x_{i,r}\, v_i$ where $ \sum_i x_{i,r}\, x_{i,s} = \delta_{rs} / || \lambda_r||^2$.
Crucially it is necessary to obtain \eqref{lrel} that $C_{ijk} = \sum_r \, x_{i,r} x_{j,r} x_{k,r} ||\lambda_r ||^2  $ where
$C_{0ij} = \delta_{ij}/||\lambda ||$. Such a diagonalisation by essentially orthogonal matrices is not possible
for arbitrary symmetric $C_{ijk}$ but depends on additional restrictions \cite{aberth}.

\section{Potentials and Symmetry Groups for $N=4$}
\label{appV4}

In \cite{Brezin2} various potentials corresponding to subgroups of $O(4)$ were identified. We revisit these
from a different perspective. It is convenient here to adopt complex coordinates where $\vphi= \phi_1+ i \phi_2$,
$\psi = \phi_3 + i \phi_4$.

First
\be
V_{\rm dipentagonal} = \tfrac18 \lambda \big ( | \vphi |^2 + |\psi |^2 \big )^2 + \tfrac18 v \big (  | \vphi |^4 + |\psi |^4 \big )
+ \tfrac{1}{48} i\,  w\big ( \vphi^3 \psi+  \vphi\,  {\bar \psi}^3 -   {\bar \vphi} \, \psi^3  -  {\bar  \vphi }^3 {\bar  \psi} \big ) \, .
\ee
The symmetries which are subgroups of $O(4)$ are generated by
\begin{align}
& a: \ \big ( \vphi, \, {\bar \vphi}, \, \psi, \, {\bar \psi} \big ) \to \big  ( e^{\frac15 \pi \, i} \vphi, \,e^{-\frac15 \pi \, i}  {\bar \vphi}, \, e^{-\frac35 \pi \, i} \psi, \,
e^{\frac35 \pi \, i} {\bar \psi}\big  ) \, , \nn \\
& b: \ \big ( \vphi, \, {\bar \vphi}, \, \psi, \, {\bar \psi} \big ) \to \big  ( i\,{\bar \psi}, \, -i \,{\psi}, \, i\,  \vphi, \, -i\,{\bar  \vphi}\big  ) \, ,
\label{ptran}
\end{align}
where
\be
a^{10} = b^4= e \, ,  \qquad b^{-1} a \hskip 0.5 pt b  = a^3 \, .
\ee
Clearly $a\in {\mathbb Z}_{10}=  {\mathbb Z}_{5}\times  {\mathbb Z}_{2} , \, b\in {\mathbb Z}_{4}$ and $b$ generates the automorphisms of ${\mathbb Z}_{5}$.
The symmetry group is then $( {\mathbb Z}_{5}\rtimes  {\mathbb Z}_{4})\times  {\mathbb Z}_{2}$ with $ {\mathbb Z}_{2}  = \{ e, \, a^5\}$.
For $w =\pm 4 v$ the symmetry is enhanced since
\be
V_{\rm dipentagonal} \big |_{w=4v} = \tfrac18( \lambda+  2v ) \big ( \phi^2)^2  -\tfrac{5}{12} \, v  \, {\ts \sum_{\alpha=1}^5} \big ( \Phi^\alpha)^4 \, ,
\label{pen}
\ee
where
\begin{align}
 \Phi^1 = {}& \tfrac{1}{2 \sqrt{2}}\big ( 1 + \tfrac{1}{\sqrt{5}} \big )\, \phi_1  +\big  ( 5 + \sqrt{5} \big )^{-\frac12} \, \phi_2 +  \tfrac12 \big ( 1 + \tfrac{1}{\sqrt{5}} \big )^{\frac12} \, \phi_3 -
 \tfrac{1}{2 \sqrt{2}} \big ( 1  - \tfrac{1}{\sqrt{5}} \big )\, \phi_4 \, ,  \nn  \\
  \Phi^2 = {}& \tfrac{1}{2 \sqrt{2}}\big ( 1 + \tfrac{1}{\sqrt{5}} \big )\, \phi_1  - \big  ( 5 + \sqrt{5} \big )^{-\frac12} \, \phi_2 -  \tfrac12 \big ( 1 + \tfrac{1}{\sqrt{5}} \big )^{\frac12} \, \phi_3 -
 \tfrac{1}{2 \sqrt{2}} \big ( 1  - \tfrac{1}{\sqrt{5}} \big )\, \phi_4 \, , \nn \\
  \Phi^3 = {}& -\tfrac{1}{2 \sqrt{2}}\big ( 1 - \tfrac{1}{\sqrt{5}} \big )\, \phi_1 - \tfrac12 \big ( 1 + \tfrac{1}{\sqrt{5}} \big )^{\frac12} \, \phi_2  +\big  ( 5 + \sqrt{5} \big )^{-\frac12} \, \phi_3 +
 \tfrac{1}{2 \sqrt{2}} \big ( 1  + \tfrac{1}{\sqrt{5}} \big )\, \phi_4 \, ,\nn  \\
  \Phi^4 = {}&- \tfrac{1}{2 \sqrt{2}}\big ( 1 - \tfrac{1}{\sqrt{5}} \big )\, \phi_1  + \tfrac12\big ( 1+ \tfrac{1}{\sqrt{5}} \big )^{\frac12} \, \phi_2  -\big  ( 5 + \sqrt{5} \big )^{-\frac12} \, \phi_3 +
 \tfrac{1}{2 \sqrt{2}} \big ( 1  + \tfrac{1}{\sqrt{5}} \big )\, \phi_4 \, , \nn\\
  \Phi^5 = {}& - \sqrt{\tfrac25} \big ( \phi_1+\phi_4 \big ) \, ,
\end{align}
which are solutions of \eqref{tetra} for $N=4$.
In terms of \eqref{ptran}
\begin{align}
& a : \ \big ( \Phi^1 ,\,\Phi^2,\, \Phi^3, \, \Phi^4, \, \Phi^5 \big ) \to  \big ( - \Phi^5 ,\, -\Phi^4,\, -\Phi^1, \, -\Phi^3, \,  -\Phi^2 \big )  \, , \nn \\
& b : \ \big ( \Phi^1 ,\,\Phi^2,\, \Phi^3, \, \Phi^4, \, \Phi^5 \big ) \to  \big (  \Phi^3 ,\, \Phi^4,\, \Phi^2, \, \Phi^1 , \,  \Phi^5 \big )  \, ,
\end{align}
but the full symmetry in \eqref{pen} extends to $\S_5 \times {\mathbb Z}_2$, with $a$ corresponding to a 5-cycle combined with a reflection and $b$
a $4$-cycle.

Secondly
\begin{align}
V_{\rm orthotetragonal} = {}&  \tfrac18 \lambda \big ( | \vphi |^2 + |\psi |^2 \big )^2 + \tfrac18 v \big (  | \vphi |^4 + |\psi |^4 \big )
+ \tfrac{1}{48}  w\big ( \vphi^2 \, {\bar \psi}^2 + {\bar  \vphi } ^2 \psi^2 \big ) \nn \\
&{} + \tfrac{1}{48}  i\,  p ( \vphi \, \psi^3+ \vphi  ^3 \psi -  {\bar \vphi}^3  {\bar \psi} -  {\bar  \vphi }\, {\bar  \psi}^3 \big ) \, .
\end{align}
The symmetries are generated by
\begin{align}
& a: \ \big ( \vphi, \, {\bar \vphi}, \, \psi, \, {\bar \psi} \big ) \to \big  ( e^{\frac14 \pi \, i} \vphi, \,e^{-\frac14 \pi \, i}  {\bar \vphi}, \, e^{\frac54 \pi \, i} \psi, \,
e^{-\frac54 \pi \, i} {\bar \psi}\big  ) \, , \nn \\
& b: \ \big ( \vphi, \, {\bar \vphi} , \, \psi, \, {\bar \psi} \big ) \to \big  ( e^{\frac14 \pi \, i}  {\bar \vphi}, \, e^{-\frac14 \pi \, i} \vphi , \, e^{\frac14 \pi \, i}{\bar \psi}, \,  \,
 e^{-\frac14 \pi \, i} \psi \big  ) \, , \nn \\
&  c : \ \big (  \vphi, \, {\bar \vphi}, \, \psi, \, {\bar \psi}  \big )\to \big (\psi, \, {\bar \psi} , \,  \vphi, \, {\bar \vphi}  \big ) \, ,
\end{align}
with
\be
a^8 = b^2 = c^2 = e  \, ,  \qquad  b\hskip 0.5 pt c \hskip 0.5 ptb = c \, , \qquad b\hskip 0.5 pt a\hskip 0.5 pt  b = a^7 \, ,  \qquad c \hskip 0.5 pt a\hskip 0.5 pt c = a^5 \, .
\ee
where  $a\in {\mathbb Z}_{8}, \, b,c\in {\mathbb Z}_{2}$. $b,c$ generate ${\rm Aut}( {\mathbb Z}_{8}) =  {\mathbb Z}_{2}\times  {\mathbb Z}_{2}$ and the
symmetry group is $ {\mathbb Z}_{8} \rtimes ( {\mathbb Z}_{2}\times  {\mathbb Z}_{2}) $.

For a potential containing cubic symmetry
\begin{align}
 V_{\rm trigonal \, cubic} = {}&  \tfrac18 \lambda \big ( \phi^2  \big )^2  + \tfrac{1}{24} v \big (  \phi_1{\!}^4  +   \phi_2{\!}^4  +\phi_3{\!}^4 + \phi_4{\!}^4\big ) \nn \\
& \hskip - 0.5cm {} + \tfrac{1}{24}  \,  w \big (  \phi_1 \phi_2 ( \phi_3{\!}^2 -\phi_4{\!}^2 ) - \phi_2 \phi_3 ( \phi_4{\!}^2 -\phi_1{\!}^2 )  - \phi_3\phi_4 ( \phi_1{\!}^2 -\phi_2{\!}^2 ) -
\phi_4 \phi_1 ( \phi_2{\!}^2 -\phi_3{\!}^2 ) \nn \\
\noalign{\vskip -2pt}
&\qquad {}+ \phi_1 \phi_3 ( \phi_2{\!}^2 - \phi_4{\!}^2 )- \phi_2 \phi_4 ( \phi_3{\!}^2 -\phi_1{\!}^2 ) \big ) \, .
\end{align}
The symmetry group is a subgroup of the permutations and reflections comprising the cubic symmetry group $B_4$ and is generated
by
\begin{align}
& \hskip -0.3cm x: \ ( \phi_1, \phi_2 ,\phi_3, \phi_4) \to ( -\phi_3, \phi_4 ,\phi_1,- \phi_2)  \, , \quad y: \ ( \phi_1, \phi_2 ,\phi_3, \phi_4) \to ( \phi_2, -\phi_1 ,\phi_4, -\phi_3)  \, , \nn \\
& \hskip -0.3cm a: \ ( \phi_1, \phi_2 ,\phi_3, \phi_4) \to ( \phi_1, -\phi_4 , - \phi_2, \phi_3)  \, ,  \quad b: \ ( \phi_1, \phi_2 ,\phi_3, \phi_4) \to ( -\phi_1, -\phi_4 ,\phi_3, -\phi_2) \, ,
\end{align}
where
\begin{align}
&x^4 = y^4= a^3= b^2 = e\, , \quad x^2=y^2 \, , \quad y \hskip 0.5 pt x \hskip 0.5 pty^{-1} = x^{-1} \, , \quad
 b\hskip 0.5 pt a \hskip 0.5 pt  b = a^2 \, , \nn \\
& b\hskip 0.5 pt x \hskip 0.5 pt  b = x^3  \, , \quad b\hskip 0.5 pt y \hskip 0.5 pt  b = x\hskip0.5pt y \, , \quad     a\hskip 0.5 pt x \hskip 0.5 pt a^{-1} = y \, , \quad
 a\hskip 0.5 pt y \hskip 0.5 pt a^{-1} = x  \, .
\end{align}
$x,\, y $  generate the quaternion group ${\rm Q}_8$  and $a,b$, $\S_3 < {\rm Aut}({\rm Q}_8)$ and the symmetry group is then ${\rm Q}_8\rtimes \S_3$.

For the remaining case with diorthorhombic symmetry the symmetry group is obtained by combining two 2-cycles with reflections and can be generated by
\begin{align}
& \hskip -0.3cm  a: \ ( \phi_1, \phi_2 ,\phi_3, \phi_4) \to ( -\phi_2, \phi_1 ,-\phi_4, \phi_3)  \, , \quad r: \ ( \phi_1, \phi_2 ,\phi_3, \phi_4) \to ( -\phi_1, \phi_2 ,-\phi_3, \phi_4)  \, , \nn \\
& \hskip -0.3cm  s: \ ( \phi_1, \phi_2 ,\phi_3, \phi_4) \to ( \phi_4, -\phi_3 , - \phi_2, \phi_1)  \, ,  \quad t: \ ( \phi_1, \phi_2 ,\phi_3, \phi_4) \to ( \phi_1, -\phi_2 ,-\phi_3, \phi_4)  \, ,
\end{align}
where
\be
a^4 =  r^2 = s^2 = t^2= e \, , \quad  r\hskip 0.5 pt a \hskip 0.5 pt  r = a^3 \, ,  \ s\hskip 0.5 pt t = t \hskip 0.5 pt s \, ,  \   s \hskip 0.5 pt a \hskip 0.5 pt  s = a \, , \
s \hskip 0.5 pt  r \hskip 0.2 pt  s = r a^2  \, , \ t \hskip 0.5 pt a \hskip 0.5 pt  t = a^3  \, , \
t \hskip 0.5 pt s \hskip 0.5 pt t = s  \, .
\ee
The symmetry group is then $D_4\rtimes ({\mathbb Z}_2\times {\mathbb Z}_2)$.

\newpage

\begin{landscape}
\section{Results for Six Indices and \texorpdfstring{$N=5$}{N=5}}
\label{appFPs}
\pagestyle{empty}
\setlength\LTcapwidth{\textwidth}
\setlength\LTleft{0pt}
\setlength\LTright{0pt}
\begin{longtable}{l c c c c c c c c c } $N=5$ & $||\lambda||^2$ & $a_0$ &  $a_2$ & $a_4$ &
$a_6$ & Symmetry & \makecell{\# different $\gamma$\\[-4pt] and degeneracies} & $ \# {\kappa<0},\,{=0}$ \\ \hline \noalign{\vskip 5pt}
Tetrahedral & 0.059928  & 0.60697 & 0 & 0.029631 & 0.008195 & $\S_6 \times {\mathbb Z}_2$ &  1(5) &  $180,\, 10$    \\
\noalign{\vskip 5pt}
& 0.086604 & 1.13032 & 0.001295 & 0.020277 & 0.001779 & & 5(1,1,1,1,1) & $161, \, 10$\\
\noalign{\vskip 5pt}
& 0.086786 & 1.13087 & 0.001470 & 0.020024 & 0.002115 & & 5(1,1,1,1,1) & $160, \, 10$\\
\noalign{\vskip 5pt}
& 0.086897 & 1.14497 & 0.002557 & 0.018825 & 0.001587 & & 5(1,1,1,1,1) & $161, \, 10$\\
\noalign{\vskip 5pt}
& 0.086907 & 1.13058 & 0.000499 & 0.020599 & 0.002045 & & 2(1,4) & $159, \, 10$\\
\noalign{\vskip 5pt}
& 0.086979 & 1.14649 & 0.002464 & 0.018787 & 0.001589 & & 5(1,1,1,1,1) & $160, \, 10$\\
\noalign{\vskip 5pt}
& 0.086985 & 1.14824 & 0.003249 & 0.018207 & 0.001715 & & 5(1,1,1,1,1) & $160, \, 10$\\
\noalign{\vskip 5pt}
& 0.087111 & 1.15711 & 0.003759 & 0.017509 & 0.001442 & & 5(1,1,1,1,1) & $159, \, 10$\\
\noalign{\vskip 5pt}
& 0.087490 & 1.16241 & 0.002244 & 0.018096 & 0.001248 & & 5(1,1,1,1,1) & $159, \, 10$\\
\noalign{\vskip 5pt}
& 0.087604 & 1.17008 & 0.002640 & 0.017501 & 0.001016 & & 5(1,1,1,1,1) & $159, \, 10$\\
\noalign{\vskip 5pt}
& 0.087637 & 1.17227 & 0.003587 & 0.016773 & 0.001215 & & 5(1,1,1,1,1) & $158, \, 10$\\
\noalign{\vskip 5pt}
& 0.087754 & 1.16267 & 0.001418 & 0.018462 & 0.001436 & & 4(1,1,1,2) & $159, \, 10$\\
\noalign{\vskip 5pt}
& 0.088058 & 1.17617 & 0.003603 & 0.016345 & 0.001685 & & 5(1,1,1,1,1) & $158, \, 10$\\
\noalign{\vskip 5pt}
& 0.088081 & 1.16953 & 0.000824 & 0.018411 & 0.001328 & & 3(2,2,1) & $159, \, 10$\\
\noalign{\vskip 5pt}
& 0.088285 & 1.17785 & 0.001260 & 0.017697 & 0.001229 & & 5(1,1,1,1,1) & $158, \, 10$\\
\noalign{\vskip 5pt}
& 0.088292 & 1.17337 & 0.000824 & 0.018139 & 0.001425 & & 5(1,1,1,1,1) & $160, \, 10$\\
\noalign{\vskip 5pt}
& 0.088293 & 1.17361 & 0.000863 & 0.018105 & 0.001422 & & 5(1,1,1,1,1) & $159, \, 10$\\
\noalign{\vskip 5pt}
& 0.088304 & 1.17498 & 0.001204 & 0.017822 & 0.001451 & & 5(1,1,1,1,1) & $158, \, 10$\\
\noalign{\vskip 5pt}
& 0.088322 & 1.17327 & 0.000940 & 0.018046 & 0.001522 & & 3(2,2,1) & $159, \, 10$\\
\noalign{\vskip 5pt}
& 0.088362 & 1.17530 & 0.001145 & 0.017813 & 0.001511 & & 5(1,1,1,1,1) & $159, \, 10$\\
\noalign{\vskip 5pt}
& 0.0883660 & 1.17522 & 0.001074 & 0.017860 & 0.001502 & & 5(1,1,1,1,1) & $158, \, 10$\\
\noalign{\vskip 5pt}
& 0.0883661 & 1.17554 & 0.000756 & 0.018060 & 0.001379 & & 4(1,2,1,1) & $158, \, 10$\\
\noalign{\vskip 5pt}
& 0.088368 & 1.17526 & 0.001010 & 0.017899 & 0.001483 & & 4(1,2,1,1) & $157, \, 10$\\
\noalign{\vskip 5pt}
& 0.088373 & 1.17742 & 0.001607 & 0.017423 & 0.001525 & & 5(1,1,1,1,1) & $158, \, 10$\\
\noalign{\vskip 5pt}
& 0.088407 & 1.17794 & 0.000685 & 0.017996 & 0.001256 & & 4(2,1,1,1) & $157, \, 10$\\
\noalign{\vskip 5pt}
& 0.088436 & 1.17880 & 0.000948 & 0.017771 & 0.001330 & & 4(1,2,1,1) & $159, \, 10$\\
\noalign{\vskip 5pt}
& 0.088438 & 1.17952 & 0.001025 & 0.017693 & 0.001306 & & 5(1,1,1,1,1) & $158, \, 10$\\
\noalign{\vskip 5pt}
& 0.088486 & 1.18006 & 0.001868 & 0.017080 & 0.001618 & & 5(1,1,1,1,1) & $158, \, 10$\\
\noalign{\vskip 5pt}
& 0.088489 & 1.18148 & 0.002051 & 0.016905 & 0.001580 & & 5(1,1,1,1,1) & $157, \, 10$\\
\noalign{\vskip 5pt}
& 0.088497 & 1.18007 & 0.001816 & 0.017107 & 0.001620 & & 5(1,1,1,1,1) & $157, \, 10$\\
\noalign{\vskip 5pt}
& 0.088828 & 1.18761 & 0.000364 & 0.017581 & 0.001215 & & 5(1,1,1,1,1) & $158, \, 10$\\
\noalign{\vskip 5pt}
& 0.088834 & 1.18810 & 0.000343 & 0.017573 & 0.001183 & & 5(1,1,1,1,1) & $157, \, 10$\\
\noalign{\vskip 5pt}
& 0.088922 & 1.19024 & 0.000733 & 0.017174 & 0.001311 & & 4(1,1,1,2) & $157, \, 10$\\
\noalign{\vskip 5pt}
& 0.088924 & 1.18914 & 0.000503 & 0.017367 & 0.001321 & & 5(1,1,1,1,1) & $156, \, 10$\\
\noalign{\vskip 5pt}
& 0.089025 & 1.19268 & 0.000427 & 0.017218 & 0.001227 & & 3(1,2,2) & $156, \, 10$\\
\noalign{\vskip 5pt}
& 0.089050 & 1.19471 & 0.000993 & 0.016747 & 0.001308 & & 5(1,1,1,1,1) & $156, \, 10$\\
\noalign{\vskip 5pt}
& 0.089405 & 1.20459 & 0.000711 & 0.016313 & 0.001161 & & 5(1,1,1,1,1) & $155, \, 10$\\
\noalign{\vskip 5pt}
& 0.089658 & 1.21752 & 0.001131 & 0.015335 & 0.000861 & & 5(1,1,1,1,1) & $156, \, 10$\\
\noalign{\vskip 5pt}
& 0.089664 & 1.21837 & 0.001347 & 0.015151 & 0.000883 & & 5(1,1,1,1,1) & $155, \, 10$\\
\noalign{\vskip 5pt}
& 0.089809 & 1.22199 & 0.001082 & 0.015078 & 0.000812 & & 5(1,1,1,1,1) & $156, \, 10$\\
\noalign{\vskip 5pt}
& 0.089812 & 1.22241 & 0.001163 & 0.015003 & 0.000816 & & 5(1,1,1,1,1) & $155, \, 10$\\
\noalign{\vskip 5pt}
& 0.089829 & 1.22319 & 0.001236 & 0.014910 & 0.000816 & & 5(1,1,1,1,1) & $155, \, 10$\\
\noalign{\vskip 5pt}
& 0.089836760 & 1.2236759 & 0.001227 & 0.0148897 & 0.0007946 & & 5(1,1,1,1,1) & $155, \, 10$\\
\noalign{\vskip 5pt}
& 0.089836763 & 1.2236756 & 0.001228 & 0.0148892 & 0.0007949 & & 5(1,1,1,1,1) & $154, \, 10$\\
\noalign{\vskip 5pt}
& 0.089853 & 1.22183 & 0.001188 & 0.014984 & 0.000934 & & 4(1,1,1,2) & $154, \, 10$\\
\noalign{\vskip 5pt}
& 0.089874 & 1.22602 & 0.001784 & 0.014392 & 0.000879 & & 5(1,1,1,1,1) & $153, \, 10$\\
\noalign{\vskip 5pt}
& 0.089891 & 1.22504 & 0.001058 & 0.014907 & 0.000747 & & 4(1,1,1,2) & $153, \, 10$\\
\noalign{\vskip 5pt}
& 0.090085 & 1.22324 & 0.000673 & 0.015112 & 0.001088 & & 4(1,2,1,1) & $154, \, 10$\\
\noalign{\vskip 5pt}
$S_5$ & 0.090280 & 1.20824 & 0.000040 & 0.016030 & 0.002249 & & 2(1,4) & $156, \, 10$\\
\noalign{\vskip 5pt}
Tetrahedral & 0.090282 & 1.20657 & 0 & 0.016123 & 0.002354 &
$\mathcal{S}_6\times\mathbb{Z}_2$ & 1(5) & $155, \, 10$\\
\noalign{\vskip 5pt}
$S_5$ & 0.090331 & 1.22279 & 0.000387 & 0.014936 & 0.001123 &
& 2(1,4) & $155, \, 10$\\
\noalign{\vskip 5pt}
& 0.090333 & 1.22305 & 0.000406 & 0.015133 & 0.001457 & & 4(1,2,1,1) & $155, \, 10$\\
\noalign{\vskip 5pt}
& 0.090335 & 1.22388 & 0.000391 & 0.015106 & 0.001400 & & 4(1,1,1,2) & $155, \, 10$\\
\noalign{\vskip 5pt}
& 0.090336 & 1.22408 & 0.000407 & 0.015087 & 0.001393 & & 5(1,1,1,1,1) & $154, \, 10$\\
\noalign{\vskip 5pt}
& 0.0903408 & 1.22873 & 0.000484 & 0.014829 & 0.001116 & & 4(1,2,1,1) & $154, \, 10$\\
\noalign{\vskip 5pt}
& 0.09034186 & 1.22880 & 0.000474 & 0.014832 & 0.001110 & & 4(1,1,1,2) & $154, \, 10$\\
\noalign{\vskip 5pt}
& 0.09034189 & 1.22881 & 0.000477 & 0.014830 & 0.001111 & & 5(1,1,1,1,1) & $153, \, 10$\\
\noalign{\vskip 5pt}
& 0.09034793 & 1.22925 & 0.000611 & 0.014716 & 0.001134 & & 4(1,1,2,1) & $153, \, 10$\\
\noalign{\vskip 5pt}
& 0.09034799 & 1.22959 & 0.000609 & 0.014703 & 0.001111 & & 4(1,1,2,1) & $152, \, 10$\\
\noalign{\vskip 5pt}
& 0.090349 & 1.22981 & 0.000650 & 0.014665 & 0.001112 & & 3(1,3,1) & $152, \, 10$\\
\noalign{\vskip 5pt}
& 0.090400 & 1.21586 & 0.000240 & 0.015504 & 0.002006 & & 2(4,1) & $154, \, 10$\\
\noalign{\vskip 5pt}
& 0.090433 & 1.23571 & 0.001065 & 0.014069 & 0.001002 & & 3(1,1,3) & $152, \, 10$\\
\noalign{\vskip 5pt}
& 0.0904737 & 1.22328 & 0.000210 & 0.015160 & 0.001628 & & 2(3,2) & $154, \, 10$\\
\noalign{\vskip 5pt}
& 0.0904742 & 1.22430 & 0.000351 & 0.015022 & 0.001605 & & 3(3,1,1) & $154, \, 10$\\
\noalign{\vskip 5pt}
& 0.090482 & 1.22470 & 0.000301 & 0.015033 & 0.001577 & & 4(2,1,1,1) & $153, \, 10$\\
\noalign{\vskip 5pt}
& 0.090507 & 1.22593 & 0.000381 & 0.014909 & 0.001564 & & 4(1,2,1,1) & $153, \, 10$\\
\noalign{\vskip 5pt}
$S_5$ & 0.090534 & 1.24451 & 0.001430 & 0.013355 & 0.000722 & & 2(1,4) & $151, \, 10$\\
\noalign{\vskip 5pt}
& 0.090595883 & 1.235964 & 0.0007011 & 0.0141916 & 0.001159 & & 5(1,1,1,1,1) & $152, \, 10$\\
\noalign{\vskip 5pt}
& 0.090595889 & 1.235956 & 0.0007014 & 0.0141918 & 0.001160 & & 4(1,2,1,1) & $151, \, 10$\\
\noalign{\vskip 5pt}
& 0.090613 & 1.24187 & 0.001398 & 0.013446 & 0.001024 & & 5(1,1,1,1,1) & $153, \, 10$\\
\noalign{\vskip 5pt}
& 0.090707 & 1.23593 & 0.000565 & 0.014210 & 0.001316 & & 3(1,2,2) & $153, \, 10$\\
\noalign{\vskip 5pt}
& 0.090712 & 1.23681 & 0.000661 & 0.014102 & 0.001297 & & 5(1,1,1,1,1) & $152, \, 10$\\
\noalign{\vskip 5pt}
& 0.090760 & 1.24918 & 0.001474 & 0.012955 & 0.000835 & & 5(1,1,1,1,1) & $152, \, 10$\\
\noalign{\vskip 5pt}
& 0.090794 & 1.24541 & 0.001198 & 0.013294 & 0.001051 & & 5(1,1,1,1,1) & $153, \, 10$\\
\noalign{\vskip 5pt}
& 0.090795 & 1.24513 & 0.001188 & 0.013313 & 0.001067 & & 5(1,1,1,1,1) & $152, \, 10$\\
\noalign{\vskip 5pt}
& 0.090804 & 1.24404 & 0.001021 & 0.013469 & 0.001102 & & 5(1,1,1,1,1) & $152, \, 10$\\
\noalign{\vskip 5pt}
& 0.090813 & 1.24536 & 0.001108 & 0.013343 & 0.001060 & & 5(1,1,1,1,1) & $151, \, 10$\\
\noalign{\vskip 5pt}
& 0.090823 & 1.25279 & 0.001705 & 0.012586 & 0.000788 & & 4(1,1,2,1) & $151, \, 10$\\
\noalign{\vskip 5pt}
& 0.090844 & 1.24470 & 0.001067 & 0.013381 & 0.001144 & & 5(1,1,1,1,1) & $152, \, 10$\\
\noalign{\vskip 5pt}
& 0.090849 & 1.24569 & 0.001190 & 0.013249 & 0.001129 & & 5(1,1,1,1,1) & $151, \, 10$\\
\noalign{\vskip 5pt}
& 0.090850 & 1.24582 & 0.001214 & 0.013227 & 0.001129 & & 5(1,1,1,1,1) & $151, \, 10$\\
\noalign{\vskip 5pt}
& 0.090862 & 1.24704 & 0.001341 & 0.013076 & 0.001112 & & 5(1,1,1,1,1) & $150, \, 10$\\
\noalign{\vskip 5pt}
& 0.090904 & 1.25430 & 0.001693 & 0.012467 & 0.000832 & & 5(1,1,1,1,1) & $152, \, 10$\\
\noalign{\vskip 5pt}
& 0.090909 & 1.25468 & 0.001740 & 0.012414 & 0.000831 & & 5(1,1,1,1,1) & $151, \, 10$\\
\noalign{\vskip 5pt}
& 0.090940 & 1.25472 & 0.001923 & 0.012270 & 0.000940 & & 5(1,1,1,1,1) & $151, \, 10$\\
\noalign{\vskip 5pt}
& 0.090944 & 1.25487 & 0.001938 & 0.012250 & 0.000943 & & 5(1,1,1,1,1) & $150, \, 10$\\
\noalign{\vskip 5pt}
& 0.091096 & 1.25570 & 0.002404 & 0.011797 & 0.001307 & & 4(1,1,2,1) & $149, \, 10$\\
\noalign{\vskip 5pt}
& 0.091113 & 1.26449 & 0.003398 & 0.010689 & 0.001095 & & 3(2,2,1) & $151, \, 8$\\
\noalign{\vskip 5pt}
& 0.091432 & 1.26389 & 0.000905 & 0.012169 & 0.000913 & & 3(2,1,2) & $150, \, 9$\\
\noalign{\vskip 5pt}
& 0.091764 & 1.28127 & 0.001267 & 0.010812 & 0.000540 & & 3(2,2,1) & $151, \, 9$\\
\noalign{\vskip 5pt}
& 0.091790 & 1.28346 & 0.001458 & 0.010550 & 0.000512 & & 3(1,2,2) & $150, \, 10$\\
\noalign{\vskip 5pt}
& 0.091908 & 1.28365 & 0.001213 & 0.010625 & 0.000633 & & 5(1,1,1,1,1) & $149, \, 10$\\
\noalign{\vskip 5pt}
& 0.091949 & 1.28351 & 0.001121 & 0.010666 & 0.000684 & & 5(1,1,1,1,1) & $148, \, 10$\\
\noalign{\vskip 5pt}
& 0.092077 & 1.29365 & 0.002008 & 0.009439 & 0.000582 & & 5(1,1,1,1,1) & $147, \, 10$\\
\noalign{\vskip 5pt}
& 0.092297 & 1.29283 & 0.000808 & 0.010138 & 0.000641 & & 4(1,2,1,1) & $148, \, 9$\\
\noalign{\vskip 5pt}
& 0.092329 & 1.29725 & 0.001581 & 0.009357 & 0.000677 & & 5(1,1,1,1,1) & $146, \, 10$\\
\noalign{\vskip 5pt}
& 0.092398 & 1.29378 & 0.000012 & 0.010550 & 0.000512 & & 2(4,1) & $151, \, 8$\\
\noalign{\vskip 5pt}
& 0.092407 & 1.29236 & 0.000034 & 0.010606 & 0.000621 & & 4(1,2,1,1) & $149, \, 9$\\
\noalign{\vskip 5pt}
& 0.0924161 & 1.29074 & 0.000054 & 0.010676 & 0.000739 & & 3(2,1,2) & $147, \, 10$\\
\noalign{\vskip 5pt}
& 0.0924165 & 1.29201 & 0.000007 & 0.010638 & 0.000649 & & 2(2,3) & $148, \, 9$\\
\noalign{\vskip 5pt}
& 0.092432 & 1.29016 & 0.000042 & 0.010705 & 0.000799 & & 3(1,3,1) & $146, \, 10$\\
\noalign{\vskip 5pt}
Cubic & 0.092476 & 1.28936 & 0 & 0.010747 & 0.000911 & $B_5$ & 1(5) & $145, \, 10$\\
\noalign{\vskip 5pt}
& 0.092925 & 1.31105 & 0.000589 & 0.008836 & 0.000612 & & 2(1,4) & $145, \, 10$\\
\noalign{\vskip 5pt}
& 0.092939 & 1.30896 & 0.000614 & 0.008931 & 0.000766 & & 2(1,4) & $144, \, 10$\\
\noalign{\vskip 5pt}
& 0.092986 & 1.32144 & 0.000475 & 0.008258 & 0.000090 & & 2(1,4) & $146, \, 8$\\
\noalign{\vskip 5pt}
& 0.093047 & 1.32271 & 0.000643 & 0.008029 & 0.000178 & & 3(2,1,2) & $146, \, 8$\\
\noalign{\vskip 5pt}
& 0.093258 & 1.32803 & 0.001126 & 0.007244 & 0.000404 & & 4(2,1,1,1) & $144, \, 9$\\
\noalign{\vskip 5pt}
& 0.093473 & 1.33092 & 0.001014 & 0.006999 & 0.000587 & & 2(2,3) & $143, \, 9$\\
\noalign{\vskip 5pt}
& 0.094124 & 1.35768 & 0.001065 & 0.004824 & 0.000298 & & 2(3,2) & $142, \, 7$\\
\noalign{\vskip 5pt}
$S_5$ & 0.094738 & 1.37896 & 0.000936 & 0.003056 & 0.000237 & $O(4)$ & 2(4,1) & $141, \, 4$\\
\noalign{\vskip 5pt}
$O_5$ & $\frac{525}{5476}$ & $\frac{105}{74}$ & 0 & 0 & 0 & $O(5)$ & 1(5) & $140, \, 0$
\end{longtable}
\end{landscape}

\begin{figure}[H]
  \centering
  \includegraphics{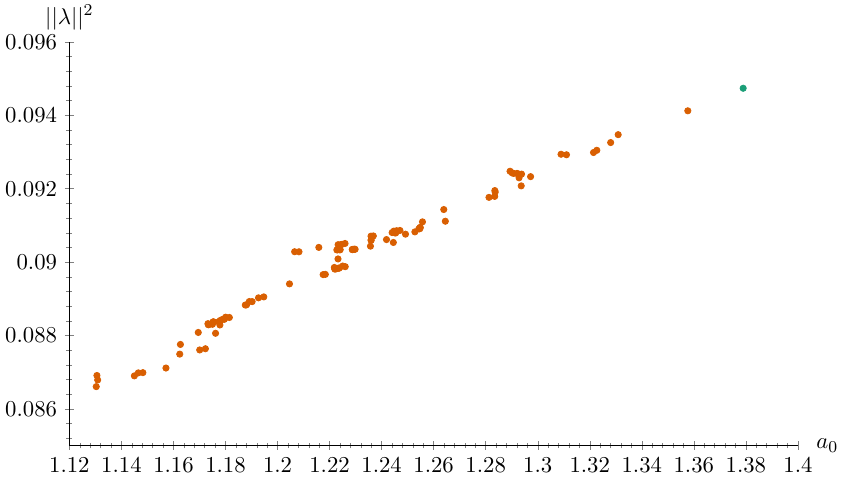}
  \caption{Fully-interacting fixed points for $N=5$ in $d=3-\vep$.}
  \label{fig:N5_d3}
\end{figure}

A pictorial representation of these fixed points (excluding the first one)
is given in Figure \ref{fig:N5_d3}. The distribution of fixed points in
$d=3-\vep$ is similar to that of Figure \ref{fig:N4} of the $d=4-\vep$
case.

\end{appendices}

\bibliographystyle{utphys}
\bibliography{Phi}{}

\providecommand{\href}[2]{#2}\begingroup\raggedright\begin{thebibliography}{10}

\bibitem{RychkovS}
S.~Rychkov and A.~Stergiou, ``{General Properties of Multiscalar RG Flows in
  $d=4-\varepsilon$},''
  \href{http://dx.doi.org/10.21468/SciPostPhys.6.1.008}{{\em SciPost Phys.}
  {\bfseries 6} no.~1, (2019) 008},
  \href{http://arxiv.org/abs/1810.10541}{{\ttfamily arXiv:1810.10541
  [hep-th]}}.

\bibitem{RGrev}
A.~Pelissetto and E.~Vicari, ``{Critical phenomena and renormalization group
  theory},'' \href{http://dx.doi.org/10.1016/S0370-1573(02)00219-3}{{\em Phys.
  Rept.} {\bfseries 368} (2002) 549--727},
\href{http://arxiv.org/abs/cond-mat/0012164}{{\ttfamily arXiv:cond-mat/0012164
  [cond-mat]}}.

\bibitem{Seeking}
H.~Osborn and A.~Stergiou, ``{Seeking fixed points in multiple coupling scalar
  theories in the $\epsilon$ expansion},''
  \href{http://dx.doi.org/10.1007/JHEP05(2018)051}{{\em JHEP} {\bfseries 05}
  (2018) 051},
\href{http://arxiv.org/abs/1707.06165}{{\ttfamily arXiv:1707.06165 [hep-th]}}.

\bibitem{Schnetz}
O.~Schnetz, ``{Numbers and Functions in Quantum Field Theory},''
  \href{http://dx.doi.org/10.1103/PhysRevD.97.085018}{{\em Phys. Rev. D}
  {\bfseries 97} no.~8, (2018) 085018},
  \href{http://arxiv.org/abs/1606.08598}{{\ttfamily arXiv:1606.08598
  [hep-th]}}.

\bibitem{Ryttov}
T.~A. Ryttov, ``{Properties of the $\epsilon$-expansion, Lagrange inversion and
  associahedra and the $O(1)$ model},''
  \href{http://dx.doi.org/10.1007/JHEP04(2020)072}{{\em JHEP} {\bfseries 04}
  (2020) 072}, \href{http://arxiv.org/abs/1910.12631}{{\ttfamily
  arXiv:1910.12631 [hep-th]}}.

\bibitem{SixLoop}
M.~V. Kompaniets and E.~Panzer, ``{Minimally subtracted six loop
  renormalization of $O(n)$-symmetric $\phi^4$ theory and critical
  exponents},'' \href{http://dx.doi.org/10.1103/PhysRevD.96.036016}{{\em Phys.
  Rev. D} {\bfseries 96} no.~3, (2017) 036016},
  \href{http://arxiv.org/abs/1705.06483}{{\ttfamily arXiv:1705.06483
  [hep-th]}}.

\bibitem{SixLoop2}
M.~Kompaniets, A.~Kudlis, and A.~Sokolov, ``{Six-loop $\epsilon$ expansion
  study of three-dimensional $O(n)\times O(m)$ spin models},''
  \href{http://dx.doi.org/10.1016/j.nuclphysb.2019.114874}{{\em Nucl. Phys. B}
  {\bfseries 950} (2020) 114874},
  \href{http://arxiv.org/abs/1911.01091}{{\ttfamily arXiv:1911.01091
  [cond-mat.stat-mech]}}.

\bibitem{SixLoop3}
L.~Adzhemyan, E.~Ivanova, M.~Kompaniets, A.~Kudlis, and A.~Sokolov, ``{Six-loop
  $\varepsilon$ expansion study of three-dimensional $n$-vector model with
  cubic anisotropy},''
  \href{http://dx.doi.org/10.1016/j.nuclphysb.2019.02.001}{{\em Nucl. Phys. B}
  {\bfseries 940} (2019) 332--350},
  \href{http://arxiv.org/abs/1901.02754}{{\ttfamily arXiv:1901.02754
  [cond-mat.stat-mech]}}.

\bibitem{Bootstrap}
F.~Kos, D.~Poland, D.~Simmons-Duffin, and A.~Vichi, ``{Precision Islands in the
  Ising and $O(N)$ Models},''
  \href{http://dx.doi.org/10.1007/JHEP08(2016)036}{{\em JHEP} {\bfseries 08}
  (2016) 036}, \href{http://arxiv.org/abs/1603.04436}{{\ttfamily
  arXiv:1603.04436 [hep-th]}}.

\bibitem{Stergiou:2019dcv}
A.~Stergiou, ``{Bootstrapping MN and Tetragonal CFTs in Three Dimensions},''
  \href{http://dx.doi.org/10.21468/SciPostPhys.7.1.010}{{\em SciPost Phys.}
  {\bfseries 7} (2019) 010}, \href{http://arxiv.org/abs/1904.00017}{{\ttfamily
  arXiv:1904.00017 [hep-th]}}.

\bibitem{Henriksson:2020fqi}
J.~Henriksson, S.~R. Kousvos, and A.~Stergiou, ``{Analytic and Numerical
  Bootstrap of CFTs with $O(m)\times O(n)$ Global Symmetry in 3D},''
  \href{http://dx.doi.org/10.21468/SciPostPhys.9.3.035}{{\em SciPost Phys.}
  {\bfseries 9} (2020) 035}, \href{http://arxiv.org/abs/2004.14388}{{\ttfamily
  arXiv:2004.14388 [hep-th]}}.

\bibitem{Gufan}
Y.~M. Gufan and V.~Sakhnenko, ``Features of phase transitions associated with
  two-and three-component order parameters,'' {\em Soviet Physics JETP}
  {\bfseries 36} (1973) 1009--1014.
  \url{{http://jetp.ac.ru/cgi-bin/dn/e_036_05_1009.pdf}}. Zh. Eksp. Teor. Fiz,
  63, 1909-1918 (1972).

\bibitem{ZiaW2}
R.~K.~P. Zia and D.~J. Wallace, ``{On the Uniqueness of $\phi^4$ Interactions
  in Two and Three-Component Spin Systems},''
\href{http://dx.doi.org/10.1088/0305-4470/8/7/012}{{\em J. Phys.} {\bfseries
  A8} (1975) 1089--1096}.

\bibitem{Safari}
A.~Codello, M.~Safari, G.~Vacca, and O.~Zanusso, ``{Epsilon-expansion for
  multi-scalar QFTs},''.
  \url{{https://indico.ectstar.eu/event/50/contributions/1514/attachments/1094/1413/safari-frgim-2019.pdf}}.
  {Talk by M. Safari at Functional and Renormalization Group Methods, Trento,
  September 20 2019}.

\bibitem{Codello4}
A.~Codello, M.~Safari, G.~Vacca, and O.~Zanusso, ``{Critical models with $N
  \leq $4 scalars in $d=4-\epsilon$},''
  \href{http://dx.doi.org/10.1103/PhysRevD.102.065017}{{\em Phys. Rev. D}
  {\bfseries 102} no.~6, (2020) 065017},
  \href{http://arxiv.org/abs/2008.04077}{{\ttfamily arXiv:2008.04077
  [hep-th]}}.

\bibitem{Michel}
L.~Michel, ``{The Symmetry and Renormalization Group Fixed Points of Quartic
  Hamiltonians},'' in {\em Symmetries in Particle Physics, Proceedings of a
  symposium celebrating Feza Gursey's sixtieth birthday}, B.~Bars, A.~Chodos,
  and C.-H. Tze, eds., pp.~63--92.
\newblock Plenum Press, 1984.

\bibitem{Michel3}
L.~Michel, ``Renormalization-group fixed points of general $n$-vector models,''
  \href{http://dx.doi.org/10.1103/PhysRevB.29.2777}{{\em Phys. Rev. B}
  {\bfseries 29} (1984) 2777--2783}.

\bibitem{MichelT}
L.~Michel and J.-C. Toledano, ``Symmetry criterion for the lack of a stable
  fixed point in the renormalization-group recursion relations,''
  \href{http://dx.doi.org/10.1103/PhysRevLett.54.1832}{{\em Phys. Rev. Lett.}
  {\bfseries 54} (Apr, 1985) 1832--1835}.

\bibitem{VicariZ}
E.~Vicari and J.~Zinn-Justin, ``{Fixed point stability and decay of
  correlations},'' \href{http://dx.doi.org/10.1088/1367-2630/8/12/321}{{\em New
  J. Phys.} {\bfseries 8} (2006) 321},
\href{http://arxiv.org/abs/cond-mat/0611353}{{\ttfamily arXiv:cond-mat/0611353
  [cond-mat]}}.

\bibitem{Hogervorst}
M.~Hogervorst, ``{Bounds on the epsilon expansion},''.
  \url{{https://www.sissa.it/tpp/activity/string/slides/900027.pdf}}.
  {ICTP/SISSA seminar, September 18 2019}.

\bibitem{HogervorstToldo}
M.~Hogervorst and C.~Toldo, ``{Bounds on multiscalar CFTs in the $\varepsilon$
  expansion},''. To appear.

\bibitem{WallaceG}
D.~J. Wallace and R.~K.~P. Zia, ``{Gradient Properties of the Renormalization
  Group Equations in Multicomponent Systems},''
\href{http://dx.doi.org/10.1016/0003-4916(75)90267-5}{{\em Annals Phys.}
  {\bfseries 92} (1975) 142}.

\bibitem{Michel2}
L.~Michel, J.-C. Toledano, and P.~Toledano, ``{Landau free energies for $n=4$
  and the subgroups of $O(4)$},'' in {\em Symmetries and Broken Symmetries in
  Condensed Matter Physics}, N.~Boccara, ed., pp.~261--274.
\newblock John Wiley \& Sons, Ltd, 1981.

\bibitem{Brezin2}
J.-C. Toledano, L.~Michel, P.~Toledano, and E.~Brezin, ``Renormalization-group
  study of the fixed points and of their stability for phase transitions with
  four-component order parameters,''
  \href{http://dx.doi.org/10.1103/PhysRevB.31.7171}{{\em Phys. Rev. B}
  {\bfseries 31} (1985) 7171--7196}.

\bibitem{Hatch}
D.~M. Hatch, H.~T. Stokes, J.~S. Kim, and J.~W. Felix, ``{Selection of stable
  fixed points by the Toledano-Michel symmetry criterion: Six-component
  example},'' \href{http://dx.doi.org/10.1103/PhysRevB.32.7624}{{\em Phys. Rev.
  B} {\bfseries 32} (1985) 7624--7627}.

\bibitem{Hatch1}
J.~S. Kim, D.~M. Hatch, and H.~T. Stokes, ``{Classification of continuous phase
  transitions and stable phases. I. Six-dimensional order parameters},''
  \href{http://dx.doi.org/10.1103/PhysRevB.33.1774}{{\em Phys. Rev. B}
  {\bfseries 33} (1986) 1774--1788}.

\bibitem{Hatch2}
D.~M. Hatch, J.~S. Kim, H.~T. Stokes, and J.~W. Felix, ``Renormalization-group
  classification of continuous structural phase transitions induced by
  six-component order parameters,''
  \href{http://dx.doi.org/10.1103/PhysRevB.33.6196}{{\em Phys. Rev. B}
  {\bfseries 33} (1986) 6196--6209}.

\bibitem{Platonic}
R.~B.~A. Zinati, A.~Codello, and G.~Gori, ``{Platonic Field Theories},''
  \href{http://dx.doi.org/10.1007/JHEP04(2019)152}{{\em JHEP} {\bfseries 04}
  (2019) 152}, \href{http://arxiv.org/abs/1902.05328}{{\ttfamily
  arXiv:1902.05328 [hep-th]}}.

\bibitem{ipopt}
A.~W{\"a}chter and L.~Biegler, ``On the implementation of an interior-point
  filter line-search algorithm for large-scale nonlinear programming,''
  \href{http://dx.doi.org/10.1007/s10107-004-0559-y}{{\em Mathematical
  Programming} {\bfseries 106} (2006) 25--57}.

\bibitem{pagmo}
F.~Biscani and D.~Izzo, ``A parallel global multiobjective framework for
  optimization: pagmo,'' \href{http://dx.doi.org/10.21105/joss.02338}{{\em
  Journal of Open Source Software} {\bfseries 5} no.~53, (2020) 2338}.

\bibitem{Komar}
N.~Chai, S.~Chaudhuri, C.~Choi, Z.~Komargodski, E.~Rabinovici, and M.~Smolkin,
  ``{Thermal Order in Conformal Theories},''
  \href{http://arxiv.org/abs/2005.03676}{{\ttfamily arXiv:2005.03676
  [hep-th]}}.

\bibitem{Biconical}
D.~R. Nelson, J.~M. Kosterlitz, and M.~E. Fisher, ``Renormalization-group
  analysis of bicritical and tetracritical points,''
  \href{http://dx.doi.org/10.1103/PhysRevLett.33.813}{{\em Phys. Rev. Lett.}
  {\bfseries 33} (1974) 813--817}.

\bibitem{VicariB}
P.~Calabrese, A.~Pelissetto, and E.~Vicari, ``{Multicritical phenomena in
  $O(n_1) \oplus O(n_2)$ symmetric theories},''
  \href{http://dx.doi.org/10.1103/PhysRevB.67.054505}{{\em Phys. Rev.}
  {\bfseries B67} (2003) 054505},
\href{http://arxiv.org/abs/cond-mat/0209580}{{\ttfamily
  arXiv:cond-mat/0209580}}.

\bibitem{Folk1}
R.~Folk, Y.~Holovatch, and G.~Moser, ``Field theory of bicritical and
  tetracritical points. i. statics,''
  \href{http://dx.doi.org/10.1103/PhysRevE.78.041124}{{\em Phys. Rev. E}
  {\bfseries 78} (Oct, 2008) 041124},
  \href{http://arxiv.org/abs/0808.0314}{{\ttfamily arXiv:0808.0314
  [cond-mat.stat-mech]}}.

\bibitem{Folk2}
R.~Folk, Y.~Holovatch, and G.~Moser, ``Field theory of bicritical and
  tetracritical points. ii. relaxational dynamics,''
  \href{http://dx.doi.org/10.1103/PhysRevE.78.041125}{{\em Phys. Rev. E}
  {\bfseries 78} (Oct, 2008) 041125},
  \href{http://arxiv.org/abs/0809.3146}{{\ttfamily arXiv:0809.3146
  [cond-mat.stat-mech]}}.

\bibitem{Folk3}
R.~Folk, Y.~Holovatch, and G.~Moser, ``{Field Theoretical Approach to
  Bicritical and Tetracritical Behavior: Static and Dynamics},'' {\em J. Phys.
  Stud.} {\bfseries 13} (2009) 4003--4003,
  \href{http://arxiv.org/abs/0906.3624}{{\ttfamily arXiv:0906.3624
  [cond-mat.stat-mech]}}.

\bibitem{Antipin}
O.~Antipin and J.~Bersini, ``{Spectrum of anomalous dimensions in hypercubic
  theories},'' \href{http://dx.doi.org/10.1103/PhysRevD.100.065008}{{\em Phys.
  Rev. D} {\bfseries 100} no.~6, (2019) 065008},
  \href{http://arxiv.org/abs/1903.04950}{{\ttfamily arXiv:1903.04950
  [hep-th]}}.

\bibitem{3coupling1}
D.~Mukamel and S.~Krinsky, ``{Physical realizations of
  $n\ensuremath{\ge}4$-component vector models. II.
  $\ensuremath{\epsilon}$-expansion analysis of the critical behavior},''
  \href{http://dx.doi.org/10.1103/PhysRevB.13.5078}{{\em Phys. Rev. B}
  {\bfseries 13} (1976) 5078--5085}.

\bibitem{MudrovV1}
A.~I. Mudrov and K.~B. Varnashev, ``{Three-loop renormalization-group analysis
  of a complex model with stable fixed point: Critical exponents up to
  ${\ensuremath{\epsilon}}^{3}$ and ${\ensuremath{\epsilon}}^{4}$},''
  \href{http://dx.doi.org/10.1103/PhysRevB.57.3562}{{\em Phys. Rev. B}
  {\bfseries 57} (1998) 3562--3576}.

\bibitem{MudrovV2}
A.~I. Mudrov and K.~B. Varnashev, ``{Stability of the three-dimensional fixed
  point in a model with three coupling constants from the
  $\ensuremath{\epsilon}$ expansion: Three-loop results},''
  \href{http://dx.doi.org/10.1103/PhysRevB.57.5704}{{\em Phys. Rev. B}
  {\bfseries 57} (1998) 5704--5710}.

\bibitem{Eichhorn}
A.~Eichhorn, T.~Helfer, D.~Mesterh{\'a}zy, and M.~M. Scherer, ``{Discovering
  and quantifying nontrivial fixed points in multi-field models},''
  \href{http://dx.doi.org/10.1140/epjc/s10052-016-3921-3}{{\em Eur. Phys. J. C}
  {\bfseries 76} no.~2, (2016) 88},
  \href{http://arxiv.org/abs/1510.04807}{{\ttfamily arXiv:1510.04807
  [cond-mat.stat-mech]}}.

\bibitem{duval}
P.~Du~Val, {\em Homographies, Quaternions, and Rotations.}
\newblock Oxford mathematical monographs. Clarendon Press, Oxford, 1964.

\bibitem{Conway}
J.~H. Conway and S.~D. A., {\em On Quaternions and Octonions}.
\newblock A.K. Peters, 2003.

\bibitem{Farrill}
P.~de~Medeiros and J.~Figueroa-O'Farrill, ``{Half-BPS M2-brane Orbifolds},''
  \href{http://dx.doi.org/10.4310/ATMP.2012.v16.n5.a1}{{\em Adv. Theor. Math.
  Phys.} {\bfseries 16} no.~5, (2012) 1349--1408},
  \href{http://arxiv.org/abs/1007.4761}{{\ttfamily arXiv:1007.4761 [hep-th]}}.

\bibitem{Pelissetto}
A.~Pelissetto, P.~Rossi, and E.~Vicari, ``{Large $n$ critical behavior of $O(n)
  \times O(m)$ spin models},''
  \href{http://dx.doi.org/10.1016/S0550-3213(01)00223-1}{{\em Nucl. Phys.}
  {\bfseries B607} (2001) 605--634},
\href{http://arxiv.org/abs/hep-th/0104024}{{\ttfamily arXiv:hep-th/0104024
  [hep-th]}}.

\bibitem{Gracey}
J.~Gracey, ``{Asymptotic freedom from the two-loop term of the $\beta$ function
  in a cubic theory},''
  \href{http://dx.doi.org/10.1103/PhysRevD.101.125022}{{\em Phys. Rev. D}
  {\bfseries 101} no.~12, (2020) 125022},
  \href{http://arxiv.org/abs/2004.14208}{{\ttfamily arXiv:2004.14208
  [hep-th]}}.

\bibitem{Giombi}
S.~Giombi, I.~R. Klebanov, and G.~Tarnopolsky, ``{Bosonic Tensor Models at
  Large $N$ and Small $\epsilon$},''
\href{http://arxiv.org/abs/1707.03866}{{\ttfamily arXiv:1707.03866 [hep-th]}}.

\bibitem{Benedetti}
D.~Benedetti, N.~Delporte, S.~Harribey, and R.~Sinha, ``{Sextic tensor field
  theories in rank $3$ and $5$},''
  \href{http://dx.doi.org/10.1007/JHEP06(2020)065}{{\em JHEP} {\bfseries 06}
  (2020) 065}, \href{http://arxiv.org/abs/1912.06641}{{\ttfamily
  arXiv:1912.06641 [hep-th]}}.

\bibitem{Benedetti2}
D.~Benedetti, R.~Gurau, and S.~Harribey, ``{Trifundamental quartic model},''
  \href{http://dx.doi.org/10.1103/PhysRevD.103.046018}{{\em Phys. Rev. D}
  {\bfseries 103} no.~4, (2021) 046018},
  \href{http://arxiv.org/abs/2011.11276}{{\ttfamily arXiv:2011.11276
  [hep-th]}}.

\bibitem{aberth}
O.~Aberth, ``The transformation of tensors into diagonal form,''
  \href{http://dx.doi.org/10.1137/0115106}{{\em SIAM Journal on Applied
  Mathematics} {\bfseries 15} no.~5, (1967) 1247--1252},
  \href{http://arxiv.org/abs/https://doi.org/10.1137/0115106}{{\ttfamily
  https://doi.org/10.1137/0115106}}. \url{https://doi.org/10.1137/0115106}.

\end{thebibliography}\endgroup

\end{document}